\documentclass[twocolumn]{aastex631}
\usepackage{hhline}
\usepackage{enumitem}

\newcommand\astrosat{\textit{AstroSat}}
\newcommand\nustar{\textit{NuSTAR}}
\newcommand\nicer{\textit{NICER}}

\newcommand\s{{\rm~s}}
\newcommand\ks{{\rm~ks}}

\newcommand\kev{{\rm~keV}}

\newcommand\ev{{\rm~eV}}

\newcommand\angstrom{{\rm~\AA}}

\begin{document}

\title{A multi-wavelength study of the hard and soft states of MAXI~J1820+070 during its 2018 outburst}
\correspondingauthor{Srimanta Banerjee}
\email{srimanta.banerjee4@gmail.com}

\author[0000-0002-6051-6928]{Srimanta Banerjee}
\affiliation{Inter-University Centre for Astronomy and Astrophysics (IUCAA), PB No.4, Ganeshkhind, Pune-411007, India}

\author[0000-0003-1589-2075]{Gulab C. Dewangan}
\affiliation{Inter-University Centre for Astronomy and Astrophysics (IUCAA), PB No.4, Ganeshkhind, Pune-411007, India}

\author{Christian Knigge}
\affiliation{School of Physics and Astronomy, University of Southampton, Highfield, Southampton SO17 1BJ, United Kingdom}

\author{Maria Georganti}
\affiliation{School of Physics and Astronomy, University of Southampton, Highfield, Southampton SO17 1BJ, United Kingdom}

\author[0000-0003-3105-2615]{Poshak Gandhi}
\affiliation{School of Physics and Astronomy, University of Southampton, Highfield, Southampton SO17 1BJ, United Kingdom}

\author[0000-0003-3431-6110]{N. P. S. Mithun}
\affiliation{Physical Research Laboratory, Navrangpura, Ahmedabad, Gujarat-380 009, India}

\author[0000-0002-5319-6620]{Payaswini Saikia}
\affiliation{Center for Astrophysics and Space Science (CASS), New York University Abu Dhabi, PO Box 129188, Abu Dhabi, UAE}
\affiliation{New York University Abu Dhabi, PO Box 129188, Abu Dhabi, United Arab Emirates}

\author{Dipankar Bhattacharya}
\affiliation{Ashoka University, Department of Physics, Sonepat, Haryana, 131029, India}

\author[0000-0002-3500-631X]{David M. Russell}
\affiliation{Center for Astrophysics and Space Science (CASS), New York University Abu Dhabi, PO Box 129188, Abu Dhabi, UAE}
\affiliation{New York University Abu Dhabi, PO Box 129188, Abu Dhabi, United Arab Emirates}


\author[0000-0003-3352-2334]{F. Lewis}
\affiliation{Faulkes Telescope Project, School of Physics and Astronomy, Cardiff University, The Parade, Cardiff, CF24 3AA, Wales, UK}
\affiliation{Astrophysics Research Institute, Liverpool John Moores University, 146 Brownlow Hill, Liverpool L3 5RF, UK}

\author[0000-0002-0333-2452]{Andrzej A. Zdziarski}
\affiliation{Nicolaus Copernicus Astronomical Center, Polish Academy of Sciences, Bartycka 18, PL-00-716 Warszawa, Poland}

\begin{abstract}
We present a comprehensive multi-wavelength spectral analysis of the black hole X-ray binary MAXI~J1820+070 during its 2018 outburst, utilizing {\it AstroSat} far UV, soft and hard X-ray data, along with (quasi-)simultaneous optical and X-ray data from \textit{Las Cumbres Observatory} and \textit{NICER}, respectively. In the soft state, we detect soft X-ray and UV/optical excess components over and above the intrinsic accretion disk emission ($kT_{\rm in}\sim 0.58$ keV) and a steep X-ray power-law component. The soft X-ray excess is consistent with a high-temperature blackbody ($kT\sim 0.79$ keV), while the UV/optical excess is described by UV emission lines and two low-temperature blackbody components ($kT\sim 3.87$ eV and $\sim 0.75$ eV). Employing continuum spectral fitting, we determine the black hole spin parameter ($a=0.77\pm0.21$), using the jet inclination angle of $64^{\circ}\pm5^{\circ}$ and a mass spanning $5-10M_{\sun}$. In the hard state, we observe a significantly enhanced optical/UV excess component, indicating a stronger reprocessed emission in the outer disk. 
Broad-band X-ray spectroscopy in the hard state reveals a two-component corona, each associated with its reflection component, in addition to the disk emission ($kT_{\rm in}\sim 0.19$ keV). The softer coronal component dominates the bolometric X-ray luminosity and produces broader relativistic reflection features, while the harder component gets reflected far from the inner disk, yielding narrow reflection features. Furthermore, our analysis in the hard state suggests a substantial truncation of the inner disk ($\gtrsim 51$ gravitational radii) and a high disk density ($\sim 10^{20}\ \rm cm^{-3}$).
\end{abstract}



\section{Introduction} \label{sec:intro}
A low-mass X-ray binary (LMXB) containing a black hole (BH) is 
a binary stellar system in which the BH accretes matter from a low-mass companion star ($\lesssim 1 M_{\sun}$) via the Roche-lobe overflow. Most of the Galactic BH-LMXBs are of transient nature, spending most of their lifetime in a faint quiescent state.
This inactive period is sporadically interrupted by short outbursts (usually lasting for weeks to months), in which the accretion rate increases by several orders of magnitude.
During an outburst, a BH-LMXB often evolves through a sequence of different X-ray spectral states: Hard State (HS), Soft State (SS) and Hard/Soft Intermediate state (HIMS/SIMS) \citep{Belloni2010,remillard2006}, displaying a `q' shaped track in the  hardness-intensity diagram (HID) \citep{homan2001,homan2005}. These spectral states differ in their spectral and timing properties, and are possibly due to a change in the accretion geometry \citep{belloni2016}.

A typical outburst of BH-LMXBs generally starts and ends in the HS. The HS X-ray spectrum is dominated by an optically thin component exhibiting roughly a power-law shape (with a photon index of $1.4<\Gamma<2.1$), 
from a few keV to several hundred keV, followed by an exponential cutoff \citep{remillard2006}. Although the accretion geometry of the HS is highly debated, this component is thought to arise due to the Compton up-scattering of soft seed photons by a hot optically thin electron cloud (`corona') located somewhere in the vicinity of the compact object \citep{sunyaev1980,done2007,gilfanov2010,banerjee2020}. Additionally, there could be a weak contribution to the X-ray spectrum from an optically thick and geometrically thin accretion disk with low temperature, with a typical temperature of $\sim0.2$ keV \citep{reis2009,basak2016,zhang2020}. The lower disk temperature generally suggests that the disk is truncated far from the innermost stable circular orbit (ISCO) \citep{zdziarski2021,done2007,gilfanov2010} (however, for other scenarios, see \citealt{reis2010,kara2019}). After starting in the HS, a source usually transitions to the SS, and returns to the HS towards the end of the outburst. The SS spectrum is described primarily by a multi-temperature black-body component originating in the accretion disk \citep{shakura1973}, often accompanied by a weak and steep ($\Gamma\geq 2.1$) power-law  tail extending out beyond 500 keV \citep{belloni2016}.
The black-body component typically peaks $\sim 1$ keV, and the disk, contrary to the HS case, reaches close to the ISCO radius \citep{done2007,gilfanov2010}. 
While evolving from the HS to SS (or SS to HS), a BH-LMXB goes through HIMS and SIMS (or SIMS and HIMS) successively. The soft thermal disk component becomes more dominant (accompanied by a decrease in power-law flux) in these intermediate states than the HS. 

Apart from the hard Comptonized and soft thermal components, the X-ray spectrum of BH-LMXBs also exhibits reflection features, generated as a fraction of Compton up-scattered photons gets reprocessed in the inner accretion disk. These features mainly include a reflection hump around 20-40 keV, and an iron K$\alpha$ line around $6.4-6.97$ keV that are modified by special and general relativistic effects near the BH 
\citep{fabian1989,fabian2005}. Besides providing detailed information on the composition and ionization state of the disk, the relativistic X-ray reflection  provides a way to determine the spin of the BH and also sheds light on the structure of the corona \citep{garcia2013,garcia2014}.

By contrast with this wealth of data in the X-ray band, the emission in the optical/UV band from these systems is poorly understood. It has been widely accepted that the reprocessing of X-rays in the outer accretion disk is the dominant source of optical/UV emission for BH-LMXBs in the SS \citep{vanparadijs1994}. In the HS, the synchrotron emission from the jet can also contribute significantly in the optical/UV band \citep{russell2006}. 
Although it is difficult to disentangle the effect of these two components from an individual optical/UV spectrum, theoretical models have predicted that the UV/optical luminosity ($L_{\rm optical/UV}$) is correlated differently with the X-ray luminosity if the emission process is irradiation heating ($L_{\rm optical/UV}\propto L_X^{0.5}$) or radiation from the jet ($L_{\rm optical/UV}\propto L_X^{0.7}$) \citep{done2009}. However, optical/infrared and X-ray (in 2-10 keV band) observations of a large sample of X-ray binaries in the hard state suggest that both the X-ray reprocessing in the disk and the jet emission show a slope close to 0.6 (holds over from B to K band) for BH-LMXBs \citep{russell2006}. 
Besides, the intrinsic thermal emission (due to viscous heating) from the outer disk may also provide significant optical/UV photons in both states.  

The transient LMXB MAXI~J1820+070 was discovered with the Monitor of All-Sky X-ray Image (MAXI) on March 11 2018 \citep{kawamuro2018}, and five days prior in the optical band with the All-Sky Automated Survey for SuperNovae (ASSAS-SN) project \citep{denisenko2018}. Soon after its discovery, X-ray flux and optical G magnitude rose to $\sim 4$ Crab \citep{shidatsu2019} and $\sim11.2$ \citep{torres2019}, respectively, making it one of the brightest X-ray transients ever observed. The source remained bright and active for several months and underwent re-brightening episodes before fading into quiescence in February 2019 \citep{russell2019}. Thanks to its low Galactic interstellar absorption ($N_{\rm H}\sim1.3\times10^{21}\ \rm cm^{-2}$,  \cite{hi4pi2016}), and relatively nearby location ($2.96\pm0.33$ kpc, \cite{atri2020}), the source has been extensively monitored across several wavelengths: from radio \citep{trushkin2018,bright2018} to infrared \citep{casella2018,mandal2018} to optical \citep{baglio2018, littlefield2018,sako2018,gandhi2018,bahramian2018,russell2019}  to X-rays \citep{uttley2018,homan2018}, producing a wealth of information about this accreting system.

The object was dynamically confirmed as a BH with a mass of $M=(5.95\pm0.22)M_{\sun}/\rm sin^3(i_b)$ ($i_b$: binary inclination angle) with a K-type companion star \citep{torres2020} of mass  $0.49\pm0.10M_{\sun}$ \citep{joanna2022}. 
The binary inclination angle was estimated as $66^{\circ}<i_b<81^{\circ}$ \citep{torres2020}, although the jet inclination (which is generally assumed to be parallel to the BH spin axis) was found to be $64^{\circ}\pm5^{\circ}$ \citep{woods2021}. On the other hand, \cite{buisson2019} found the same to be  $30_{-5}^{+4}$ degree in their reflection analysis with \textit{NuSTAR} data.
However, the detection of X-ray dips confirms that the inclination of the outer disk is indeed high  \citep{kajava2019}.  

The source went through all canonical states of a BH-LMXBs during the 2018 outburst, following roughly the typical 'q' pattern on HID \citep{buisson2019,chakraborty2020}. It initially stayed in the HS over three months and transitioned to the SS in July 2018 \citep{homan2020}.  A compact jet was observed in the HS \citep{bright2018}. Also, a strong radio flare associated with the launch of bipolar superluminal ejecta was detected at the beginning of the transition to the SS \citep{bright2020}. The long-term optical and X-ray monitoring of the source during the outburst phase suggests that the jet contributed significantly to the optical emission in the HS state, while the outer disk emission through irradiation provides the dominant optical flux in the intermediate states, and SS \citep{shidatsu2019}.     

The initial spectral analysis of the HS data implied that two Comptonization components are required to offer a satisfactory fit to the data, and the disk extends close to the ISCO radius \citep{buisson2019,chakraborty2020}. The inclination angle was reported low ($\sim 30$ degree), and iron abundance ($\sim 5$ times solar abundance) was found to be high in these studies. The spectro-timing analysis of \cite{kara2019} supported their claim related to the extent of the disk. \cite{kara2019} found out that the corona reduced in spatial extent as the source moves toward the SS from the HS, while the inner disk stays stable roughly at $\sim 2R_g$. On the contrary, the disk was observed to be truncated far from the source in the HS in the spectral analysis of \cite{zdziarski2021,zdziarski2021hybrid,zdziarski2022} with a two-component corona. Besides, the spectral-timing analysis of \cite{demarco2021} and \cite{veledina2021} validate the above observation.
Interestingly, the inclination in these works \citep{zdziarski2021,zdziarski2022} was found to be close to the jet inclination angle, and the iron abundance was roughly solar. 
Finally, the SS observations of this source show an excess emission component in the X-ray spectrum which was proposed to originate from the plunging region \citep{fabian2020}. \cite{fabian2020} also found (using \textit{NICER} SS data) that the mass of the BH is $\sim5-10M_{\sun}$ and the spin lies in the range of 0.5 to -0.5 (for an inclination in the range of $30^{\circ}-40^{\circ}$). However, \cite{zhao2021} reported a spin of $a=0.14\pm0.09\ (1\sigma)$ using the \textit{Insight}-HXMT SS data, assuming $M=8.48^{+0.79}_{-0.72}M_{\sun}$, inclination $=63\pm3$ degree, and distance $=2.96\pm0.33$ kpc.  \cite{bhargava2021} studied the characteristic frequencies of several power-density spectral components within the framework of the relativistic precession model, and obtained a spin of $0.799_{-0.015}^{+0.016}$. Thus, there is a discrepancy in the measurement of the spin of this BH.

The multi-wavelength spectral analysis of MAXI J1820+070 during its 2018-2019 outburst have only been considered in a few works, e.g., \cite{rodi2021, arabaci2022, trujillo2023}. \cite{rodi2021} primarily delved into studying the jet properties in the hard state on April 12 within the \texttt{JetSet} framework. In contrast, \cite{trujillo2023} focused on the evolution of jet spectral properties and their connection to accretion flow parameters. On the other hand, \cite{arabaci2022} analyzed two multi-wavelength observations (near infrared to hard X-ray) in the hard state using \textit{SWIFT}, \textit{INTEGRAL}, \textit{SMARTS}, and \textit{TUG}, with one observation during the outburst decay and the other close to the mini-outburst peak. However, the latter work lacked a detailed reflection analysis of hard X-ray spectra, similar to \cite{trujillo2023}, and was constrained by low data quality. In contrast, our work conducts a comprehensive spectral analysis of MAXI J1820+070 in both hard and soft states, leveraging high-quality spectroscopic UV/X-ray data from \textit{AstroSat} and \textit{NICER} missions, along with photometric optical data from Las Cumbres Observatory (hereafter, \textit{LCO}). Our investigation explores the evolution of spectral parameters related to X-ray emission from the inner disk and UV/optical emission from the outer disk. We study how the emission from the inner accretion flow influences the outer disk and constrain the global geometry of the accretion disk. It should be noted that our work marks the first case where data from all \textit{AstroSat} instruments are employed for studying a source.

The paper is organized as follows.  We describe the observations and data reduction in Section \ref{sec:data}. The results from the spectral analysis are presented in Section \ref{sec:result}. We summarize and discuss our results in Section \ref{sec:discuss} and draw conclusions in  \ref{sec:conclusion}.  

\section{Observations and Data Reduction} \label{sec:data}
We use the UV, soft and hard X-ray data acquired with the first dedicated Indian multi-wavelength space observatory 
\astrosat{} \citep{singh2014}. It carries four co-aligned scientific payloads:  the Ultraviolet Imaging Telescopes (UVIT; \citealt{tandon2017orbit,tandon2020additional}), the Soft X-Ray Telescope (SXT; \citealt{singh2016orbit,singh2017soft}),  the Large Area X-ray Proportional Counters (LAXPC; \citealt{yadav2016large, antia2017}, and the Cadmium-Zinc-Telluride Imager (CZTI; \citealt{czti,bhalerao2017}). \astrosat{} observed MAXI~J1820+070 twice during its 2018 outburst. To support our \astrosat{} observations, we also use (quasi-)simultaneous \nicer{} X-ray and \textit{LCO} optical data. We list all the observations in Table~\ref{tab:obs1} and Table~\ref{tab:obs2}, and provide more details below. In this work, we use the \texttt{HEASOFT} version 6.30.1 for data processing and spectral analysis.

\begin{figure}
\includegraphics[width=0.45\textwidth]{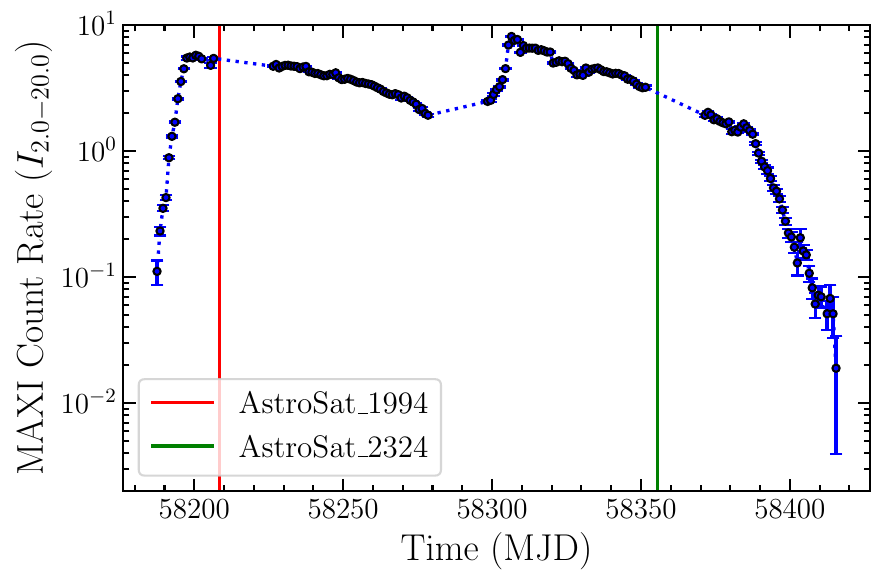}
\includegraphics[width=0.45\textwidth]{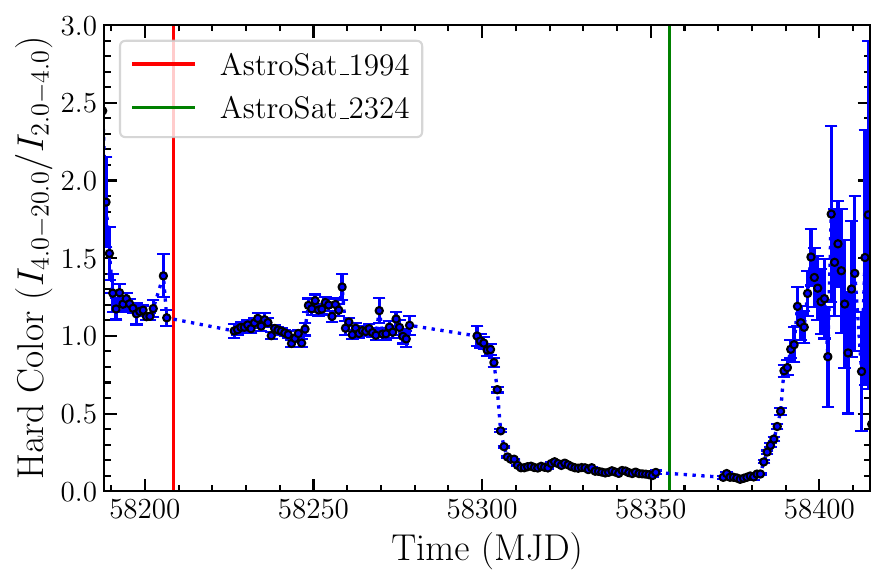}
\caption{MAXI light-curve in the energy band 2.0 -- 20.0 keV (upper panel), and 
hardness-time diagram (lower panel) of MAXI~J1820+070 during 2018 outburst. The hardness is defined as the ratio between the MAXI count-rate in the $4.0-20.0$ keV and $2.0-4.0$ energy bands.
Two vertical lines on each panel designate the \textit{AstroSat} observations.\label{maxi}}
\end{figure}

To identify the spectral states during our \astrosat{} observations, we use the daily averaged \textit{MAXI} light curves in the energy bands 2.0 -- 4.0 keV, 4.0 -- 20.0 keV, and 2.0 -- 20.0 keV and derive the hardness-time and intensity-time diagrams for MAXI~J1820+070, which we show in Fig.~\ref{maxi}. 
We define hardness as a ratio between the MAXI count rate in the 4.0 -- 20.0 keV and 2.0 -- 4.0 keV energy bands. We find that the source was in the hard state during the first \textit{AstroSat} observation in March 2018 (AstroSat\_1994) and in the soft state during the second observation in August 2018 (AstroSat\_2324). We employ all the \textit{AstroSat} instruments from far UV to hard X-ray band (FUV, SXT, LAXPC, and CZTI) to observe the source in the hard state. Additionally, we use data from two nearly simultaneous \textit{NICER} and one simultaneous \textit{LCO} observations for the hard state observation. The two \textit{NICER} observations combinedly cover a slightly longer time period than the \textit{AstroSat} observation period.
For the soft state observation, we use data from three \textit{AstroSat} instruments and one quasi-simultaneous \textit{LCO} observation. 
In the following subsections, we briefly describe the instruments used for the observations and the data reduction process.

\subsection{\astrosat{}/SXT}\label{sxt}
The SXT \citep{singh2016orbit,singh2017soft} is equipped with X-ray optics and a CCD camera, and operates in the photon counting mode. It is well suited for medium resolution spectroscopy (FWHM $\sim 150\ev$ at $6\kev$) in the $0.5-7\kev$ band, and is also capable of low resolution imaging (FWHM $\sim 2{\rm~arcmin}$, HPD $\sim 11{\rm~arcmin}$). We process the level-1 data using the SXT pipeline (AS1SXTLevel2-1.4b) available at the SXT payload operation center (POC) website\footnote{\url{https://www.tifr.res.in/~astrosat_sxt/sxtpipeline.html}}, and generate level-2 clean event files for individual orbits. For each observation, we merge the orbit-wise clean event files using the {\it Julia} SXT event merger tool SXTMerger.jl\footnote{\url{https://github.com/gulabd/SXTMerger.jl}}. We obtain the processed and cleaned level-2 data for a net SXT exposure time of $\sim 17.97\ks$  and $\sim 7.93\ks$ for the first and second \astrosat{} observations.  

\begin{figure}
\includegraphics[width=0.45\textwidth]{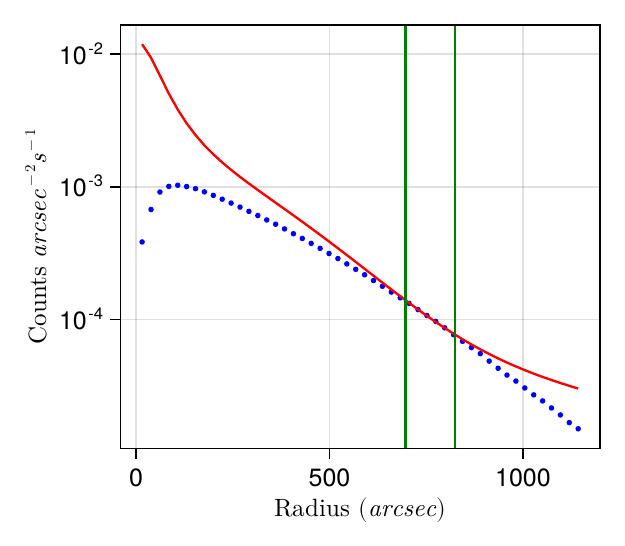}
\caption{The SXT radial profile of MAXI~J1820+070 in the soft state (shown in blue dots) compared with the SXT PSF (represented by a red line). The vertical lines mark the inner and outer radii within which the radial profile matches well with the SXT PSF, and therefore free of significant pile-up. See Section \ref{sxt} for details. \label{sxt_pileup}}
\end{figure}

MAXI~J1820+070 was very bright, exceeding the Crab flux in the $2-10\kev$ band in both the hard and soft states, and thus causing severe pile-up in the SXT data.  To correct for the pile-up, we first generate SXT radial profile of a blazar Mrk~421 that is bright but not affected with any pile-up. We then model the radial profile with a Moffat function and a Gaussian profile, and derive the PSF of the SXT as,
\begin{equation}   
PSF(r) = A\left(\frac{1}{\left\{1 + (\frac{r}{r_c})^2\right\}^{1.04}} + 0.08 \times \exp{\left\{-\frac{r^2}{2\sigma^2}\right\}} \right),
\end{equation}
with $r_c=65.3{\rm~arcsec}$, $\sigma=294.2{\rm~arcsec}$.
We then fit this PSF  with a variable amplitude $A$ to the radial profile of MAXI~J1820+070 and obtain the range of radii where the radial profile is well described by the SXT PSF, and the corresponding annular region is free of pile-up.  We find that the annular regions with inner and outer radii of $r_{i}=800{\rm~arcsec}$ (194 pixels) and $r_{o}=900{\rm~arcsec}$ (218 pixels) for the soft state, and $r_{i}=618{\rm~arcsec}$ (150 pixels) and $r_{o}=907{\rm~arcsec}$ (220 pixels) for the hard state are not 
affected with photon pile-up significantly. In Figure~\ref{sxt_pileup}, we show the radial profile of MAXI~J1820+070 (blue dots) in the soft state and the SXT PSF (red line). The vertical lines mark the annular region between $r_{i}$ and $r_{o}$ that is free from significant pile-up. The deficit of counts in the outermost regions is caused by the loss of events beyond the detector boundary due to the off-axis observations and large PSF/HPD of the SXT.

We use the \textsc{xselect} tool and extract the source spectrum from the  merged level-2 event files using an annular region with $r_{i}$ and $r_{o}$ inferred above. Clearly, the ancillary response file (ARF) made available for a circular extraction region centred on the source is inappropriate for a heavily piled-up source like in this case, where a large fraction of counts from the inner region are excluded. Therefore, we derive the corrected ARF as follows. We use the SXT observation of Crab which is a standard X-ray calibrator. We first extract the SXT spectra of Crab using the same annular regions we used for MAXI~J1820+070 in the soft and hard states. We fit these Crab spectra with an absorbed power-law model with fixed parameters ($N_{\rm H} = 3.1\times 10^{21}{\rm~cm^{-2}}$, $\Gamma = 2.1$, $f_X(2-10\kev) = 2.4\times 10^{-8}{\rm~ergs~cm^{-2}~s^{-1}}$; see \cite{2010ApJ...713..912W})  using the ARF/RMF provided by the SXT POC, and derive the
data-to-model ratio. Using these ratios,  we correct the ARF and derive separate ARFs appropriate for the soft and hard states. As a cross-check, we use the corrected ARFs and fit the Crab spectra and obtain spectral parameters similar to those already known.

\subsection{\astrosat{}/LAXPC}
The X-ray instrument LAXPC consists of three proportional counters (LAXPC10, LAXPC20, and LAXPC30) operating in the energy range of 3 -- 80 keV with a temporal resolution of 10 $\mu s$ \citep{antia2017}. Out of these three detectors, LAXPC10 has been showing unpredictable high-voltage variations since March 2018, and LAXPC30 was switched off due to a gas leakage \citep{antia2021}. Thus, we only use data acquired with the detector LAXPC20 in this work. We extract the LAXPC source and background light curves using the software \texttt{laxpcsoftv3.4.3} \footnote{\url{https://www.tifr.res.in/~astrosat_laxpc/LaxpcSoft_v1.0/antia/laxpcsoftv3.4.3_07May2022.tar.gz}}. The response files \texttt{lx20v1.0.rmf} and \texttt{lx20cshm01v1.0.rmf} are used for the hard and soft observations, respectively. We group the LAXPC spectra using the FTOOLS package \texttt{ftgrouppha} to have a signal-to-noise ratio of at least 25 per bin.

\subsection{\astrosat{}/CZTI}
Cadmium Zinc Telluride Imager (CZTI) is a hard X-ray instrument onboard \textit{AstroSat} providing spectroscopic observations in 22 -- 200 keV energy range and in-direct imaging by employing coded aperture mask \citep{bhalerao2017}. CZTI consists of four independant quadrants each having an array of 16 CZT detectors and data are available for each quadrant separately.
For the analysis of CZTI data, we use CZTI data analysis pipeline version 3.0 along with the associated
CALDB \footnote{\url{http://astrosat-ssc.iucaa.in/cztiData}}. Following the standard pipeline procedure, clean event list are filtered out from the raw event file. From the clean event files, background subtracted source spectra for each quadrant along with associated response matrices are obtained by using \texttt{cztbindata} task of the data analysis pipeline,
which employs the mask-weighting technique. We use the optimal binning scheme of \cite{kaastra2016} to group the CZTI spectra with minimum 25 counts per bin.

\subsection{\astrosat{}/UVIT}
The UVIT \citep{tandon2017orbit,tandon2020additional} consists of three channels providing sensitivity in three different bands -- far ultraviolet ($1200-1800$\angstrom{})  (FUV channel),  near ultraviolet ($2000-3000$\angstrom{}; NUV channel),  and the visible ($3200-5500$\angstrom{}; VIS channel) bands.  The FUV and NUV channels are used for scientific observations, while the VIS channel is mainly used for tracking satellite pointing.  Both FUV and NUV channels  are equipped with a number of broadband filters for imaging with a point spread function (PSF) in the range of $1-1.5{\rm~arcsec}$ and slit-less gratings for low resolution spectroscopy. 
The FUV channel has two slit-less gratings, FUV-Grating1 and FUV-Grating2 (hereafter FUV-G1 and FUV-G2), that are arranged orthogonal to each other to avoid possible contamination along the dispersion direction, due to the presence of neighboring sources in the dispersed image. These two channels operate in the photon counting mode. More details on the performance and calibration of the UVIT gratings can be found in \cite{dewangan2021}.

We obtain the level1 data on MAXI~J1820$+$070 from the \astrosat{} archive\footnote{\url{https://astrobrowse.issdc.gov.in/astro_archive/archive/Home.jsp}}, and process them using the CCDLAB pipeline \citep{postma2017}. We generate orbit-wise drift-corrected, dispersed images for each observation.
We then align the orbit-wise images and merge them into a single image for each observation. We use the \textsc{UVITTools.jl}\footnote{\url{https://github.com/gulabd/UVITTools.jl}} package for spectral extraction following the procedures described in \citet{dewangan2021} and \citet{kumar2023}. We first locate the position of the zeroth order image of the source in the grating images, and then use the centroids, along the spatial direction at each pixel, along the dispersion direction for the $-2$ order. 
We use a 50-pixel width along the cross-dispersion direction and extract the one-dimensional count spectra for the FUV gratings in the $-2$ order. Following a similar procedure, we also extract background  count spectra from source-free regions, and correct the source spectra for the background contribution. 
We use an updated version of grating responses which are adjusted to match a simultaneous hard state \textit{HST} spectrum of this source (a detailed analysis of the \textit{HST} spectrum will be discussed on the forthcoming paper, Georganti et. al. 2023). These files are thus generated following the procedures described in  \cite{dewangan2021}.

\subsection{NICER}
\begin{figure}
\includegraphics[width=0.45\textwidth]{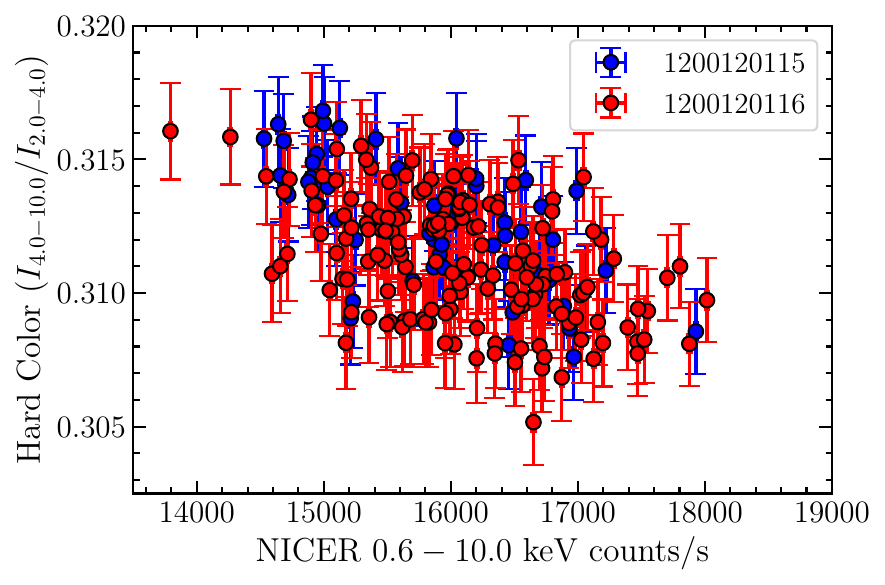}
\caption{\textit{NICER} hardness-intensity diagram for the two observations: 1200120115 and 1200120116. The hardness is defined as the ratio between the \textit{NICER} count-rate in the $2.0-4.0$ keV and $4.0-10.0$ energy bands.\label{nicer}}
\end{figure}

In this work, we consider two \textit{NICER} observations, which were quasi-simultaneous with the \textit{AstroSat} hard state observation (see Table~\ref{tab:obs1}). The \textit{NICER} data are reduced and calibrated using the \texttt{NICERDAS} 2022-01-17\_V009 and \texttt{CALDB} version xti20210707.
The \textit{NICER} X-ray Timing Instrument \citep{gendreau2016} consists of an array of 56 co-aligned X-ray concentrator optics, each of which is paired with a single-pixel silicon drift detector working in the 0.2 –- 12 keV energy band (with a spectral resolution of $\sim85$ eV FWHM at 1 keV and $\sim137$ eV at 6 keV). 
Although 52 detectors were working at the time of the observation, we exclude data from the detectors numbered 14 and 34 as they sometimes exhibit periods of increased noise.  We generate cleaned event files of the \textit{NICER} observations using the script \texttt{nicerl2} (with default criteria) and employ the background estimator \texttt{nibackgen3C50} \citep{remillard2021} to generate the source and background spectra. The scripts \texttt{nicerarf} and \texttt{nicerrmf} are used to obtain the ARF and RMF files for all the \textit{NICER} observations. We further obtain background uncorrected \textit{NICER} light-curves in three energy bands: 2.0 -- 4.0 keV, 4.0 -- 10.0 keV, and 0.6 -- 10.0 keV (with 64s bin time) to produce the \textit{NICER} hardness-intensity (see Fig.~\ref{nicer}) diagram. In our work, the \textit{NICER} hardness is the ratio between the \textit{NICER} count-rate in the 4.0 -- 10.0 keV and 2.0 -- 4.0 keV bands. We do not find any significant hardness variation between the two \textit{NICER} observations (see Fig.~\ref{nicer}), and also within a single observation. Hence, we consider these two observations entirely in the present work. We group the \textit{NICER} spectra using the FTOOLS package \texttt{ftgrouppha} to  a signal-to-noise ratio of at least 50 per bin.

\subsection{LCO}
MAXI~J1820+070 was monitored during its 2018 outburst in the optical wavelengths by the Las Cumbres Observatory (\textit{LCO}), as part of an on-going monitoring campaign of $\sim$50 low mass X-ray binaries coordinated by the Faulkes Telescope Project \citep{lewis2018}. For this study, we use the optical observations obtained with the 1-m robotic telescopes at the LCO nodes of Cerro Tololo Inter-American Observatory (Chile) and South African Astronomical Observatory, Sutherland (South Africa), that were simultaneous/quasi-simultaneous with the hard state and the soft state observations of \textit{Astrosat}, respectively (see Tables~\ref{tab:obs1} and \ref{tab:obs2}). The observations were performed in the SDSS $g^{'}$, $i^{'}$ and $r^{'}$ bands, with 20 second exposure times on each filter for hard state, and 40 second exposure times for the soft state. We use the ``X-ray Binary New Early Warning System (XB-NEWS)'' data analysis pipeline \citep{russel2019,pirbhoy2020} for calibrating the data, computing an astrometric solution for each image using Gaia DR2 positions, performing aperture photometry of all the stars in the image, solving for zero-point calibrations between epochs, and flux calibrating the photometry using the ATLAS All-Sky Stellar Reference Catalog (ATLAS-REFCAT2) \citep{tonry2018}.

\begin{table*}
\centering
\caption{Details of \textit{AstroSat} and \textit{NICER} observations used in this study. \label{tab:obs1}}
\medskip
  \begin{tabular}{ccccccc}
    \hline
Mission &  Obs ID & Instrument & Exposure (ks) & Start Time (UT) & End Time (UT) & State\\
          &         &            &               & yyyy-mm-dd/hh-mm-ss  & yyyy-mm-dd/hh-mm-ss &\\ 
 \hline 
 \textit{AstroSat}  & T02$\_$038T01$\_$900001994 & SXT & 17.97 & 2018-03-30/10:45:52 & 2018-03-31/14:13:04 & Hard\\
           & (AstroSat$\_1994$) & LAXPC & 37.78 & 2018-03-30/10:45:51 & 2018-03-31/14:13:05 & \\                           
           &                    & CZTI-Quad0 & 37.17 & 2018-03-30/10:46:36 & 2018-03-31/14:13:04 &\\
           &                    & CZTI-Quad1 & 37.04 & ,, & ,, &\\
           &                    & CZTI-Quad2 & 37.18 & ,, & ,, &\\
           &                    & CZTI-Quad3 & 34.69 & ,, & ,, &\\
           &                    & UVIT/FUV-G1 & 11.39 & 2018-03-30/12:02:29 & 2018-03-31/14:11:17 &\\
\textit{NICER}  & 1200120115 (NICER-1) & XTI & 3.34 & 2018-03-30/09:15:40 & 2018-03-30/11:30:20 & Hard\\ 
       & 1200120116 (NICER-2) & XTI & 10.71 & 2018-03-31/08:28:18 & 2018-03-31/19:33:09  &,,\\
\hline
\textit{AstroSat}  & T02$\_$066T01$\_$900002324 & SXT & 7.929 & 2018-08-25/11:10:10 & 2018-08-25/16:37:28 & Soft \\ 
        & (AstroSat$\_2324$) & LAXPC & 11.5 & ,, & 2018-08-25/16:37:29 &\\ 
        & & UVIT/FUV-G1 & 2.845 & 2018-08-25/11:18:01 & 2018-08-25/13:20:09 &\\ 
        &  & UVIT/FUV-G2 & 2.739 & 2018-08-25/14:32:54 & 2018-08-25/16:35:03 &\\ 
       \hline
 \end{tabular}
 \end{table*}

\begin{table*}
\centering
\caption{Details of \textit{LCO} observations used in this study. \label{tab:obs2}}
\medskip
  \begin{tabular}{ccccccc}
    \hline
 Filter & Wavelength  & Wavelength  & Exposure (s) & Date & Time (UT) & State\\
        &  center (nm) & width (nm) & & yyyy-mm-dd & hh-mm-ss & \\ 
       \hline
\hline 
       SDSS-i' & 754.5 & 129 & 20 & 2018-03-31 & 7:32:59 & Hard\\
       SDSS-r' & 621.5 & 139 & 20 & & 7:37:17 &\\
       SDSS-g' & 477.0 & 150 & 20 & & 7:34:17 &\\
\hline
       SDSS-i' & 754.5 & 129 & 40 & 2018-08-24 & 21:16:53  & Soft\\
       SDSS-r' & 621.5 & 139 & 40 & & 21:22:44 &\\
       SDSS-g' & 477.0 & 150 & 40 & & 21:18:29 &\\
       \hline
 \end{tabular}
 \end{table*}

\section{Spectral Analysis} \label{sec:result}

All spectral analyses presented in the following sections are performed using \texttt{XSPEC} \citep{arnaud1996} version 12.12.1. A multiplicative cross-normalization constant (implemented using \texttt{constant} in \texttt{XSPEC}) is allowed to vary freely for LAXPC, CZTI, XTI/\textit{NICER}, and fixed to unity for SXT. We consider an energy range of $6.9-9.5$ eV for FUV-G1/FUV-G2, $0.6-7.0$ keV for SXT, $4.0-60.0$ keV for LAXPC (for the hard state observation), $25.0-150.0$ keV for CZTI, and $0.6-10.0$ keV for XTI/\textit{NICER}. 
We limit the LAXPC data to $40$ keV for the soft state observation as the spectrum is background dominated beyond that.
A systematic error of $2\%$ for all the \textit{AstroSat} instruments (and \textit{LCO} filters) and $1\%$ for \textit{NICER} (as suggested by the their respective instrument teams) is considered in this work. We apply a gain correction to the SXT data using the \texttt{XSPEC} command \texttt{gain fit} with the slope fixed to unity. The best-fit offset is found to be $\sim 0.24$ eV for both the hard state and soft state observations.

We use \texttt{tbabs} \citep{wilms2000} to take account for the absorption of X-rays in the interstellar medium along the line of sight. We adopt abundances from \cite{wilms2000} (\texttt{wilm} in \texttt{XSPEC}), and the photoelectric absorption cross-sections from \cite{vern1996} (\texttt{vern} in \texttt{XSPEC}). Furthermore, we assume a distance to the source of 2.96 kpc \citep{atri2020}. The uncertainties reported in this work corresponds to $90\%$ confidence level for a single parameter of interest. 
\subsection{Hard State}
\subsubsection{X-ray spectral Analysis}\label{sec:hardx}
\begin{figure}
\includegraphics[width=0.45\textwidth]{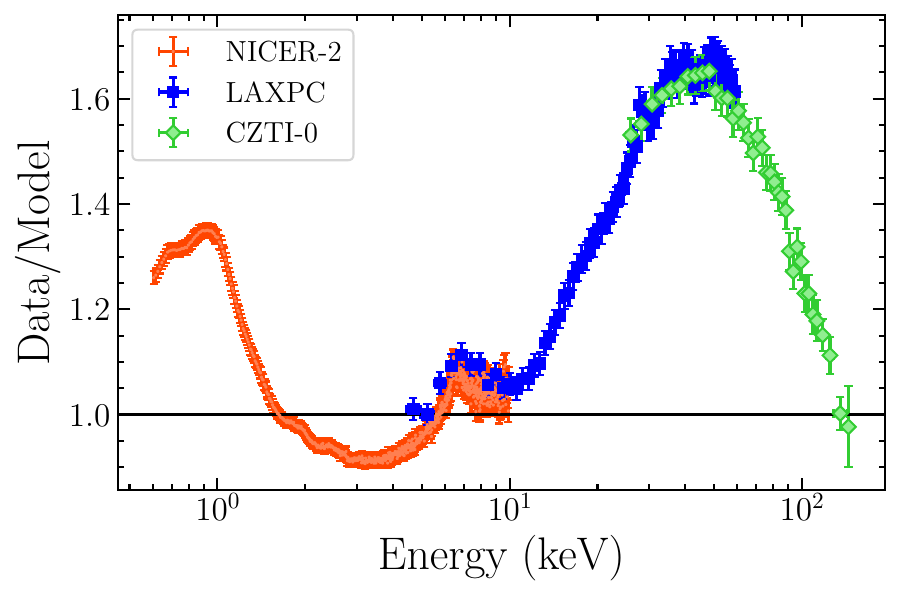}
\caption{Ratio (Data/Model) of the $0.6-150.0$ keV \textit{NICER} and \textit{AstroSat} data to the fiducial model \texttt{tbabs*constant*cutoffpl} (hard state observation). This exercise is performed just to show different spectral features in the hard state spectra. We clearly see a soft excess, the Fe K-$\alpha$ emission line, and a Compton hump with a peak around $\sim 40$ keV in the hard state spectra. Only data from the second \textit{NICER} observation, LAXPC data, and CZTI Quadrant-0 data are used for this investigation.  Data are rebinned for plotting purpose. See Section \ref{sec:hardx} for more details. \label{rat1}}
\end{figure}
We begin our investigation with the hard state X-ray data in the energy range of $0.6-150.0$ keV with an absorbed cutoff power-law (i.e., \texttt{tbabs*constant*cutoffpl} in \texttt{XSPEC} notation). We perform this analysis just to demonstrate different spectral features in the data and consider only the data from the second \textit{NICER} observation, the LAXPC data, and CZTI Quadrant-0 data for this purpose. We note a soft-excess, a broad iron emission line around 6.4 keV, and a Compton hump with a peak at approximately 40 keV in the hard state spectra (see Fig.~\ref{rat1}).

\begin{figure}
\includegraphics[width=0.45\textwidth]{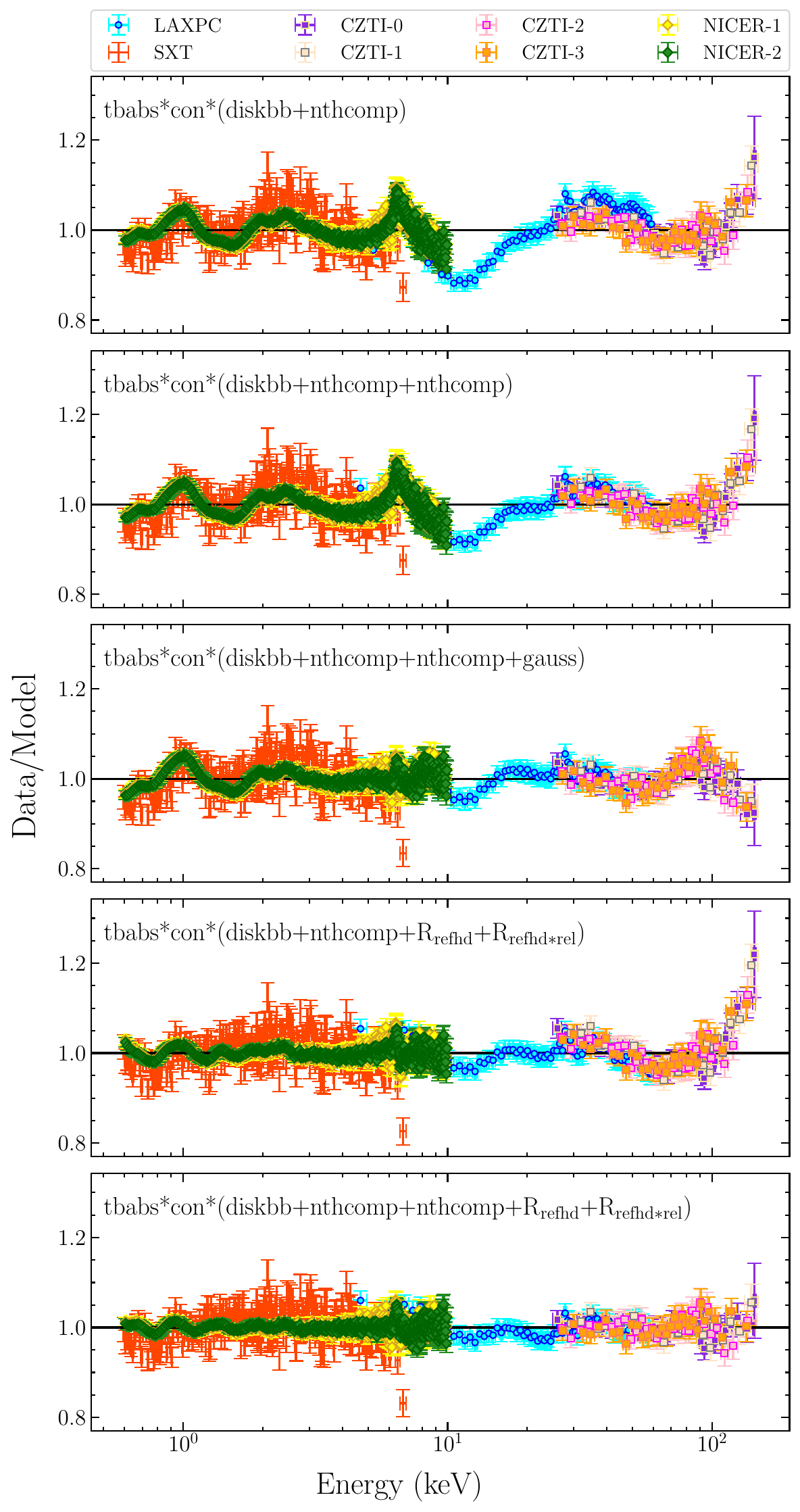}
\caption{Ratio (Data/Model) of the $0.6-150.0$ keV X-ray multi-instrument (\textit{AstroSat}+\textit{NICER}) data to several models for the hard state observation. Here, $R_{\rm refhd}$ and $R_{\rm refhd*rel}$ stand for \texttt{reflionxhd} and \texttt{relconv*reflionxhd}. 
In this plot, the model components are mentioned on top of each panels. The panels, from top to bottom, illustrate
the improvement in residuals as further model components are added. Data are
rebinned for plotting purpose. See Section \ref{sec:hardx} for more details.\label{rat2}}
\end{figure}

We first fit the broadband continuum of the hard-state X-ray data (in the energy range $0.6-150.0$ keV) of MAXI~J1820+070 with a simple model (Model: \texttt{tbabs*constant*(diskbb+nthcomp)}) composed of a \texttt{diskbb} \citep{mitsuda1984,makishima1986} component (for describing the multicolored black-body emission from a geometrically thin and optically thick accretion disk), and a thermal Comptonized component \texttt{nthcomp} \citep{zdziarski1996,zycki1999} (accounting for the hard coronal emission) in which the soft seed photons are provided by the accretion disk (i.e., \texttt{inp\_type} of \texttt{nthcomp} has been set to 1). The seed photon temperature of \texttt{nthcomp} is tied to the inner disk temperature of \texttt{diskbb}.
This model poorly describes the data ($\chi^2/\text{d.o.f}=7627.7/2398$, where $\text{d.o.f}$ stands for degrees of freedom). To obtain a better fit and explore the possibility of another coronal component, we add a second \texttt{nthcomp} to the above model (Model : \texttt{tbabs*constant*(diskbb+nthcomp(1)+nthcomp(2))}). We tie the seed photon temperature of the two Comptonization components to that of the \texttt{diskbb} component. The fit quality improved significantly, giving a $\chi^2/\text{d.o.f}=7081.3/2395$ (F-test probability of chance improvement $\sim10^{-38}$). This improvement is not surprising, as earlier studies have shown \citep{buisson2019,chakraborty2020,zdziarski2021,zdziarski2022} that the double coronae model offers a better fit to the data. In both these models, the neutral hydrogen column density ($N_{\rm H}$) lies in the range of $1.2-1.4\times 10^{21}\ \rm atoms\ cm^{-2}$, which is close to the Galactic absorption column along the direction of the source, $1.3\times10^{21}\ \rm atoms\ cm^{-2}$ \citep{hi4pi2016}. So, hereafter, we will fix $N_{\rm H}$ to the Galactic value for all subsequent spectral analyses performed on the hard state observation. The residuals for both these models are depicted in Fig.~\ref{rat2}. Although the \texttt{diskbb} component has already taken care of the soft excess (seen in Fig.~\ref{rat1}), we still observe significant residuals around 1 keV in both of these models, which could be indicative of a reflection feature (see Fig.~\ref{rat2}). 

However, still, we could not obtain a reasonably good fit, probably due to the strong presence of reflection features (see Fig.~\ref{rat1}). Thus, we add a Gaussian component to our double Comptonization model to represent the Fe-K$\alpha$ line observed in the spectra and obtain a huge improvement in the spectral fit (Model: \texttt{tbabs*constant*(diskbb+nthcomp(1)+nthcomp(2)+
gauss)}) with $\chi^2/\text{d.o.f}$ of 3921.7/2392. 
The residuals around 6.5 keV reduces significantly in this model (see Fig.~\ref{rat2}). But the low and high-energy residuals do not improve much. The cross-normalization constant between SXT and LAXPC or CZTI (Quadrant-0/1/2/3) or NICER-1/NICER-2 is found to lie in the range of $\sim 0.74-0.91$ ($\sim 9\% - 26\%$), which is within the acceptable limit. The peak energy and width of the iron K$\alpha$ line come out to be $6.58\pm0.02$ keV and $0.77\pm0.02$ keV, respectively. The equivalent width corresponding to this line for the \textit{NICER} data is found to be $0.18\pm0.33$ keV. The inner disk temperature is $0.27\pm0.05$, which is close to the value of $\sim0.2$ keV obtained earlier for the hard state \textit{NICER} observations \citep{dzielak2021,wang2020}. 
The two Comptonization components are well separated in power-law indices ($\Gamma$) and electron temperatures ($kT_{\rm e}$) in this model, with the best-fit values of them being $\Gamma=1.72\pm 0.01$ and $kT_{\rm e}=243.1_{-65.6}^{+46.9}$ and $\Gamma=1.18\pm 0.04$ and $kT_{\rm e}=13.40\pm0.69$ keV, respectively. Although adding a Gaussian line to our single Comptonization model provides a huge improvement like the previous case, it still could not produce an acceptable fit ($\chi^2/\text{d.o.f}=5507.3/2395$). Since other reflection features like the Compton hump are quite prominent in this observation (see Fig.~\ref{rat1}), we perform an in-depth reflection analysis of joint \textit{AstroSat} and \textit{NICER} data.

For a detailed investigation of the broadband spectra and the reflection features, we use a self-consistent reflection model \texttt{reflionxhd}, the latest model from the reflection suite \texttt{reflionx} \citep{ross2005,ross2007}. The \texttt{reflionx} based reflection models generate an angle-averaged reflection spectrum for an optically-thick atmosphere (such as the surface of an accretion disk) with constant density irradiated by hard Comptonized emission. This new model \texttt{reflionxhd}  assumes that the illuminating continuum (responsible for ionizing the disk) is based on the \texttt{nthcomp} Comptonization model (as opposed to a cutoff power-law), with the soft seed photons being provided by the accretion disk \citep{jiang2020,riley2021,chakraborty2021}. Besides, the density of the disk ($n_{\rm e}$) is a model parameter in the range of $15\leq \log{(n_{e}/{\rm cm^{-3}})}\leq 22$. 
We further convolve this component with \texttt{relconv} \citep{dauser2010}, which is part of the \texttt{relxill} distribution of models \citep{dauser2014,garcia2014}, to consider the effect of relativistic blurring. We add them to our double Comptonization model to represent the relativistically smeared reflected emission. We tie the power-law index ($\Gamma$), electron temperature ($kT_{\rm e}$), and seed photon temperature $kT_{\rm seed}$ of the \texttt{reflionxhd} component (i.e., parameters related to the input continuum) with one of our external \texttt{nthcomp} components. 
The ionization parameter ($\xi$), iron-abundance ($A_{\rm Fe}$), and density ($\rm log$$(n_{\rm e})$) are left as free parameters. On the other hand,
in \texttt{relconv}, we fix the spin of the BH (i.e., Kerr parameter) to $a=0.998$ (i.e., the ISCO radius or radius of innermost stable circular orbit around a Kerr BH, $R_{\rm ISCO}=1.237\ R_g$, where $R_g=GM/c^2$, $M$ is the mass of the BH, $G$ is the Newtonian gravitational constant, and $c$ is the speed of light in free space) to enable comparisons with previous studies, as this assumption has been made in all the works that performed reflection analysis in the hard state \citep{buisson2019,chakraborty2020,zdziarski2021,zdziarski2022}. Since the outer accretion disk could be misaligned with respect to the BH spin axis \citep{poutanen2022,thomas2022}, we assume that the inclination of the inner disk is identical to that of the jet, i.e., $i=64^{\circ}$ \citep{woods2021}, which is presumed to be in the direction of the BH spin axis.
We also assume a single emissivity index ($q=q_1=q_2$), making the break radius (which separates the inner disk with emissivity $q_1$ from the outer disk with emissivity $q_2$) redundant, and set the outer disk to $1000R_g$. 
To account for the narrow core of the Fe-K$\alpha$ line, we add another \texttt{reflionxhd} component to our existing model and tie the parameters of the internal \texttt{nthcomp} part ($\Gamma$, $kT_{\rm e}$, and $kT_{\rm seed}$) of \texttt{reflionxhd} with those of another external \texttt{nthcomp} component in our existing model. Additionally, we tie the density and iron-abundance of this reflection component to the previously added \texttt{reflionxhd}. The density of the disk associated with these two reflection components (relativistic and distant) is perhaps different. However, we tie them to keep our model simple by reducing the number of free parameters. Thus, in the present model (hereafter referred to as Model 1A), one \texttt{nthcomp} (i.e., \texttt{nthcomp(1)}) component gets reflected through the relativistic reflection component \texttt{relconv*reflionxhd(1)}, and another one, i.e., \texttt{nthcomp(2)} through the distant reflection component \texttt{reflionxhd(2)}. 
We link the seed photon temperature of these two Comptonization components to the inner disk temperature of the \texttt{diskbb} component.
Therefore, the resulting model takes the following form,
\begin{itemize}[align=parleft,left=0pt..1em]
    \item Model 1A: \texttt{tbabs}*\texttt{constant}*(\texttt{diskbb}+\texttt{nthcomp(1)}\\
    +\texttt{nthcomp(2)}+\texttt{relconv*reflionxhd(1)}\\
    +\texttt{reflionxhd(2)}).
\end{itemize}
We obtain a $\chi^2/\text{d.o.f}$ of 1904.2/2388, i.e., a huge improvement in the spectral fit compared to our previous models. The results are presented in Table~\ref{tab:hardx} and the residuals are depicted in Fig.~\ref{rat2}. 

We first notice that the residuals below 2 keV and above 10 keV diminish significantly in this model (see Fig.~\ref{rat2}). The disk temperature, in this case, $0.192\pm0.002$, takes a value almost identical to what was estimated earlier \citep{wang2020}. Besides, the disk is found to get truncated far from the source, at a distance of $62.3^{+15.2}_{-11.4}\ R_g$. 
We also estimate the inner disk radius from the \texttt{diskbb} normalization following the relation, 
\begin{equation} \label{eqn2}
    R_{\rm in}=\eta \kappa^2 \sqrt{\frac{N_{\rm disk}}{\cos{i}}} \frac{D}{10\ \rm kpc} 
\end{equation}
where, $R_{\rm in}$ is true inner radius in km, $\kappa$ is the spectral hardening factor (i.e., the ratio between the color temperature to effective temperature) \citep{shimura1995}, $\eta$ is the correction factor for the inner torque-free boundary condition for a Schwarzschild BH ($a=0$) \citep{kubota1998}, $N_{\rm disk}$ is the \texttt{diskbb} normalization, and $D$ is the distance to the source. However, it is unlikely that the zero torque condition is applicable in the hard state, where the disk is truncated far from the ISCO radius. Thus, it is perhaps incorrect to include the correction factor $\eta$ in our estimation of true inner radius (see \cite{basak2016} for more details).
Adopting $\kappa=1.7$ \citep{kubota1998}, $i=64^{\circ}$, and mass of the BH ($M$)=6.75 $M_{\sun}$ \citep{joanna2022}, we obtain $R_{\rm in}\sim73R_g$, which is consistent with the value obtained from the reflection fit. 

The power-law index and electron temperature of the two Comptonization components are $1.168\pm0.002$ and $30.55_{-0.99}^{+0.95}$ keV (\texttt{nthcomp(2)}), and $1.596\pm0.003$ and $15.39\pm0.34$ keV (\texttt{nthcomp(1)}), respectively. The reflecting part of the disk corresponding to the soft Comptonization component is found to be strongly ionized $\xi=2365_{-47}^{+64}$ compared to that  ($\xi=488_{-12}^{+14}$) illuminated by the hard Comptonization component. Similar two-component Comptonization scenarios with other reflection models (e.g., \texttt{relxilllpCp}) were investigated earlier by \cite{buisson2019,chakraborty2020,zdziarski2021,zdziarski2022}, and in all these works, it was noted that a double coronae model provides a much better fit to data than a single corona model. The values of the best-fit parameters in Model 1A are consistent with those obtained in an earlier investigation of this source with contemporaneous \textit{Insight-HXMT}, \textit{NuSTAR}, and \textit{INTEGRAL} data \citep{zdziarski2022}. 
Notably, in all the above analyses, constant density reflection models (i.e., $n_{\rm e}$ is fixed at $10^{15}\ \rm cm^{-3}$) were employed. In some of these works \citep{buisson2019,chakraborty2020}, a higher iron abundance ($>3A_{\rm Fe,solar}$) was reported.
In our work, $\rm log$$ (n_{\rm e})$ is a variable parameter, and it takes a higher value of $20.32\pm0.04$. Moreover, $A_{\rm Fe}$ is almost close to the solar abundance, which is expected as the secondary star is a weakly evolved  low-mass donor star \citep{joanna2022}. We will discuss the inner disk geometry of the hard state in detail in Section \ref{sec:discuss1}. 

In this model (Model 1A), we additionally perform fitting with leaving the inclination, and emissivity index as free parameters. But, their values remain unconstrained in their allowed ranges. We also consider a single Comptonization model for performing this reflection study, by tying $\Gamma$, $kT_{\rm e}$, and $kT_{\rm seed}$ of two \texttt{reflionxhd} components with that of a single \texttt{nthcomp} component. This model results in a poorer fit, with $\chi^2/\text{d.o.f}=2546.3/2391$. Additionally, $kT_{\rm in}$ is found to be quite low $\sim0.1$ keV and the disk is estimated to have reached almost the ISCO radius, $=2.4\pm0.3\ R_{\rm ISCO}$, inconsistent with the same calculated from the \texttt{diskbb} normalization. Furthermore, we observe significant residuals above 100 keV, which progressively increase with energy (see Fig.~\ref{rat2}). Therefore, Model 1A not only provides a better statistical description of the data but also yields physically consistent values for all model parameters. We will further consider this model for the multi-wavelength spectral analysis of this source in the hard state.

\subsubsection{UV Spectral Analysis}\label{sec:harduv}

\begin{figure}
\includegraphics[width=0.45\textwidth]{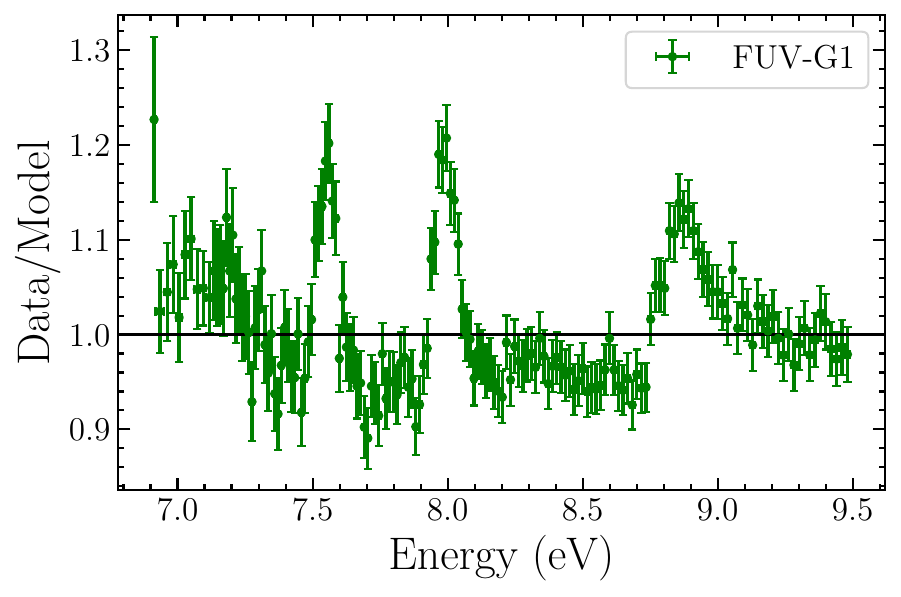}
\caption{Ratio of the 6.9 -- 9.5 eV FUV-G1 data to the model \texttt{redden}*\texttt{bbodyrad} (hard state observation). The residuals around $7.55$ eV, $8.0$ eV, and $8.89$ eV are clearly observed. See Section \ref{sec:harduv} for more details. \label{plot:harduv1}}
\end{figure}

\begin{figure}
\includegraphics[width=0.45\textwidth]{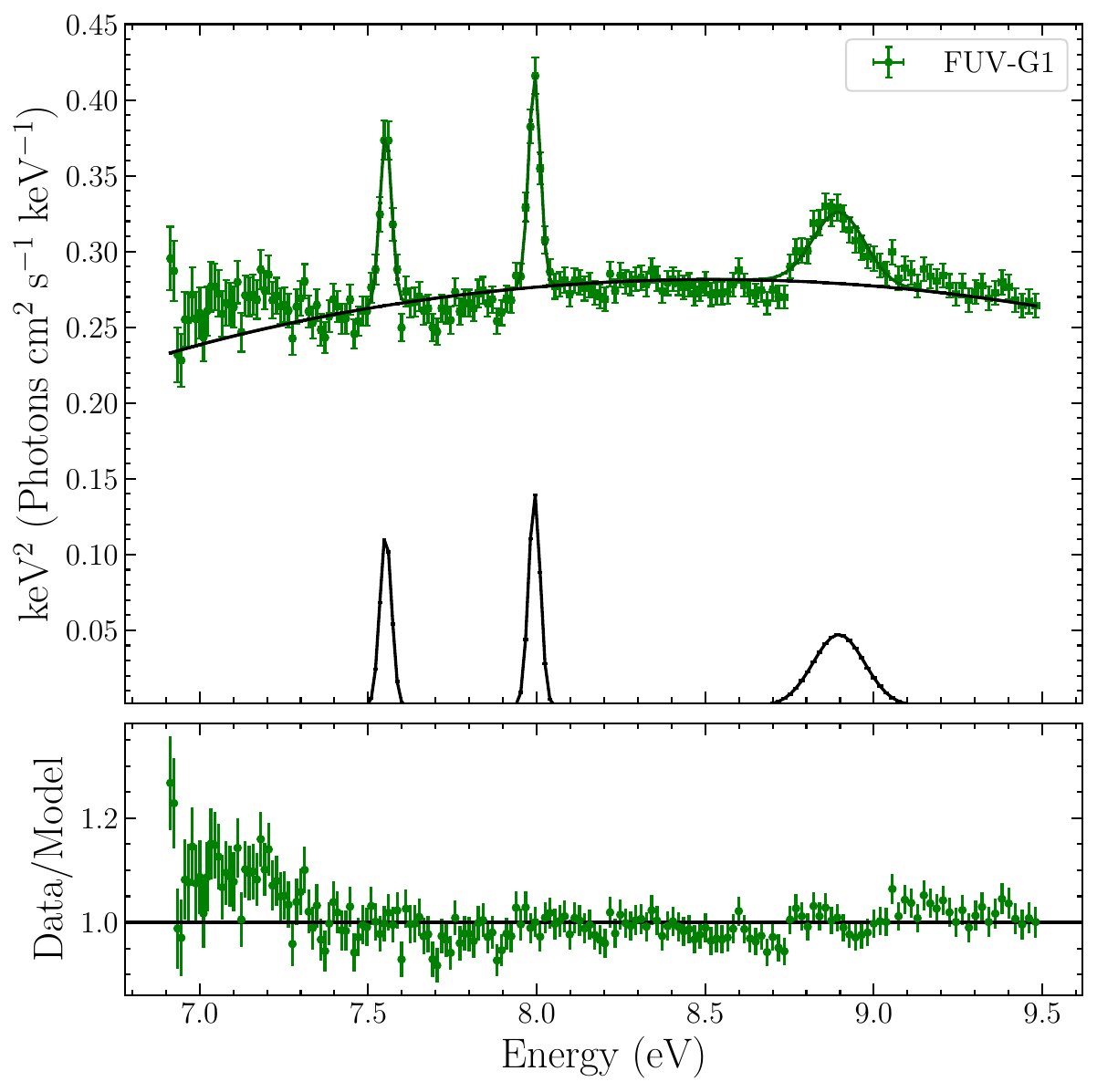}
\caption{Unfolded spectrum with model (Model 1B) components in black (upper panel) and ratio of the 6.9 -- 9.5 eV FUV-G1 data to Model 1B (lower panel) (hard state observation). See Section \ref{sec:harduv} for more details. \label{plot:harduv2}}
\end{figure}

We fit the FUV-G1 spectrum in the energy range 6.9 -- 9.5 eV with an absorbed black-body model (Model: \texttt{redden*bbodyrad}) \citep{alex2018}
and obtain a $\chi^2/\text{d.o.f}$ of $661.6/175$. Here, we fix the color excess $E(B-V)$ to 0.17 corresponding to the neutral hydrogen column density of $1.3\times10^{21}\ \rm atoms\ cm^{-2}$ along the source line of sight via equation 15 of \cite{zhu2017}, which is also close to the earlier estimated value of $E(B-V)=0.163\pm0.007$ \citep{baglio2018}. We clearly observe residuals around $7.55$ eV, $8.0$ eV, and $8.89$ eV (see Fig.~\ref{plot:harduv1}). The residuals around these three energy values most likely correspond to the emission lines:  
\ion{He}{2}~$\lambda$1640.4, \ion{C}{4}~$\lambda$1549.1, and \ion{Si}{4}~$\lambda$1396.8 \citep{berk2001}. We add three Gaussian lines (\texttt{gauss} in \texttt{XSPEC} notation) to account for these features, and notice that the Gaussian width ($\sigma$) of the emission line \ion{Si}{4} is significantly broader than the other lines. The \ion{Si}{4} line is probably a doublet, with components at $1393\rm \AA$ and $1403\rm \AA$ \citep{morton2003}, which is responsible for making this line broader. Additionally,  this feature is sometimes contaminated by or potentially even dominated by a semi-forbidden emission line 
\ion{O}{4]}~$\lambda$1402.1  \citep{berk2001}. We tie the line-widths of \ion{He}{2} and \ion{C}{4} lines, and leave that of  \ion{Si}{4} as a free parameter. Thus, our new model (hereafter, Model 1B) becomes: 
\begin{itemize}
    \item Model 1B: \texttt{redden}*(\texttt{gauss(\ion{He}{2})}+\texttt{gauss(\ion{C}{4})}\\+\texttt{gauss(\ion{Si}{4})}+\texttt{bbodyrad(UV)}).
\end{itemize} 
This model provides a $\chi^2/\text{d.o.f}$ of $222.1/165$. The results are presented in Table~\ref{tab:harduv}, and the unfolded spectrum and residuals in the form of a ratio (model/data) are depicted in Fig.~\ref{plot:harduv2}.
\subsubsection{Broadband Optical/UV/X-ray Spectral Analysis}\label{sec:hardtotal}

\begin{figure}
\includegraphics[width=0.45\textwidth]{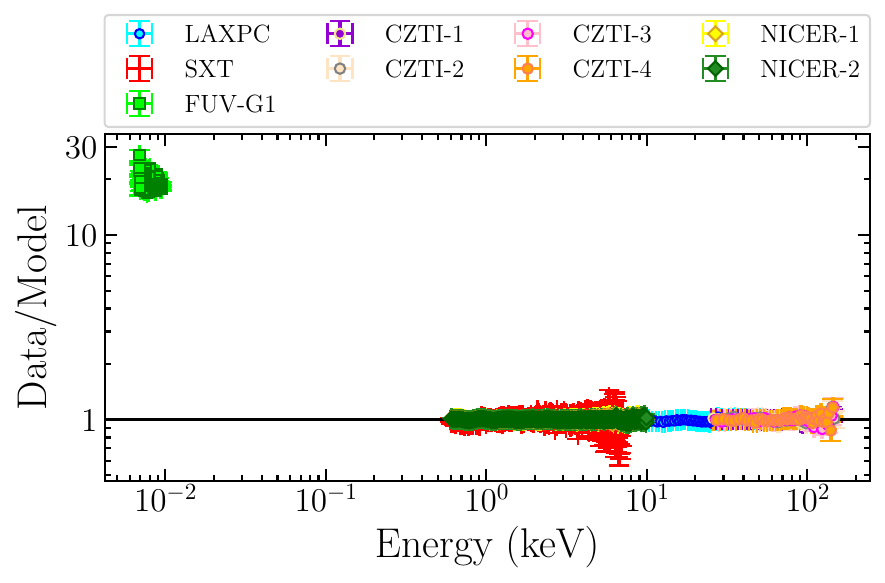}
\caption{Ratio (Data/Model) of the 0.0069 -- 150.0 keV multi-wavelength (\textit{AstroSat}+\textit{NICER}) data to the model: Model 1A (hard state observation). We see a huge UV excess below 10 eV. See Section \ref{sec:hardtotal} for more details.\label{plot:harduvexcess}}
\end{figure}

\begin{figure}
\includegraphics[width=0.45\textwidth]{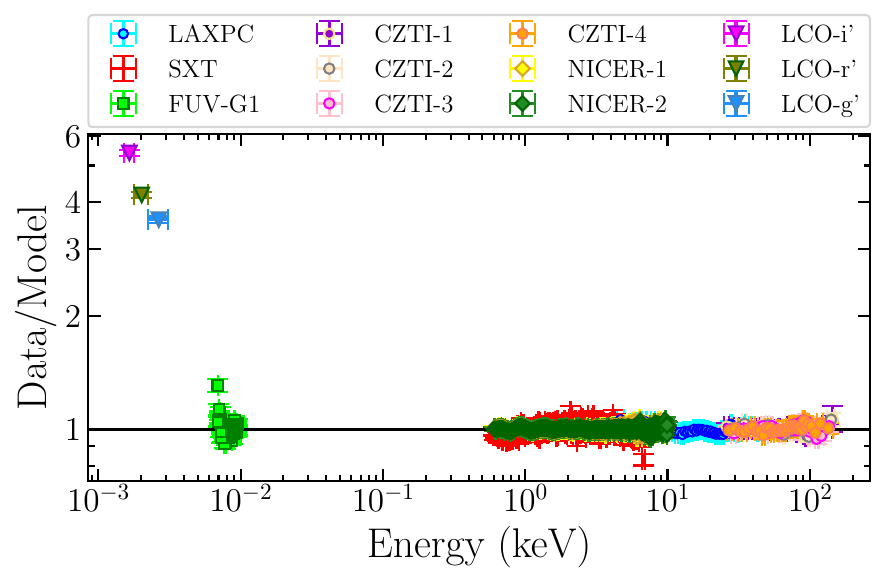}
\caption{Ratio (Data/Model) of the 0.00164 -- 150.0 keV multi-wavelength (\textit{AstroSat}+\textit{NICER}+\textit{LCO}) data to the combined model: Model 1A + Model 1B (hard state observation). We note an optical excess below 5 eV. See Section \ref{sec:hardtotal} for more details.\label{plot:hardtotalinit}}
\end{figure}

\begin{figure}
\centering
\includegraphics[width=0.5\textwidth]{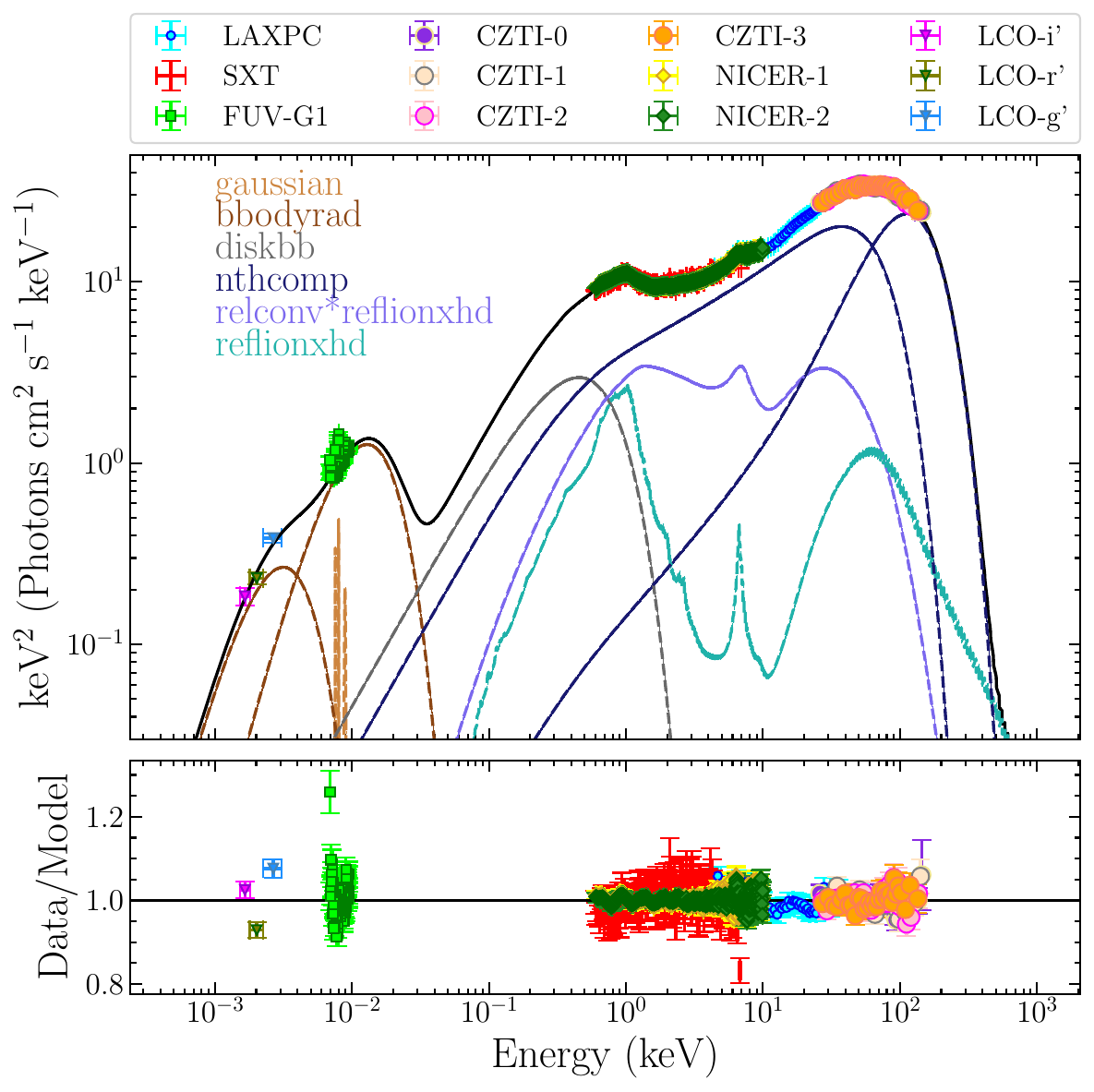}
\caption{Broad-band (Optical to hard X-ray) unabsorbed SED (upper panel) and residuals (lower panel), in the form of ratio (data/model), corresponding to Model 1C (hard state observation). The total model is represented by a solid black line in the upper panel. Data are rebinned for plotting purpose.  See Section \ref{sec:hardtotal} for more details. \label{plot:hardtotal}}
\end{figure}

We will now perform a multi-wavelength spectral study of this source in the energy range 1.64 eV -- 150 keV and investigate the correlation between the spectral parameters in the X-ray and optical/UV bands. 
We first add the FUV-G1 spectral data to our X-ray datasets and extrapolate the best-fit X-ray model (Model 1A). We observe a huge UV excess below 0.01 keV (see Fig.~\ref{plot:harduvexcess}). This indicates that our X-ray model severely underestimates the UV flux, implying the dominance of the effect of irradiation in the outer accretion disk \citep{done2009}. Thus, we add our best fit UV model, Model 1B, to the X-ray spectral model 1A to describe the UV emission in the hard state spectra.
We set $N_{\rm H}=0$ for the FUV-G1 spectrum and $E(B-V)=0$ for the X-ray part of the spectra. Additionally, we fix the Gaussian centroid energies and widths of the emission lines, and the \texttt{bbodyrad} temperature in this model to their respective values in Model 1B, and keep only the normalization of these components as free parameters. The parameter \texttt{constant} is also kept frozen to unity for the FUV spectrum.
Thus, the present model takes the form, \texttt{tbabs*redden*constant*(diskbb+nthcomp(1)
+nthcomp(2)+reflionxhd+relconv*reflionxhd+
gauss(\ion{He}{2})+gauss(\ion{C}{4})+gauss(\ion{Si}{4})+bbodyrad(UV}). We obtain a $\chi^2/\text{d.o.f}$ of 2152.5/2561. Finally, we add \textit{LCO} data to this setup and find an unacceptable fit with a $\chi^2/\text{d.o.f}$ of 6543.5/2564. We note that adding \textit{LCO} data to our previous model results in significant residuals below 5 eV (see Fig.~\ref{plot:hardtotalinit}). To circumvent the issue, we add another \texttt{bbodyrad} component to this model to take care of the excess observed in the mentioned energy band. This results in our model, Model 1C: 
\begin{itemize}
    \item Model 1C: \texttt{tbabs}*\texttt{redden}*\texttt{constant}*(\texttt{diskbb}\\
    +\texttt{nthcomp(1)}+\texttt{nthcomp(2)}+\texttt{reflionxhd(2)}\\
+\texttt{relconv}*\texttt{reflionxhd(1)}+\texttt{gauss(\ion{He}{2})}\\
+\texttt{gauss(\ion{C}{4})}+\texttt{gauss(\ion{Si}{4})}+\texttt{bbodyrad(UV)}\\
+\texttt{bbodyrad(optical})).
\end{itemize}
We observe a substantial improvement in the spectral fit, $\chi^2/\text{d.o.f}=2157.7/2562$, and the previously mentioned residuals consequently decrease significantly (see Fig.~\ref{plot:hardtotal}). Besides, all the X-ray spectral parameters in this new model take almost identical values to the same parameters of Model 1A, i.e., the previous spectral fit does not get affected due to the inclusion of \textit{LCO} data. The results are given in Table~\ref{tab:hardtotal} and the broadband unabsorbed SED (along with residuals) is provided in Fig.~\ref{plot:hardtotal}.
We estimate the reprocessed fraction in this model by taking the ratio of the flux in the 0.5 -- 10.0 eV band (the flux contribution below 0.5 eV is $\lesssim1\%$) to that in the 0.1 -- 200.0 keV band, and find this quantity to be $\sim 9\times10^{-3}$. Since the outer disk can emit a significant fraction of photons in the 0.5 -- 10.0 eV band through viscous dissipation, we consider only the flux of two \texttt{bbodyrad} components and 3 emission lines in that band for calculating the value of reprocessed fraction.
\subsection{Soft State}
\subsubsection{X-ray Spectral Analysis}\label{sec:softx}
We fit the SXT+LAXPC soft state spectra in the  $0.6-40.0$ keV band with a model comprising a multicolored disk black-body component (\texttt{diskbb}; \citealt{mitsuda1984,makishima1986}) and a thermal Comptonization component (\texttt{thcomp}; \citealt{zdziarski2020}) to describe the weak Comptonization in the soft state. We additionally require a single-temperature black-body component (\texttt{bbodyrad})
to achieve a good fit (F-test probability of chance improvement is $\sim10^{-90}$). The black-body component was earlier detected with the \nustar{} data with similar parameters, and proposed to represent the radiation from the plunging region \citep{fabian2020}. The Comptonization component \texttt{thcomp} is described by 3 parameters: $\Gamma$, $kT_{\rm e}$, and covering fraction $cov\_frac$. We convolve this component over \texttt{diskbb} and \texttt{bbodyrad}, as both of them can provide soft seed photons for Comptonization. Thus, we finally arrive at a simple three component model,  
\begin{itemize}
    \item Model 2A: \texttt{tbabs}*\texttt{constant}*\texttt{thcomp}*(\texttt{diskbb}\\
    +\texttt{bbodyrad}).
\end{itemize}
A similar model was earlier used by \cite{fabian2020} to describe the broad-band soft-state \textit{NuSTAR} spectra of this source. In their model, a \texttt{cutoffpl} component was employed to describe the weak Comptonization component, rather than a more physically meaningful \texttt{thcomp} component. Our Model 2A reasonably describes the soft-state spectrum, yielding a $\chi^2$/\text{d.o.f} of 490.1/413. The best-fit parameters for this model are listed in Table~\ref{tab:softx}. The value of the cross-normalization factor between SXT and LAXPC is found to be $1.21\pm 0.05$ ($\sim20\%$), which falls within the accepted limit \citep{antia2021}. We obtain a hydrogen column density ($N_H$) of $\sim0.8\times10^{21}\ \rm atoms\ cm^{-2}$, which is close to the Galactic column density in the direction to the source, $1.3\times10^{21}\ \rm atoms\ cm^{-2}$ \citep{hi4pi2016}. The disk ($kT_{\rm in}$) and black-body ($kT_{\rm BB}$) temperatures, and the electron temperature, $kT_{\rm e}$ ($0.58\pm0.02$ keV, $0.79\pm0.02$ keV, and $>36.5$ keV, respectively) are found to be roughly consistent with the nearest ($0.64\pm0.01$, $0.92\pm0.02$, and $64.6_{-26}^{+116}$, respectively) \textit{NuSTAR} observation \citep{fabian2020} (Observation Nu31 in their paper; the AstroSat observation was performed $\sim6$ days after the Nu31 observation). Furthermore, we find the value of $cov\_frac\sim5\times10^{-3}$ to be quite small, implying a very weak Comptonization component. The \texttt{bbodyrad} normalization implies that the X-ray emission is coming from a radius of $\simeq 42-53$ km, considering $D=2.96$ kpc and $\kappa=1.7$. This region is found to be smaller than the true inner disk radius, $\simeq88-99$ km, as estimated from the \texttt{diskbb} normalization using equation (\ref{eqn2}) (we consider $\kappa=1.7$, $\eta=0.4$, $i=64^{\circ}$ and $D=2.96$ kpc for this calculation). Since the disk fraction is $\sim85\%$ (i.e., the fraction of disk flux to the total flux in the 0.1-200.0 keV range) in this observation, the inner disk can be assumed to reach the ISCO radius \citep{mcclintock2014}. Therefore, the X-ray emission associated with the \texttt{bbodyrad} component could be related to the radiation coming from the plunging region \citep{fabian2020}.

We will now replace the \texttt{diskbb} component with a more sophisticated model \texttt{kerrbb} \citep{li2005} to describe the emission from a geometrically thin and optically thick accretion disk around a spinning BH (i.e., a Kerr BH). This model takes into account general relativistic effects such as frame-dragging, Doppler boost, gravitational redshift, and bending of light caused by the gravity of a Kerr BH. The spin and mass of the BH, along with the inclination of the inner accretion disk, serve as input parameters for the \texttt{kerrbb} model, in addition to the distance to the source, which we fix at 2.96 kpc, as determined in \cite{atri2020}. In our spectral analysis, we incorporate the effects of both limb-darkening and returning radiation by setting both \texttt{r\_flag} and \texttt{l\_flag} of \texttt{kerrbb} to 1. In addition, we set the spectral hardening factor to the default model value of $\kappa=1.7$. Thus, the new model takes the form,
\begin{itemize}
    \item Model 2B: \texttt{tbabs}*\texttt{constant}*\texttt{thcomp}*(\texttt{kerrbb}\\
    +\texttt{bbodyrad})
\end{itemize}
Fitting this model to the data gives a $\chi^2$/\text{d.o.f} of 464.1/410. The results from the fit are presented in Table~\ref{tab:softx}. 
The value of $N_{\rm H}$ ($\sim1.10\ \times 10^{21}\ \rm atoms\ cm^{-2}$) is found to be close to that of Model 2A, and the cross-normalization constant a little higher, $1.25\pm0.06$. 

The spin ($a$) and mass ($M$) of the BH in this model come out to be $>0.84$ and $9.73_{-2.52}^{+2.25}\ \rm M_{\sun}$, respectively, for an inclination of $46.8_{-10.1}^{+4.7}$. \cite{torres2020} found that the inclination of the binary ($i_b$) lies in the range of $66^{\circ}-81^{\circ}$ based on their intermediate-resolution spectroscopic analysis of the optical counterpart of MAXI~J1820+070. They also provided a prediction for the BH's mass: $M=(5.95\pm0.22)M_{\sun}/\rm sin^3i_b$. For inclinations between $66^{\circ}-81^{\circ}$, this relationship yields a mass range of $5.73-8.34\ M_{\sun}$, which is slightly smaller than our estimated value.
Interestingly, there is a significant discrepancy in the measurement of the spin of this BH between several studies. \cite{zhao2021} performed a continuum-spectral analysis of this source in the soft state, similar to our approach, but with a different model (\texttt{kerrbb2}) and using \textit{Insight-HXMT} data. They found a slowly rotating BH with a spin of $0.14\pm0.09$ ($1\sigma$), assuming a BH mass of $8.48$ and an inclination of the inner disk of $63^{\circ}$. Their analysis also indicated that the BH most likely has a prograde spin if $5.73M_{\sun}< M < 8.34M_{\sun}$ and an inclination in the range of $66^{\circ}-81^{\circ}$. On the other hand, \cite{bhargava2021} analyzed the power-density spectra obtained from \textit{NICER} high cadence observations of the source in the hard state, and employed relativistic precession model to estimate the spin of the BH. Their analysis yielded a spin value of $a=0.799_{-0.015}^{+0.016}$, which is close to our value. Additionally, our estimation of the inclination angle is significantly lower than the jet inclination angle of $64^{\circ}\pm 5^{\circ}$, which is possibly identical to the angle of the BH's spin axis (see \citealt{liska2018} for an alternative scenario). 

The values of mass and spin of a BH in the \texttt{kerrbb} model strongly depend on the inclination of the inner disk and the distance to the source \citep{mcclintock2014}. Therefore, we also perform spectral fitting with Model 2B, with the inclination angle fixed to the jet inclination angle, which could be used as a proxy for the inner disk inclination angle (assuming that the inner disk's angular momentum is aligned with the BH's spin axis; however, for other scenarios, see \cite{banerjee2019a, banerjee2019b}). We obtain a $\chi^2$/\text{d.o.f} of 467.2/411. The mass and spin values of this source are found to be greater than $5.9M_{\sun}$ and in the range of $0.60-0.95$ (See Table~\ref{tab:softx}), respectively, which are quite consistent with earlier measurements (we restrict the upper limit of the BH mass to $12M_{\sun}$ in our spectral fit).
\cite{fabian2020} employed a similar model, using \texttt{cutoffpl} instead of \texttt{thcomp} to represent the Comptonization component, in their analysis with \textit{NuSTAR} observations. They fixed the spin (a) and inclination ($i$) at $0.2$ and $34^{\circ}$, respectively, based on the results reported in \cite{buisson2019}. 
The best-fit temperatures of the black-body ($kT_{\rm BB} = 0.84\pm0.04$ keV) and the Comptonising corona ($kT_{\rm e} > 17$ keV) in our  model are roughly in agreement with those derived by \cite{fabian2020} from the nearest \textit{NuSTAR} observations.

We will finally consider our third model, where we utilize the irradiated disk model \texttt{diskir} \citep{done2009} to describe the broad-band X-ray continuum. In addition to this, we use the \texttt{bbodyrad} component mentioned earlier. Apart from describing the emissions from a disk and corona (\texttt{diskir} assumes \texttt{diskbb} and \texttt{nthcomp} routines for this purpose), \texttt{diskir} (has 9 parameters) considers the optical/UV emission resulting from the irradiation of the outer disk by the X-ray emission from the inner accretion disk and corona. Furthermore, this model not only describes the reprocessing of X-rays in the outer accretion disk, but also takes into account the illumination of the inner disk by the Compton tail. In essence, this model considers both the irradiation of the inner accretion disk and the outer accretion disk.
We fit the following 5 parameters of \texttt{diskir} (along with the parameters of \texttt{bbodyrad} and \texttt{tbabs}) to the joint SXT-LAXPC data: 1) the temperature of the accretion disk $kT_{\rm disk}$, the normalization (which is identical to the diskbb normalization), power-law index $\Gamma$, electron temperature $kT_{\rm e}$, and a ratio of luminosity in the Compton tail to the unilluminated disk $L_{\rm c}/L_{\rm d}$.
Since we are exclusively considering the X-ray part of the total spectra, we will keep the outer disk radius ($\rm log (r_{\rm out})=log (R_{\rm out}/R_{\rm in})$ (where $R_{\rm out}$ and $R_{\rm in}$ denote the outer disk and inner disk radii, respectively) and reprocessed fraction ($f_{\rm out}$: the fraction of bolometric X-ray luminosity thermalized in the outer disk) fixed at 4.5 and 0 (i.e., irradiation of the outer disk is turned off), respectively, as these parameters are constrained from the optical/UV spectrum. We additionally freeze the parameters $f_{\rm in}$ (the fraction of the Comptonized luminosity thermalized in the inner disk) and $r_{\rm irr}$ (the radius of the Compton illuminated disk as a fraction of the inner disk radius) to their default values of $0.1$ and $1.2$, respectively, since they remain unconstrained when left free. Thus, our full model is,
\begin{itemize}
    \item Model 2C: \texttt{tbabs}*\texttt{constant}*(\texttt{diskir}+\texttt{bbodyrad}).
\end{itemize}
We obtain a $\chi^2$/\text{d.o.f} of 490/413. The resulting best-fitting parameters are given in Table~\ref{tab:softx}. The value of disk and black-body temperatures, electron temperature, the power-law index in this model is almost identical to that of Model 2A. 
\begin{figure}
\includegraphics[width=0.45\textwidth]{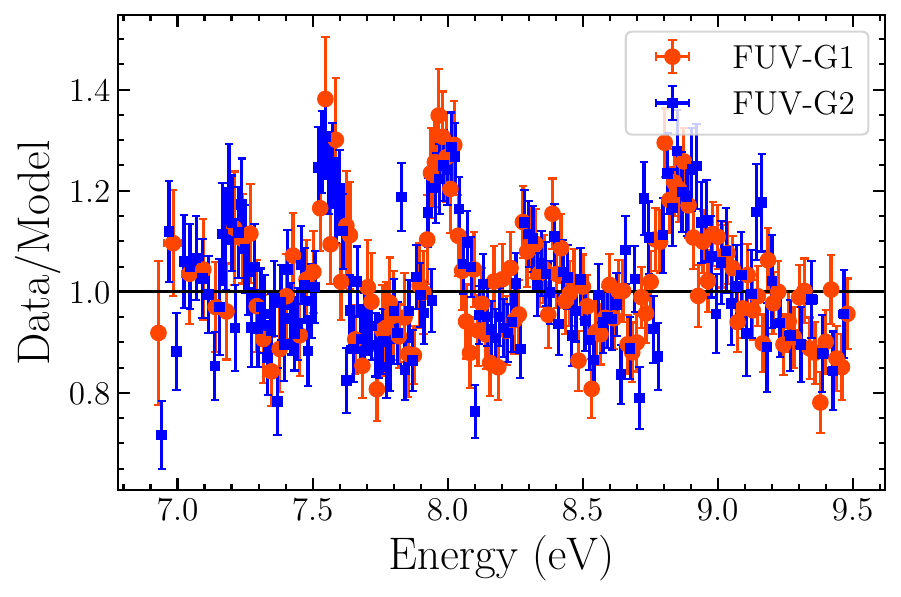}
\caption{Ratio of the 6.9 -- 9.5 eV FUV-G1 and FUV-G2 data to the model \texttt{redden}*\texttt{bbodyrad} (soft state observation). The residuals around $7.21$ eV, $7.55$ eV, $8.0$ eV, $8.34$ eV and $8.89$ eV are clearly observed. See Section \ref{sec:softuv} for more details.\label{plot:softuv1}}
\end{figure}
\begin{figure}
\includegraphics[width=0.45\textwidth]{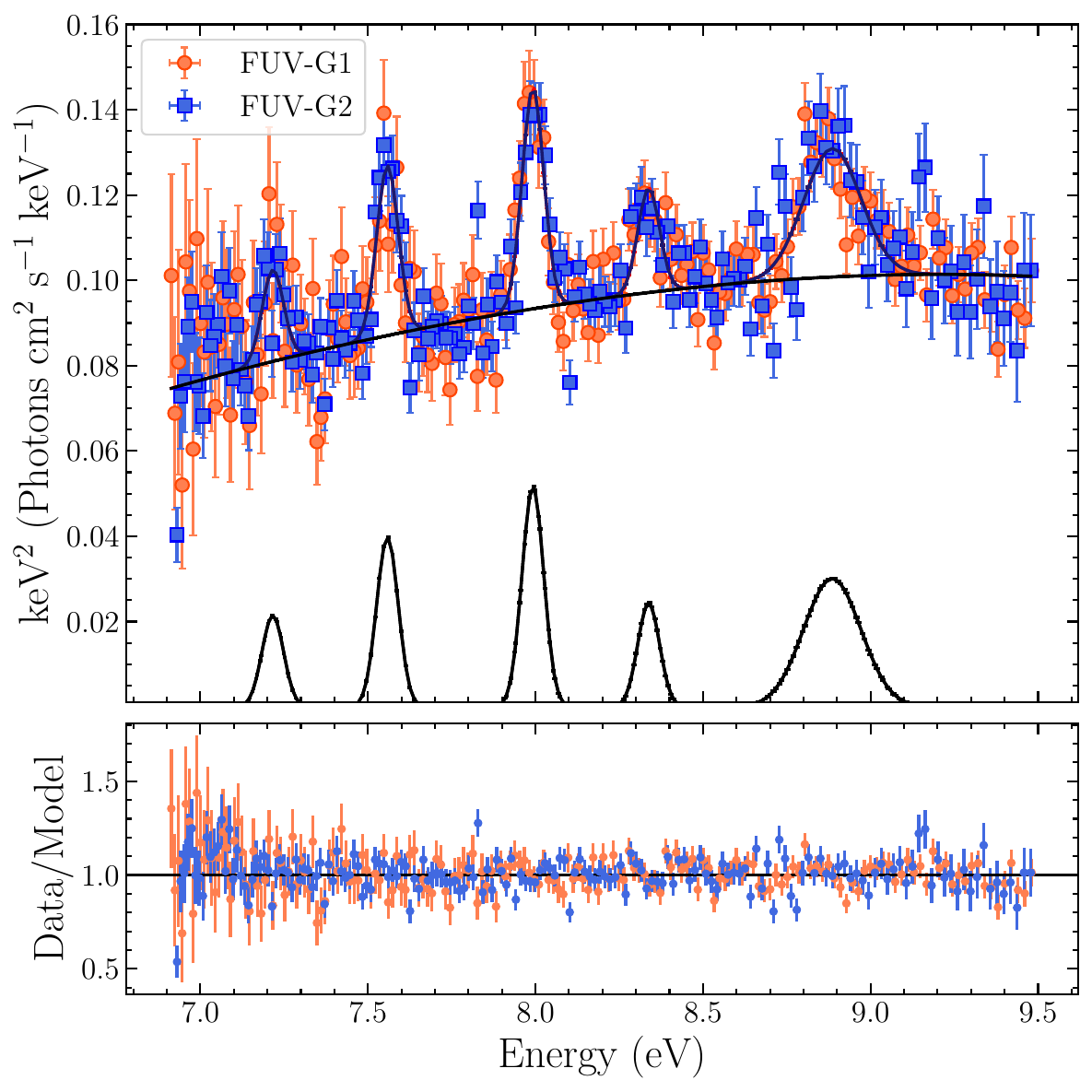}
\caption{Unfolded spectrum with model (Model 2D) components in black (upper panel) and ratio of the 6.9 -- 9.5 eV FUV-G1 and FUV-G2 data to Model 2D (lower panel) (soft state observation). See Section \ref{sec:softuv} for more details.\label{plot:softuv2}}
\end{figure}

\subsubsection{UV Spectral Analysis}\label{sec:softuv}
From the X-ray spectral fit of the joint SXT+LAXPC data, we find that the hydrogen column density along the source line of sight can be approximated as $\sim9.0 \times 10^{20}\ \rm atoms\ cm^{-2}$, which is roughly the median value in our estimated range. Hereafter, we will fix $N_{\rm H}$ to the above-mentioned value for all subsequent fits in the soft state case. This value corresponds to a color excess of $E(B-V)=0.12$ via equation 15 of \cite{zhu2017}), which is roughly consistent with the earlier estimated value of $E(B-V)=0.163\pm0.007$ \citep{baglio2018}. 

We fit the FUV-G1 + FUV-G2 spectra with an absorbed single temperature black-body (Model: \texttt{redden*bbodyrad}) \citep{alex2018} and obtain a $\chi^2$/\text{d.o.f} of 895/345. We observe large residuals around $7.21$ eV, $7.55$ eV, $8.0$ eV, $8.34$ eV and $8.89$ eV (see Fig.~\ref{plot:softuv1}). The residuals around these five energy values most likely correspond to the five emission lines: \ion{N}{4}~$\lambda$1718.5, \ion{He}{2}~$\lambda$1640.4, 
\ion{C}{4}~$\lambda$1549.1, \ion{N}{4]}~$\lambda$1486.5, and \ion{Si}{4}~$\lambda$1396.8 \citep{berk2001,harris2016}. We thus add five Gaussian lines to account for these features, and find that the width of the emission line Si IV is significantly broader than the other lines as also noted earlier for the hard state case (see Section~\ref{sec:harduv} for a discussion on this). 
Therefore, we tie the width of the Gaussians corresponding to the lines \ion{N}{4}, \ion{He}{2}, \ion{C}{4}, and \ion{N}{4]}, but leave that of \ion{Si}{4} as a free parameter. Thus, we arrive at our following model:
\begin{itemize}
    \item Model 2D: \texttt{redden}*(\texttt{gauss(N IV)}+\texttt{gauss(He II)}\\+\texttt{gauss(C IV)}+\texttt{gauss(N IV])}+\texttt{gauss(Si IV)}\\+\texttt{bbodyrad(UV)}).
\end{itemize}
This model provides a $\chi^2$/\text{d.o.f} of 408/333. The results are presented in Table~\ref{tab:softuv}, and the unfolded spectrum and residuals in the form of a ratio (model/data) are depicted in Fig.~\ref{plot:softuv2}. The temperature of the black-body component ($kT_{\rm uv}=3.87\pm0.24$ eV) is found to be slightly higher than the hard state case ($=3.27\pm0.08$ eV), although the normalization is an order of magnitude smaller than that of the hard state case (i.e., UV flux in the hard state is much higher than the soft state case).
\begin{figure}
\includegraphics[width=0.45\textwidth]{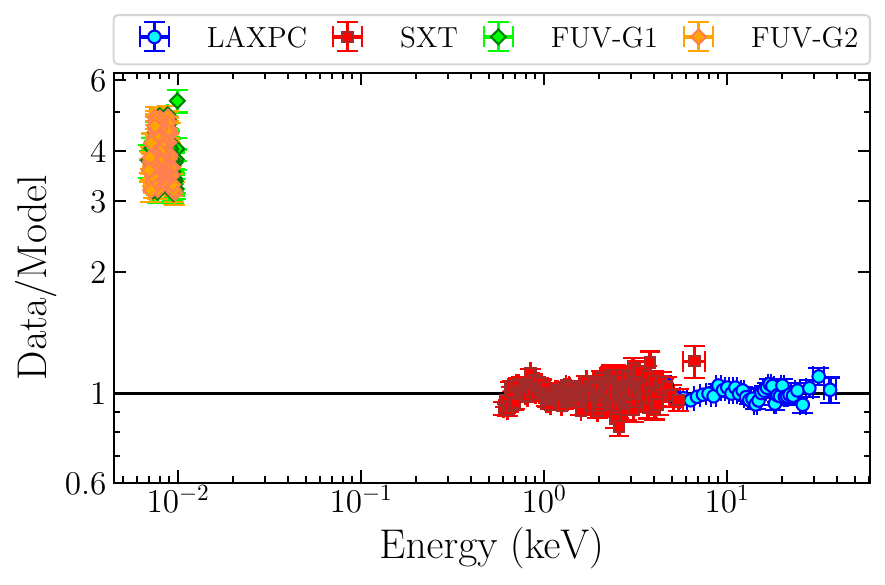}
\caption{Ratio (Data/Model) of the 0.00690 -- 40.0 keV multi-wavelength \textit{AstroSat} data to the model: Model 2A (soft state observation). We note an UV excess below 10 eV. See Section \ref{sec:softtotal} for more details.\label{plot:softuvexcess}}
\end{figure}
\begin{figure}
\includegraphics[width=0.45\textwidth]{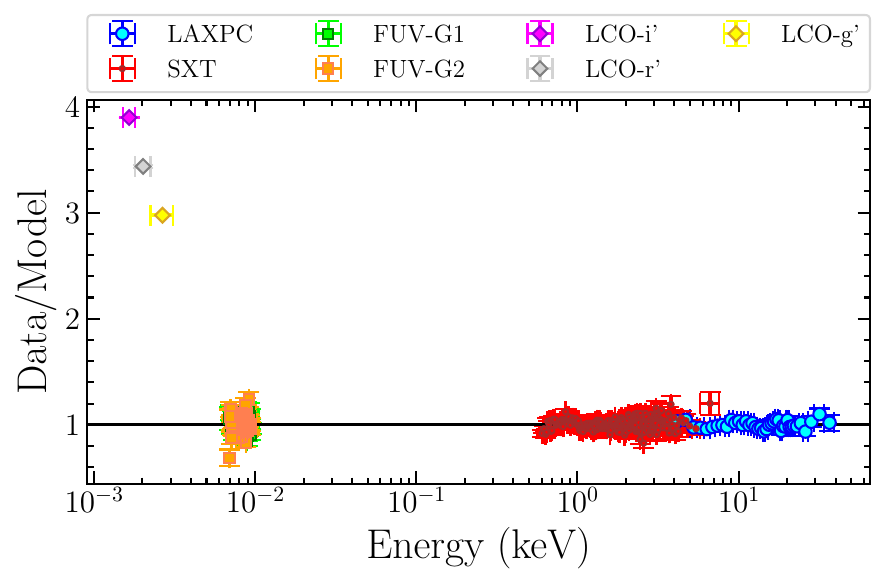}
\caption{Ratio (Data/Model) of the 0.00164 -- 40.0 keV multi-wavelength (\textit{AstroSat}+\textit{LCO}) data to the combined model: Model 2A + Model 2D (soft state observation). We see some residuals below 5 eV. See Section \ref{sec:softtotal} for more details.\label{plot:softtotalinit}}
\end{figure}
\subsubsection{Broadband Optical/UV/X-ray Spectral Analysis}\label{sec:softtotal}

\begin{figure}
\centering
\includegraphics[width=0.5\textwidth]{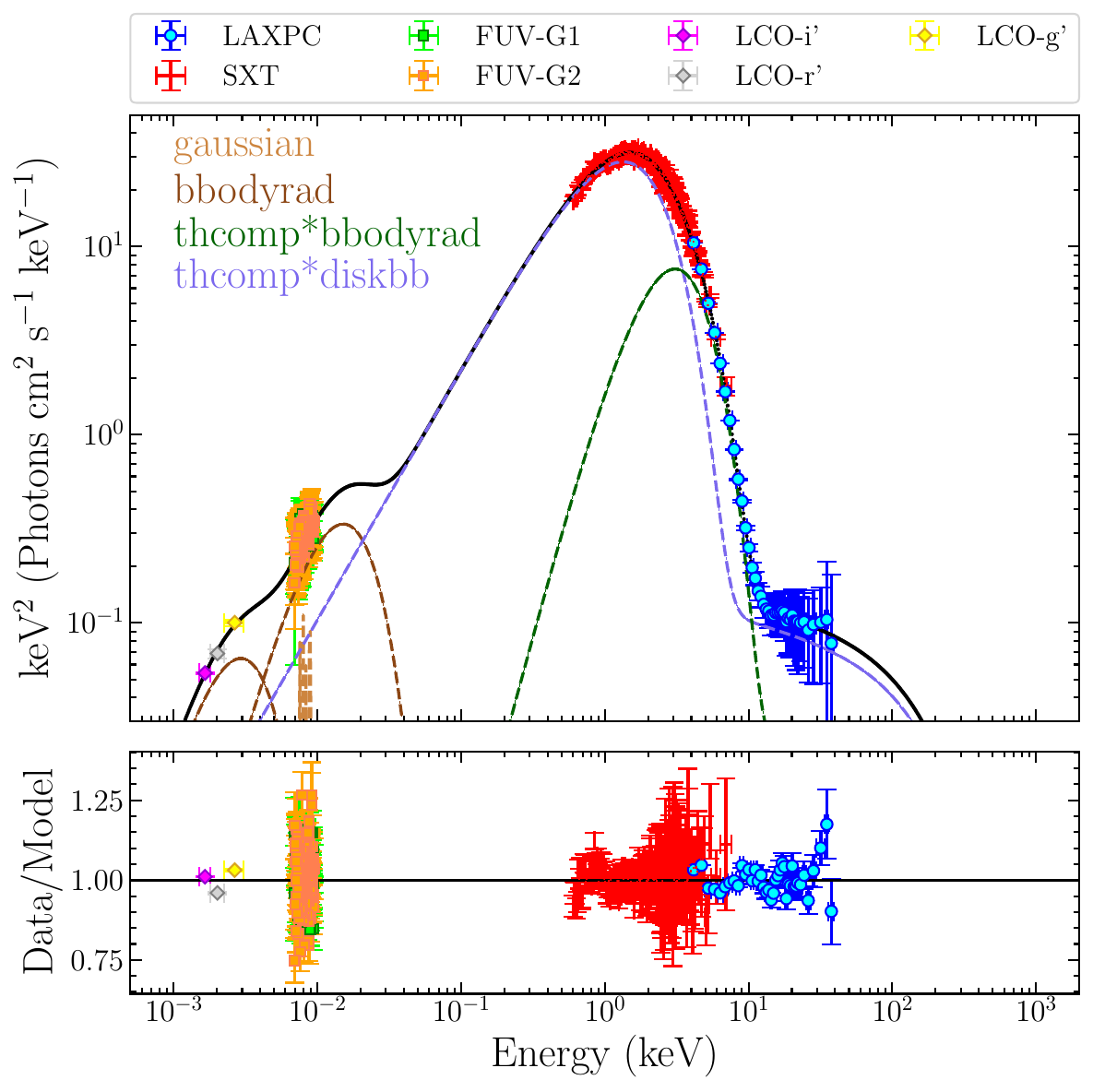}
\caption{Broad-band (Optical to hard X-ray) unabsorbed SED (upper panel) and residuals (lower panel), in the form of ratio (data/model), corresponding to Model 2E (soft state observation). The total model is represented by a solid black line in the upper panel. Data are rebinned for plotting purpose. See Section \ref{sec:softtotal} for more details. \label{total4}}
\end{figure}

We initially add the FUV-G1 and FUV-G2 spectral datasets to our X-ray datasets and extrapolate our best-fit X-ray model (Model 2A) to lower energies, and note a significant UV excess below 10 eV (see Fig.~\ref{plot:softuvexcess}). However, the UV excess in the soft state is considerably weaker than what has been observed in the hard state observation.
To account for the UV excess, we add our best fit UV model, Model 2D, to the X-ray model. Thus, we perform a joint UVIT+SXT+LAXPC spectral analysis with the combined model: Model 2A + Model 2D. We set the $N_{\rm H}=0$ for the UVIT/FUV spectra and $E(B-V)=0$ for the SXT and LAXPC spectra. Furthermore, we keep all the parameters of Model 2D fixed, except for the normalizations of the individual spectral components. This combined model yields a reasonable fit to the data, with a $\chi^2/\text{d.o.f}$ of 888.2/755. Now, \textit{LCO} data are added to this setup, resulting in a fit with $\chi^2/\text{d.o.f}$ of 1781.9/758. We observed some residuals below 5 eV in the present model (see Fig.~\ref{plot:softtotalinit}), and add another \texttt{bbodyrad} component empirically to take care of the optical excess, as we did previously for the hard state case. Our new model thus becomes, 
\begin{itemize}[align=parleft,left=0pt..1em]
    \item Model 2E: \texttt{tbabs}*\texttt{redden}*\texttt{constant}*(\texttt{bbodyrad(UV)}\\
    +\texttt{gauss(\ion{N} {4})}+\texttt{gauss(\ion{He}{2})}+\texttt{gauss(\ion{C}{4})}\\+\texttt{gauss(\ion{N}{4]})}+\texttt{gauss(\ion{Si}{4})}+ \texttt{bbodyrad(optical)}\\
    +\texttt{thcomp}*(\texttt{diskbb}+\texttt{bbodyrad})).
\end{itemize}
We achieve a significant improvement in the spectral fit, obtaining a $\chi^2/\text{d.o.f}$ of 900.5/756. Consequently, the residuals below 5 eV are also notably reduced (see Fig.~\ref{total4}).
The results are given in Table~\ref{tab:softtotal}. The broad-band unabsorbed SED and the residuals are depicted in Fig.~\ref{total4}. The temperature ($kT_{\rm optical}$) of the new \texttt{bbodyrad} component ($=0.75\pm0.04$ eV) is found to be close to the same in the hard state case ($=0.80\pm0.03$ eV). However, the corresponding normalization is substantially smaller, suggesting a higher optical flux in the hard state case. We estimate the reprocessed fraction in this model by taking a ratio of the flux in the 0.5-10.0 eV band (the flux contribution below 0.5 eV is $\lesssim1\%$) to the flux in the 0.1-200.0 keV band, and find this quantity to be quite smaller ($\sim3.5\times10^{-3}$) than the hard state case. While calculating the flux in the 0.5-10.0 eV band for determining the reprocessed fraction, we do not consider the contribution of the disk (i.e., the \texttt{diskbb} flux) since the disk can intrinsically emit a significant fraction of optical/UV photons through viscous dissipation.

\begin{figure}
\centering
\includegraphics[width=0.5\textwidth]{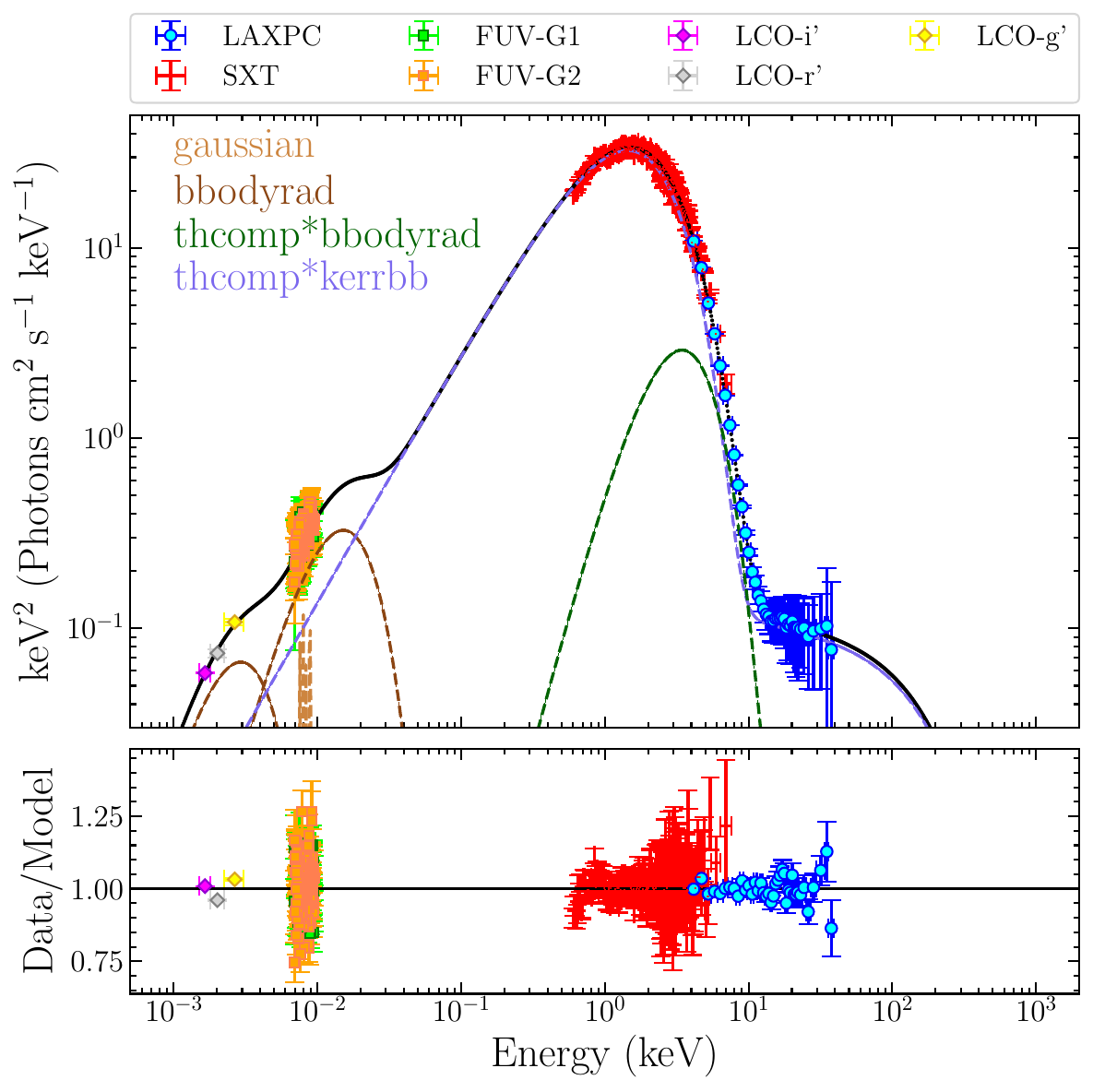}
\caption{Broad-band (Optical to hard X-ray) unabsorbed SED (upper panel) and residuals (lower panel), in the form of ratio (data/model), corresponding to Model 2F (soft state observation). The total model is represented by a solid black line in the upper panel. Data are rebinned for plotting purpose. See Section \ref{sec:softtotal} for more details. \label{total2}}
\end{figure}

Similarly, we add \textit{LCO} data to the UV/X-ray data, and fit the data with our combined model: Model 2B + Model 2D, and find significant residuals below 5 eV. Therefore, just like the previous case, we consider another \texttt{bbodyrad} component to describe the optical excess, and obtain our new model, \begin{itemize}[align=parleft,left=0pt..1em]
    \item Model 2F: \texttt{tbabs}*\texttt{redden}*\texttt{constant}*(\texttt{bbodyrad(UV)\\
    +\texttt{gauss(\ion{N} {4})}+\texttt{gauss(\ion{He}{2})}+\texttt{gauss(\ion{C}{4})}\\+\texttt{gauss(\ion{N}{4]})}+\texttt{gauss(\ion{Si}{4})}+\texttt{bbodyrad(optical)}\\+\texttt{thcomp}*(\texttt{kerrbb}+\texttt{bbodyrad})}).
\end{itemize}
We obtain a $\chi^2/\text{d.o.f}$ of 881.9/756. Similar to the previous case, the residuals below 5 eV reduces significantly (see Fig.~\ref{total2}).
The results are presented in Table~\ref{tab:softtotal}. The broad-band unabsorbed SED and the residuals are depicted in Fig.~\ref{total2}.
In this model, we set the inclination and the mass of the BH to $64^{\circ}$ and $6.75M_{\odot}$, respectively. Additionally, we fix the value of the Kerr parameter at $a=0.75$, which approximately represents the median value within our estimated range for this parameter. Notably, we observe that the spectral parameters in Models 2E and 2F generally exhibit good agreement with each other.

Now, we explore whether the necessity of two black-body components is a direct consequence of our choice of $E(B-V)$. To investigate this, we leave both the parameters $N_{\rm H}$ and $E(B-V)$ as free parameters in our Model 2E. This results in a slightly worse fit with a $\chi^2/\text{d.o.f}$ of 893.7/754 and a somewhat lower value of $E(B-V)\approx0.09$ (the value of $N_{\rm H}$ remains close to its fixed value). However, it's worth noting that the two \texttt{bbodyrad} components and the emission lines remain statistically significant. Subsequently, we fix the values of $N_{\rm H}$ and $E(B-V)$ to $0.13\times10^{21}\ \rm atoms\ cm^{-2}$ \citep{hi4pi2016} and 0.17, respectively, in our Model 2E, which are the standard values of these quantities in the literature (we consider these values in the hard state case). This results in a significantly poorer fit, with a $\chi^2/\text{d.o.f}$ of 1012.8/756. Nevertheless, both \texttt{bbodyrad} components (and the emission lines) remain statistically required to achieve a reasonable fit. In both the cases, our statements regarding the soft state inner and outer geometry do not change.

\begin{figure}
\includegraphics[width=0.5\textwidth]{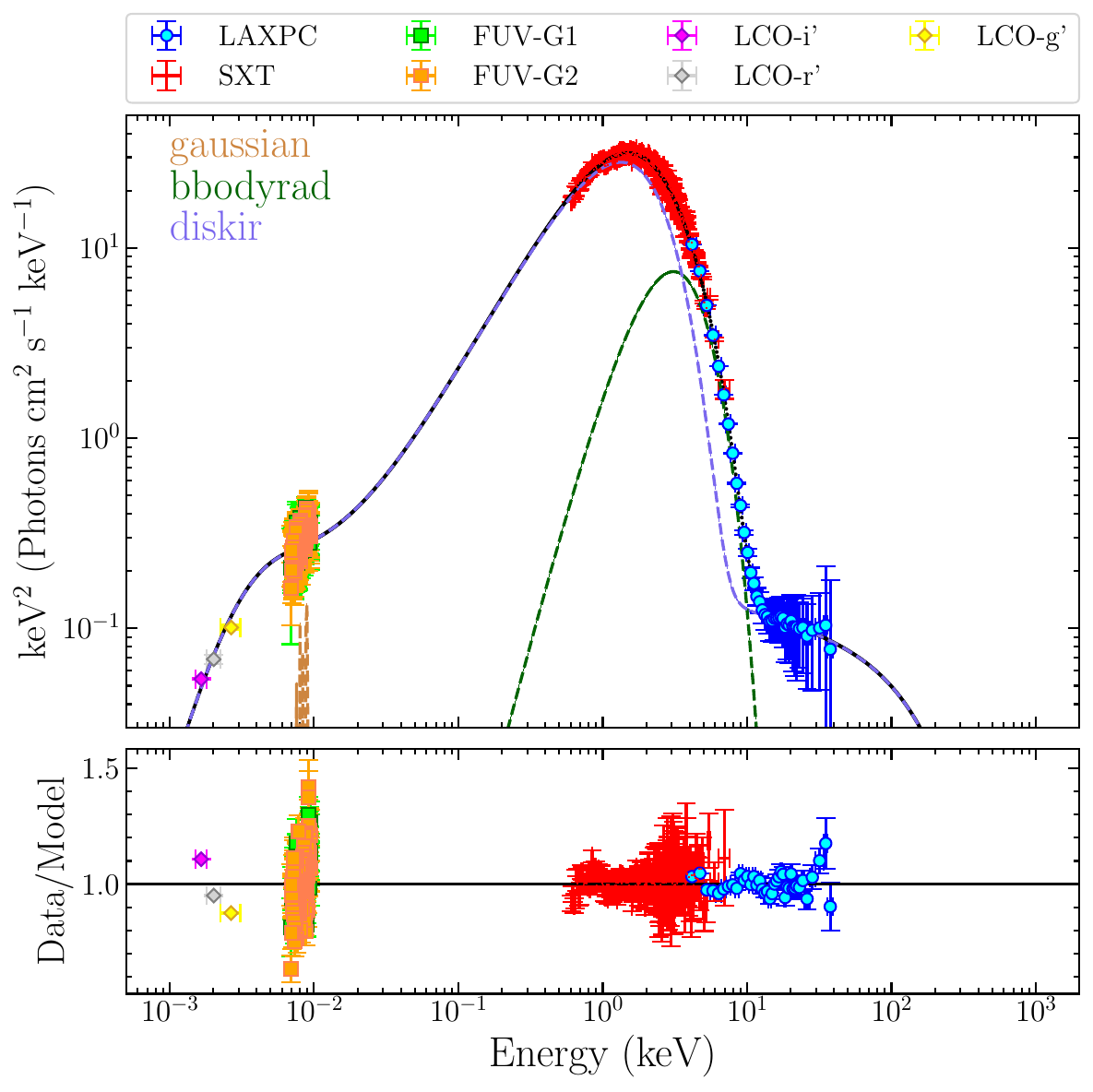}
\caption{Broad-band (Optical to hard X-ray) unabsorbed SED (upper panel) and residuals (lower panel), in the form of ratio (data/model), corresponding to the Model 2G (soft state observation). The total model is represented by a solid black line in the upper panel. Data are rebinned for plotting purpose. See Section \ref{sec:softtotal} for more details. \label{total3}}
\end{figure}

Finally, we perform a fit to the \textit{LCO}+UVIT+SXT+ LAXPC data spanning the energy range from 1.64 eV to 40 keV using the combined model: Model 2C + Model 2D. Similar to the previous case, we leave only the normalizations of Model 2D unfrozen, set $E(B-V)=0$ for the X-ray part, and $N_{\rm H}=0$ for the optical/UV part of the spectra. Since we are considering optical/UV data here, we keep the parameters $f_{\rm out}$ and $\rm log$$(r_{\rm out})$ free during the fitting. In this new model, we exclude the \texttt{bbodyrad} component from Model 2D, as the optical/UV continuum is already accounted for by the \texttt{diskir} component through the parameters $f_{\rm out}$ and $log(r_{\rm out})$. Additionally, we find that the emission line \ion{N}{4} becomes statistically insignificant in this model, possibly due to a shift in the UV continuum. So, we remove the Gaussian component corresponding to this line. Therefore,  our final model becomes, 
\begin{itemize}[align=parleft,left=0pt..1em]
    \item Model 2G: \texttt{tbabs}*\texttt{redden}*\texttt{constant}*(\texttt{diskir}\\+\texttt{gauss(\ion{N} {4})}+\texttt{gauss(\ion{He}{2})}+\texttt{gauss(\ion{C}{4})}\\+\texttt{gauss(\ion{N}{4]})}+\texttt{gauss(\ion{Si}{4})}+\texttt{bbodyrad}).
\end{itemize}
This model provides a $\chi^2/\text{d.o.f}$ of 1230.6/758. The results are given in Table~\ref{tab:softtotal}, and the unabsorbed SED and residuals are shown in Fig.~\ref{total3}. Thus, this model provides a poorer fit to the data compared to the previous phenomenological models, Model 2E and Model 2F. 

The value of the reprocessed fraction ($\sim2\times10^{-3}$) obtained from our spectral fit is consistent with that of other BH-LMXBs in the soft state \citep{done2009}. Since the \texttt{diskir} normalization is identical to the \texttt{diskbb} normalization and $r_{\rm out}=R_{\rm out}/R_{\rm in}$, we can estimate the outer disk radius ($R_{\rm out}$) from the value of $\rm log$$(r_{\rm out})$ ($=4.38\pm0.02$) using equation (\ref{eqn2}) (as $R_{\rm in}$ can be estimated from \texttt{diskir} normalization). 
Adopting $\kappa=1.7$, $\eta=0.4$, and $i=64^{\circ}$, we find that the size of the disk, $R_{\rm out}$, is $(=2.30\pm0.33\times10^{11})$ cm. The size of an accretion disk cannot be smaller than the circularization radius due to the conservation of angular momentum, and larger than the tidal truncation radius. To check the consistency of our result, we determine the values of circularization radius ($R_{\rm circ}$) and tidal truncation radius ($R_{\rm tidal}$) using equations (11) and (12) of \cite{gilfanov2005}, respectively. We find that $R_{\rm circ}\simeq0.27R_{\rm orb}$ ($R_{\rm orb}$ is the orbital separation) and $R_{\rm tidal}\simeq0.57R_{\rm orb}$, assuming a mass ratio of $q=0.072$ \citep{torres2020}. We thus use Kepler's third law of motion to calculate $R_{\rm orb}$, and obtain $R_{\rm orb}\simeq4.75\times10^{11}$ cm, considering an orbital period of 16.45 hr for this binary system \citep{torres2020}. Therefore, $R_{\rm out}$ lies in between $R_{\rm tidal}(\simeq2.71\times10^{11}\rm \ cm)$ and $R_{\rm circ}(\simeq1.28\times10^{11}\rm \ cm)$. \cite{torres2020} approximated the outer disk radius at the time of their observation as $0.6b_1/R_{\rm orb}$, where $b_1$ is the distance of the primary from the $L_1$ point. Using equation (4.9) of \cite{fkr2002}, one finds out $b_1/R_{\rm orb}\simeq0.76\Rightarrow$ $R_{\rm out}\simeq2.16\times10^{11}$ cm. Hence, our estimated value of the disk size is also consistent with earlier reported value.

\section{Summary of Main Results and Discussion} \label{sec:discuss}
During the first \astrosat{} observation in March 2018, MAXI~J1820+070 was in the hard state, emitting at an X-ray luminosity (0.1-200.0 keV; hereafter, the total/broadband X-ray flux corresponds to the 0.1-200.0 keV energy range) of $\sim 2\times10^{38}\ \rm erg\ s^{-1}$ ($\sim23.5\%$ of Eddington luminosity, calculated assuming $M=6.75M_{\sun}$).
Our main results from the multi-wavelength spectral analysis of the hard state data can be summarized as follows.
\begin{enumerate}
    \item The hard state 0.6 -- 150.0 keV X-ray spectra are found to be composed of two Comptonized emission components with different power-law indices and electron temperatures, their associated reflection components, and a disk component with a temperature of $0.19\pm0.01$ keV (see Section \ref{sec:hardx}, Figure \ref{plot:hardtotal}, and Table~\ref{tab:hardx} for more details). 
    \item The softer Comptonization component ($\Gamma=1.59\pm0.01$, $kT_{\rm e}=15.39\pm0.35$ keV) dominates the broadband X-ray luminosity, providing $\sim50\%$ of the total flux, and gets reflected from a strongly ionized disk ($\xi=2365.5_{-46.9}^{+64.5}{\rm~erg~cm~s^{-1}}$), generating relativistic reflection component. On the other hand, the harder Comptonization component ($\Gamma=1.17\pm0.01$, and $kT_{\rm e}=30.6\pm1.0$ keV) contributes $\sim30\%$ of the total flux and produces  unblurred reflection features from a weakly ionized disk ($\xi=488.1_{-12.0}^{+13.8}{\rm~erg~cm~s^{-1}}$), situated far from the ISCO radius.
    \item The inner accretion disk is truncated at far from the BH, $\simeq(51-78) R_g$ and the density of the disk is quite high, $\sim 2\times 10^{20}{\rm~cm^{-3}}$.
    \item We detect optical/UV emission in excess of the standard multi-temperature black-body disk emission (see Sections~\ref{sec:harduv} and \ref{sec:hardtotal}, Figures~\ref{plot:harduvexcess} and \ref{plot:hardtotalinit}, and Tables~\ref{tab:harduv} and \ref{tab:hardtotal} for more details).
    \item The UV excess emission is  described by a low-temperature black-body of $kT=3.27\pm0.08$ eV ($37,932\pm 928$ K), and three emission lines: \ion{Si}{4}, \ion{C}{4} and \ion{He}{2} (see Figures~\ref{plot:harduv1} and \ref{plot:harduv2}).  Another black-body component with  $kT=0.80\pm0.03$ eV ($9,200\pm 348$ K) accounts for the observed optical excess (see Figure~\ref{plot:hardtotal}).
    \item  We estimate the reprocessed fraction in the hard state by taking a ratio between the $0.1 - 200\ev$ X-ray flux and the $0.5-10\ev$ optical/UV flux \footnote{\label{foot}Since the outer disk can intrinsically emit a significant fraction of optical/UV photons in the 0.5-10.0 eV band through viscous dissipation, we consider only the flux of two \texttt{bbodyrad} components and emission lines in that band for computing reprocessed fraction.}, and find that $0.9\%$ of the bolometric X-ray flux gets reprocessed and thermalized in the outer disk.
\end{enumerate}

During the second \astrosat{} observation in the soft state, the source was found to accrete at an X-ray luminosity of $\sim 1.1\times10^{38}\ \rm erg\ s^{-1}$ ($\sim13\%$ of Eddington luminosity, estimated assuming $M=6.75M_{\sun}$). The main results of our multi-wavelength spectral study of the soft state can be summarized as follows. 
\begin{enumerate}
    \item The soft state X-ray spectrum, in the energy band 0.6 -- 40.0 keV, is comprised of a multi-temperature disk component with $kT_{\rm in} = 0.58\pm0.02\kev$, a soft excess, and a weak Comptonization component ($\Gamma=2.19\pm0.05$, and $kT_{\rm e}\gtrsim36.5$ keV; see Section~\ref{sec:softx} and Table~\ref{tab:softx}). The soft X-ray excess, which most likely arises from the plunging region \citep{fabian2020}, is well described by a black-body component with $kT=0.79\pm0.02\kev$.
    \item Using the continuum fitting method and employing the \texttt{kerrbb} model, we measure 
    the BH spin and mass for the source to be $a=0.85_{-0.25}^{+0.10}$ and  $M_{\rm BH}>5.9M_{\sun}$ for an inclination of $64^{\circ}$, which is the jet inclination angle.
    \item Similar to the hard state case, we detect optical/UV excess components in the soft state (see Figures~\ref{plot:softuvexcess} and \ref{plot:softtotalinit}), which is comprised of two low-temperature black-body components ($kT =3.87\pm0.24$ eV and $0.75\pm0.04$ eV or $kT =44,892\pm 2784$ K and $8704\pm464$ K) and 5 emission lines: \ion{Si}{4}, \ion{N}{4}], \ion{C}{4}, \ion{He}{2}, and \ion{N}{4} (see Sections~\ref{sec:softuv} and \ref{sec:softtotal}, Figures~\ref{plot:softuv1}, \ref{plot:softuv2}, \ref{total4}, \ref{total2}, and \ref{total3}, and Tables~\ref{tab:softuv} and \ref{tab:softtotal} for more details).
    \item The flux in the optical/UV band ($0.5-10$ eV) is found to be significantly smaller than that of the hard state case. Consequently, the reprocessed fraction is low ($\sim2\times10^{-3}$), which is directly estimated by fitting the multi-wavelength data to the irradiated disk model \texttt{diskir}. The reprocessed fraction is estimated to be $\sim3.5\times10^{-3}$, from the  ratio of flux$^{\ref{foot}}$ in the 
    $0.5 - 10$ eV to that in the $0.1 - 200$ keV  band. The reduction of optical/UV flux in the soft state (compared to the hard state) has also been noticed earlier for the BH-LMXB XTE J1817-330 \citep{done2009}.
    \item We estimate the outer disk radius directly from our spectral fitting with the \texttt{diskir} model, and find a radius of $(2.30\pm0.33)\times10^{11}$ cm, assuming $i=64^{\circ}$. 
\end{enumerate}

We discuss below the implications of our multi-wavelength spectral results in the hard and soft states.
\subsection{Inner accretion geometry in the hard state}\label{sec:discuss1}
The geometry of the inner accretion flow in the hard state is the subject of ongoing debate. The current paradigm suggests that the disk truncates far from the ISCO radius in the hard state and is replaced by a hot accretion flow \citep{done2007}. However, this picture has been contested in many works, and 
an alternative geometry of disk extending into the ISCO radius (or almost ISCO) has emerged \citep{reis2010,kara2019}. For example, \cite{buisson2019} and \cite{chakraborty2020} performed reflection analysis of MAXI~J1820+070 in the hard state using data from the  \textit{NuSTAR} mission, and found that the disk has reached almost the ISCO radius ($\sim 2-6\ R_g$) with their two-component Comptonization model (we have also employed a similar model in our work). \cite{chakraborty2020} also considered the \textit{AstroSat} hard state observation and obtained results similar to those from their \nustar{} analysis.
In both these works, 
the inclination was low $\sim30^{\circ}$ and the iron abundance high, $4-10$ times the solar abundance. Such  a high iron abundance is unlikely as the donor star is a low-mass weakly evolved star \citep{zdziarski2021,joanna2022}. Besides, the binary inclination of the source or the inclination of the jet was estimated to be $>59^{\circ}$. 

On the other hand, \cite{zdziarski2021,zdziarski2022} also performed reflection analyses using the \textit{NuSTAR} hard state data (along with \textit{INTEGRAL} and Insight-HXMT data in the latter work), some of which were considered in the two previously mentioned works, and found the disk to be truncated far from the ISCO radius with a similar double Comptonization model. A similar conclusion regarding the truncation of the inner accretion disk was reported with \textit{NICER}, \textit{NuSTAR}, and \textit{SWIFT} data using the JED-SAD model in \cite{marino2021}.  

Unlike the previous works \citep{buisson2019,chakraborty2020}, both the inclination value ($\gtrsim50^{\circ}$) and the iron abundance ($\sim 1-2.6\ A_{\rm Fe}$) in the studies by \cite{zdziarski2021,zdziarski2022} do not suffer from the earlier inconsistencies. Their proposed hard state geometry consists of two Comptonization components: the harder component having a larger scale-height accretion flow located downstream the truncation radius, and the softer component forming a corona over the inner part of the disk. The harder part is reflected from the remote part of a weakly ionized disk, whereas the softer component gets reflected from a highly ionized underlying disk producing narrow reflection features. However, the disk temperature ($\sim0.4-0.5$ keV) reported in \cite{zdziarski2022} is significantly higher than the inner disk temperature obtained with the \textit{NICER} data \citep{wang2020}. Finally, in all the above works, the low energy data ($<2.0$ keV) were not used to perform the analysis, and only constant density reflection models, i.e., density ($n_{\rm e}$) is fixed to  $10^{15}\ \rm cm^{-3}$,   (like \texttt{relxilllpCp}, \texttt{reflkerr}, \texttt{xillverCp}) were employed.

In our work, we include low energy X-ray data from \textit{NICER} and \astrosat{}/SXT down to 0.6 keV to obtain a robust picture of the accretion geometry in the hard state. This approach is not only helpful in consistently constraining the disk components but also in providing a clearer picture of the inner accretion flow  \citep{garcia2015}. The hard state spectra in the 0.6 - 150 keV band are well described by a structured accretion flow consisting of two Comptonization components (see Section \ref{sec:hardx} and Table~\ref{tab:hardx} for more details). The softer component is found to dominate the broadband X-ray luminosity, and is reflected from a strongly ionized disk, producing the relativistic reflection features. The inner disk responsible for the relativistic reflection is truncated far from the source, $\simeq 51-78\ R_{g}$. We calculate a reflection fraction of $\sim 0.25$ for this component as the ratio of the reflected flux in the 1 eV to 1000 keV band (the reflected spectrum of \texttt{reflionxhd} is calculated over this energy range) to the incident flux in the  0.1 to 1000 keV band \citep{fuerst2015}. The harder component is reflected from a further distant and moderately ionized disk. The corresponding reflection fraction is $\sim0.13$. One should note that the definition of reflection fraction we use differs from that of \cite{dauser2016}.  
Since, the softer component has higher $\Gamma$ and lower $kT_{\rm e}$ than the harder component, it is most likely located closer to the accretion disk \citep{haardt1991}. The relatively higher values of reflection fraction and the ionization parameter also support this picture. Besides, a hard Comptonized spectrum of $\Gamma\sim1.2$ implies that the hot Comptonizing plasma is situated away from the disk \citep{poutanen2018}. 
Furthermore, the section of the disk reflecting the harder Comptonized component exhibits moderate ionization. This suggests that the scale-height of the accretion flow emitting the harder Comptonization component is likely large. Therefore, our investigation broadly aligns with the accretion geometry of this source as described by \cite{zdziarski2021} (please refer to their Sections 3 and 4 for more detailed information on the geometry). The spectral parameters, such as the power-law index and electron temperature, associated with the two Comptonization components are close to those reported in \cite{zdziarski2022} for the nearest \textit{NuSTAR} observation (their epoch 1 observation, which was performed approximately 6 days prior to our hard state observation). However, there are differences in the values of the reflection parameters between our work and theirs. This discrepancy may be related to the fact that we consider the possibility of a higher density disk. Specifically, we leave the parameter $\rm log$ $(n_{\rm e})$ free during the spectral fitting, whereas it was fixed to a default value of $n_{\rm e}=10^{15}\ \rm cm^{-3}$ in all the other works.

It was previously suggested that a higher value of disk density can influence the thermodynamic processes in the reflection skin of the disk, i.e., the disk atmosphere. At higher densities, free-free heating becomes more dominant, leading to an increase in the temperature of the disk atmosphere. This, in turn, results in a soft excess below 2 keV in a disk with higher density \citep{garcia2016}. Furthermore, a soft excess in a higher density disk may also arise because ionization parameters fitted at different densities are of a similar order. This results in a higher irradiating X-ray flux for a disk with higher density (see \citet{zdziarskidisk} for more details). Therefore, the impact of a higher density disk on the X-ray spectra can be better understood when including low-energy data ($< 2.0$ keV). To investigate how a higher density disk could affect the spectral parameters, we fix the density to $10^{15}\ \rm cm^{-3}$ in our Model 1A, and fit the model to the data. This results in a relatively poor fit with a $\chi^2/\text{d.o.f}$ of 2574.1/2389 ($\Delta\chi^2 = +669.9$ for one less parameter). Additionally, we observe significant changes in the values of the ionization parameter, iron abundance, and the radius of the inner disk, consistent with earlier findings in \cite{tomsick2018} and \cite{chakraborty2021}. The iron abundance increases to the maximum allowed value of 5. The ionization parameter associated with the reflection of the harder Comptonization component got pegged to 0, while it increased to a much higher value (around $\sim 4000$) for the reflection of the softer Comptonization component compared to the case with a free log$(n_{\rm e})$.
Furthermore, the inner disk radius ($\mathcal{R}_{\rm in}$) becomes poorly constrained in the this model, with $\mathcal{R}_{\rm in}>110R_g$. Thus, both the physical consistency of the best-fit parameters and statistical significance of the spectral fit indicate a higher density disk in MAXI~J1820+070.
However, as emphasized in \cite{garcia2016}, the atomic physics considered in these reflection models is uncertain beyond $10^{19}\rm cm^{-3}$. Thus, more accurate determination of the rates of the pertinent atomic processes at higher densities could influence the spectrum of these reflection models, thereby our results also may get affected.

\subsection{Mass and spin of MAXI~J1820+070}\label{sec:discuss2} 
We constrain the black hole spin and mass by fitting the \texttt{kerrbb} model to the soft state X-ray spectrum of MAXI~J1820+070. This method requires the emission to be disk dominated where the disk (in general) reaches the ISCO  and it remains thin. 
These two requirements are thought to meet when the disk fraction $\geq75\%$ (substantial thermal component) and $L/L_{\rm Edd}<0.3$ (the disk scale-height grows beyond this, and the thin disk model may not hold) \citep{mcclintock2014}. We find the disk fraction to be $\sim85\%$ and $L/L_{\rm Edd}\sim0.13$ in the soft state of MAXI~J1820+070 using Model 2A, thus making our soft state X-ray spectrum suitable for the estimation of BH spin and mass.

Additionally, a meaningful estimation of the BH mass and spin  requires a proper knowledge of the distance to the source and the inclination of the inner disk.
While the distance to the source is well measured to be $2.96\pm0.33{\rm~kpc}$, the inclination is less certain (see the introduction section for more details). Besides, the outer disk could be misaligned with respect to the BH spin axis \citep{poutanen2022}, which further complicates the estimation of inclination.
But, the inclination of the inner disk is most likely the same as the jet inclination of $64^{\circ} \pm 5^{\circ}$ (which can be considered to be aligned with the BH spin axis) as measured by \cite{woods2021}. 
If we fix the inclination parameter to $64^{\circ}$, we obtain the BH spin, $a=0.85_{-0.25}^{+0.10}$ and mass, $M_{\rm BH}>5.9M_{\odot}$ (see Table~\ref{tab:softx} and Section~\ref{sec:softx} for further details).
Our spin measurement agrees well with the estimation   of \cite{bhargava2021}($a=0.799_{-0.015}^{+0.016}$) based on an independent timing-based technique utilizing the evolution of the characteristic frequencies in the power density spectra. Furthermore, the BH mass we find is consistent with that measured by  \cite{torres2020} ($M_{BH} = 5.73 - 8.34M_{\sun}$ for binary inclination in the range, $66^{\circ}-81^{\circ}$, with $95\%$ confidence level). A recent determination of mass \citep{joanna2022} of this system ($M=6.75_{-0.46}^{+0.64}M_{\sun}$ with $68\%$ confidence level) is also in line with our estimation. By constraining the inclination parameter in the range of $59^{\circ}-69^{\circ}$ and mass in the range of $5.0-10.0M_{\sun}$ in the \texttt{kerrbb} model, we measure the BH spin to be $a=0.77\pm0.21$.

\begin{figure}
\includegraphics[width=0.5\textwidth]{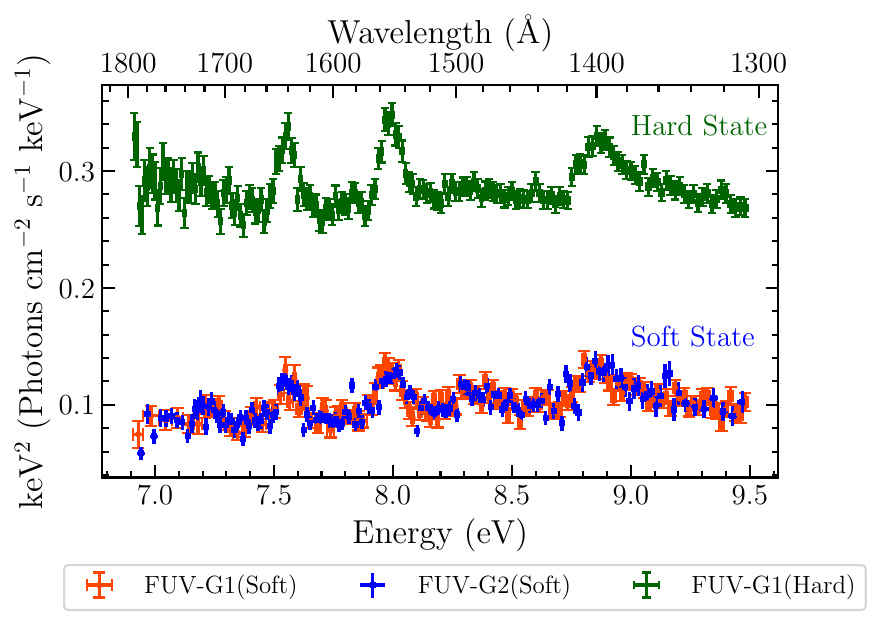}
\caption{Hard and soft state unfolded Far-UV spectra. The models 1B (hard state) and 2D (soft state) are used for this purpose. See Sections \ref{sec:harduv} and \ref{sec:softuv} for more details.\label{compare}}
\end{figure}
\subsection{X-ray irradiation and geometry of the outer disk}\label{sec:discuss3}
It is widely believed that the reprocessing of X-rays in the outer disk plays a dominant role in the optical/UV emission in BH-LMXBs  \citep{vanparadijs1994,done2009}. In the case of MAXI~J1820+070, we find clear evidence of reprocessed emission in terms of excess optical/UV continuum and strong emission lines. To describe the optical/UV excess components phenomenologically, we require two low-temperature black-body components, suggesting reprocessed emission from a range of radii rather than a narrow annulus of the disk. Furthermore, we find that the multi-wavelength spectral fit with the irradiated disk model \texttt{diskir} provides a significantly worse fit than our phenomenological models with two low-temperature black-body components. This is perhaps not unexpected as the \texttt{diskir} model assumes uniform illumination of the outer accretion disk by the inner accretion flow, captured through the constant $f_{\rm out}$. 
In this context, we calculate the ratio of two single temperature black-body fluxes (representing UV and optical excesses, respectively)
in the energy band 0.5-10.0 eV to the incoming X-ray flux in the energy band 0.1-200.0 keV in hard and soft states, and find this to be different, suggesting a non-uniform illumination of the outer accretion disk. In the hard state, the ratio of UV excess flux (0.5-10 eV) and the X-ray flux (0.1-200.0 keV) is $\sim5.8\times10^{-3}$, and the ratio of optical excess flux (0.5-10 eV) and the X-ray flux (0.1-200.0 keV) is $\sim3.2\times10^{-3}$ (Model 1C is used for this estimation). On the other hand, the same quantities in the soft state are $\sim2\times10^{-3}$ and $\sim1.3\times10^{-3}$, respectively (Model 2E is employed for this calculation). Generally, $f_{\rm out}$ should be a function of the disk aspect ratio, $H/R$ ($H$ is the disk scale-height and $R$ is the radius) and $H/R$ itself depends on $R$ \citep{fkr2002,alex2018}. However, in \texttt{diskir}, $H/R$ is assumed to be constant, which is a limiting case to the actual scenario \citep{alex2018}. For example, $H\propto R^{9/8}$ in the outer zone of the standard Shakura-Sunyaev model, whereas $H\propto R^{9/7}$ in isothermal disk model of Cunningham \citep{kimura2019}.

The observed optical/UV flux in the $0.5-10\ev$ band, calculated from our multi-wavelength spectra, is $\sim 4$ and $\sim 33$ times higher than those estimated for the intrinsic disk emission in the soft and hard states using the models 2E and 1C, respectively. This clearly implies the dominance of X-ray irradiation over the intrinsic viscous dissipation in the outer disk in both soft and hard states. 
We find that the strength of the reprocessed optical/UV emission relative to the intrinsic disk emission is much higher in the hard state than that in the soft state. This is also evident from the stronger FUV grating spectrum in the hard state, as can be observed in  Fig.~\ref{compare}. Also, the fraction  of the intrinsic disk/corona emission reprocessed in the disk is nearly a factor of three higher in  the hard state ($\sim 9\times 10^{-3}$) than in  the soft state ($\sim 3.5\times 10^{-3}$). These observations clearly demonstrate that  X-ray irradiation onto the disk is much more dominant in the hard state of MAXI~J1820+070, similar to found in BH-LMXB XTE J1817-330 \citep{done2009}. The stronger optical and possibly UV continuum in the hard state, in principle, could also arise due to the Synchrotron emission from jets \citep{russell2006}. The SED and timing studies of MAXI~J1820+070 have shown that the jet does make a contribution to the infrared band and to the optical band to some extent \citep{paice2019,markoff2020,tetarenko2022}. However, we find that the stronger far UV continuum is accompanied by strong emission lines in the hard state. The emission lines due to \ion{He}{2}, \ion{C}{4} and \ion{Si}{4} are nearly a factor of two stronger in the hard state than in the soft state, implying that most of the excess UV/optical emission in the hard state is due to X-ray reprocessing in the outer accretion disk. The stronger X-ray reprocessing in the hard state is most likely the outcome of the geometry where the X-ray corona in the innermost regions has larger scale height than the accretion disk, possibly in the form of a spherical corona or elongated along the jet axis.
In addition, the disk wind, observed during the hard state \citep{munoz2019}, can also contribute to the irradiation of the disk, increasing the optical/UV flux \citep{done2009,dubus2019,tetarenko2020}.

We also note that it is unlikely that the emission from the secondary star is contributing a significant number of photons to the optical/UV band in both states, as the surface temperature is  low, $\sim 4200$ K, i.e, $\sim0.36$ eV \citep{joanna2022}, which is much lower than the temperature of the black-body ($\sim 0.8$ eV). However, the companion, which is detected in the optical only during quiescence, is fainter than the observed fluxes in outburst.

Finally, the high observed reprocessed fraction ($\sim 10^{-2}-10^{-3}$) from MAXI~J1820+070 is unlikely to be achieved in the framework of standard thin disk prescription \citep{dubus1999}. The outer disk is most likely convex or warped in shape, making the disk more effective for X-ray reprocessing. Interestingly, \cite{thomas2022} proposed the outer disk of MAXI~J1820+070 to be warped to explain the large amplitude modulation, seen in the hard state optical light-curves. Furthermore, a significant spin-orbit misalignment has been inferred from the optical polarimetric observations of this source \citep{poutanen2022}, which could also result in a warped accretion disk.
 
\section{Conclusion}\label{sec:conclusion}
Utilizing data from the \textit{AstroSat}, \textit{NICER}, and \textit{LCO} observatories, we constrain the inner and outer geometries of the accretion flow around the BH-LMXB MAXI J1820+070 in the hard and soft states during its 2018 outburst.  In the hard state, our analysis reveals that the inner accretion disk is truncated far from the ISCO radius, and has been replaced by a structured accretion flow containing two Comptonization components with different slopes and temperatures. The softer Comptonization component dominates the X-ray emission and produces broad relativistic X-ray reflection features. Meanwhile, the harder component undergoes reflection from a distant, high-density disk ($\sim 2\times10^{20}\ \rm cm^{-3}$), resulting in unblurred reflection features.
In the soft state, the X-ray spectrum features a dominant disk component (disk fraction $\sim85\%$), a soft X-ray excess, and a weak Comptonization component. We estimate the spin of BH using continuum spectral fitting method, yielding $a=0.77\pm0.21$, for an inclination in the range of $59^{\circ}-69^{\circ}$ (i.e., the range of jet inclination angle) and a mass in the range of $5.0-10.0M_{\sun}$ (predicted mass range of this BH). Finally, we find that a significant fraction of the X-ray radiation from the inner disk and the coronal emission is reprocessed and thermalized in the outer accretion disk, with the reprocessed fraction being much higher in the hard state ($\sim 9\times 10^{-3}$) compared to the soft state ($\sim 3.5\times 10^{-3}$). A strong reprocessing in the outer accretion disk likely suggests that the outer disk could be warped or convex, as also indicated in \cite{thomas2022}.

\begin{acknowledgments}
We thank the anonymous referee for his/her constructive comments that helped to improve the manuscript.
We use data from the \textit{AstroSat} mission of the Indian Space Research Organisation (ISRO), archived at the Indian Space Science Data Centre (ISSDC). We acknowledge the POC teams of the SXT, LAXPC, CZTI, and UVIT instruments for archiving data and providing the necessary software tools. This research has made use of data and software provided by the High Energy Astrophysics Science Archive Research Center (HEASARC), which is a service of the Astrophysics Science Division at NASA/GSFC. We acknowledge the use of MAXI data provided by RIKEN, JAXA, and the MAXI team, and of public data from the \textit{NICER} data archive. This work also uses data from the Faulkes Telescope Project, which is an education partner of Las Cumbres Observatory (LCO). The Faulkes Telescopes are maintained and operated by LCO. AAZ acknowledges support from the Polish National Science Center under the grant 2019/35/B/ST9/03944. We would like to express our gratitude to John Tomsick for providing the reflionxhd model, and we extend our thanks to Michael Parker for developing this model. SB also wants to acknowledge the valuable assistance provided by Sudip Chakraborty, Nilam Navale, and Labanya K. Guha in the preparation of this manuscript.
\end{acknowledgments}

%

\vspace{5mm}

\bibliography{maxij1820}{}

\begin{thebibliography}{}
\expandafter\ifx\csname natexlab\endcsname\relax\def\natexlab#1{#1}\fi
\providecommand{\url}[1]{\href{#1}{#1}}
\providecommand{\dodoi}[1]{doi:~\href{http://doi.org/#1}{\nolinkurl{#1}}}
\providecommand{\doeprint}[1]{\href{http://ascl.net/#1}{\nolinkurl{http://ascl.net/#1}}}
\providecommand{\doarXiv}[1]{\href{https://arxiv.org/abs/#1}{\nolinkurl{https://arxiv.org/abs/#1}}}

\bibitem[{Antia {et~al.}(2017)Antia, Yadav, Agrawal, Chauhan, Manchanda,
  Chitnis, Paul, Dedhia, Shah, Gujar, Katoch, Kurhade, Madhwani, Manojkumar,
  Nikam, Pandya, Parmar, Pawar, Pahari, Misra, Navalgund, Pandiyan, Sharma, \&
  Subbarao}]{antia2017}
Antia, H.~M., Yadav, J.~S., Agrawal, P.~C., {et~al.} 2017, The Astrophysical
  Journal Supplement Series, 231, 10, \dodoi{10.3847/1538-4365/aa7a0e}

\bibitem[{{Antia} {et~al.}(2021){Antia}, {Agrawal}, {Dedhia}, {Katoch},
  {Manchanda}, {Misra}, {Mukerjee}, {Pahari}, {Roy}, {Shah}, \&
  {Yadav}}]{antia2021}
{Antia}, H.~M., {Agrawal}, P.~C., {Dedhia}, D., {et~al.} 2021, Journal of
  Astrophysics and Astronomy, 42, 32, \dodoi{10.1007/s12036-021-09712-8}

\bibitem[{{Arnaud}(1996)}]{arnaud1996}
{Arnaud}, K.~A. 1996, in Astronomical Society of the Pacific Conference Series,
  Vol. 101, Astronomical Data Analysis Software and Systems V, ed. G.~H.
  {Jacoby} \& J.~{Barnes}, 17

\bibitem[{{Atri} {et~al.}(2020){Atri}, {Miller-Jones}, {Bahramian}, {Plotkin},
  {Deller}, {Jonker}, {Maccarone}, {Sivakoff}, {Soria}, {Altamirano},
  {Belloni}, {Fender}, {Koerding}, {Maitra}, {Markoff}, {Migliari}, {Russell},
  {Russell}, {Sarazin}, {Tetarenko}, \& {Tudose}}]{atri2020}
{Atri}, P., {Miller-Jones}, J.~C.~A., {Bahramian}, A., {et~al.} 2020, \mnras,
  493, L81, \dodoi{10.1093/mnrasl/slaa010}

\bibitem[{{Axelsson} \& {Veledina}(2021)}]{veledina2021}
{Axelsson}, M., \& {Veledina}, A. 2021, \mnras, 507, 2744,
  \dodoi{10.1093/mnras/stab2191}

\bibitem[{{Baglio} {et~al.}(2018){Baglio}, {Russell}, \& {Lewis}}]{baglio2018}
{Baglio}, M.~C., {Russell}, D.~M., \& {Lewis}, F. 2018, The Astronomer's
  Telegram, 11418, 1

\bibitem[{{Bahramian} {et~al.}(2018){Bahramian}, {Strader}, \&
  {Dage}}]{bahramian2018}
{Bahramian}, A., {Strader}, J., \& {Dage}, K. 2018, The Astronomer's Telegram,
  11424, 1

\bibitem[{{Banerjee} {et~al.}(2019{\natexlab{a}}){Banerjee}, {Chakraborty}, \&
  {Bhattacharyya}}]{banerjee2019a}
{Banerjee}, S., {Chakraborty}, C., \& {Bhattacharyya}, S. 2019{\natexlab{a}},
  \apj, 870, 95, \dodoi{10.3847/1538-4357/aaf102}

\bibitem[{{Banerjee} {et~al.}(2019{\natexlab{b}}){Banerjee}, {Chakraborty}, \&
  {Bhattacharyya}}]{banerjee2019b}
---. 2019{\natexlab{b}}, \mnras, 487, 3488, \dodoi{10.1093/mnras/stz1518}

\bibitem[{{Banerjee} {et~al.}(2020){Banerjee}, {Gilfanov}, {Bhattacharyya}, \&
  {Sunyaev}}]{banerjee2020}
{Banerjee}, S., {Gilfanov}, M., {Bhattacharyya}, S., \& {Sunyaev}, R. 2020,
  \mnras, 498, 5353, \dodoi{10.1093/mnras/staa2788}

\bibitem[{{Basak} \& {Zdziarski}(2016)}]{basak2016}
{Basak}, R., \& {Zdziarski}, A.~A. 2016, \mnras, 458, 2199,
  \dodoi{10.1093/mnras/stw420}

\bibitem[{{Belloni}(2010)}]{Belloni2010}
{Belloni}, T.~M. 2010, in Lecture Notes in Physics, Berlin Springer Verlag, ed.
  T.~{Belloni}, Vol. 794, 53, \dodoi{10.1007/978-3-540-76937-8_3}

\bibitem[{{Belloni} \& {Motta}(2016)}]{belloni2016}
{Belloni}, T.~M., \& {Motta}, S.~E. 2016, in Astrophysics and Space Science
  Library, Vol. 440, Astrophysics of Black Holes: From Fundamental Aspects to
  Latest Developments, ed. C.~{Bambi}, 61, \dodoi{10.1007/978-3-662-52859-4_2}

\bibitem[{{Bhalerao} {et~al.}(2017){Bhalerao}, {Bhattacharya}, {Vibhute},
  {Pawar}, {Rao}, {Hingar}, {Khanna}, {Kutty}, {Malkar}, {Patil}, {Arora},
  {Sinha}, {Priya}, {Samuel}, {Sreekumar}, {Vinod}, {Mithun}, {Vadawale},
  {Vagshette}, {Navalgund}, {Sarma}, {Pandiyan}, {Seetha}, \&
  {Subbarao}}]{bhalerao2017}
{Bhalerao}, V., {Bhattacharya}, D., {Vibhute}, A., {et~al.} 2017, Journal of
  Astrophysics and Astronomy, 38, 31, \dodoi{10.1007/s12036-017-9447-8}

\bibitem[{{Bhargava} {et~al.}(2021){Bhargava}, {Belloni}, {Bhattacharya},
  {Motta}, \& {Ponti.}}]{bhargava2021}
{Bhargava}, Y., {Belloni}, T., {Bhattacharya}, D., {Motta}, S., \& {Ponti.}, G.
  2021, \mnras, 508, 3104, \dodoi{10.1093/mnras/stab2848}

\bibitem[{{Bright} {et~al.}(2018){Bright}, {Fender}, \& {Motta}}]{bright2018}
{Bright}, J., {Fender}, R., \& {Motta}, S. 2018, The Astronomer's Telegram,
  11420, 1

\bibitem[{{Bright} {et~al.}(2020){Bright}, {Fender}, {Motta}, {Williams},
  {Moldon}, {Plotkin}, {Miller-Jones}, {Heywood}, {Tremou}, {Beswick},
  {Sivakoff}, {Corbel}, {Buckley}, {Homan}, {Gallo}, {Tetarenko}, {Russell},
  {Green}, {Titterington}, {Woudt}, {Armstrong}, {Groot}, {Horesh}, {van der
  Horst}, {K{\"o}rding}, {McBride}, {Rowlinson}, \& {Wijers}}]{bright2020}
{Bright}, J.~S., {Fender}, R.~P., {Motta}, S.~E., {et~al.} 2020, Nature
  Astronomy, 4, 697, \dodoi{10.1038/s41550-020-1023-5}

\bibitem[{{Buisson} {et~al.}(2019){Buisson}, {Fabian}, {Barret}, {F{\"u}rst},
  {Gandhi}, {Garc{\'\i}a}, {Kara}, {Madsen}, {Miller}, {Parker}, {Shaw},
  {Tomsick}, \& {Walton}}]{buisson2019}
{Buisson}, D.~J.~K., {Fabian}, A.~C., {Barret}, D., {et~al.} 2019, \mnras, 490,
  1350, \dodoi{10.1093/mnras/stz2681}

\bibitem[{{Casella} {et~al.}(2018){Casella}, {Vincentelli}, {O'Brien}, {Testa},
  {Maccarone}, {Uttley}, {Fender}, \& {Russell}}]{casella2018}
{Casella}, P., {Vincentelli}, F., {O'Brien}, K., {et~al.} 2018, The
  Astronomer's Telegram, 11451, 1

\bibitem[{{Chakraborty} {et~al.}(2020){Chakraborty}, {Navale}, {Ratheesh}, \&
  {Bhattacharyya}}]{chakraborty2020}
{Chakraborty}, S., {Navale}, N., {Ratheesh}, A., \& {Bhattacharyya}, S. 2020,
  \mnras, 498, 5873, \dodoi{10.1093/mnras/staa2711}

\bibitem[{{Chakraborty} {et~al.}(2021){Chakraborty}, {Ratheesh},
  {Bhattacharyya}, {Tomsick}, {Tombesi}, {Fukumura}, \&
  {Jaisawal}}]{chakraborty2021}
{Chakraborty}, S., {Ratheesh}, A., {Bhattacharyya}, S., {et~al.} 2021, \mnras,
  508, 475, \dodoi{10.1093/mnras/stab2530}

\bibitem[{{Connors} {et~al.}(2021){Connors}, {Garc{\'\i}a}, {Tomsick}, {Hare},
  {Dauser}, {Grinberg}, {Steiner}, {Mastroserio}, {Sridhar}, {Fabian}, {Jiang},
  {Parker}, {Harrison}, \& {Kallman}}]{riley2021}
{Connors}, R. M.~T., {Garc{\'\i}a}, J.~A., {Tomsick}, J., {et~al.} 2021, \apj,
  909, 146, \dodoi{10.3847/1538-4357/abdd2c}

\bibitem[{{Dauser} {et~al.}(2014){Dauser}, {Garcia}, {Parker}, {Fabian}, \&
  {Wilms}}]{dauser2014}
{Dauser}, T., {Garcia}, J., {Parker}, M.~L., {Fabian}, A.~C., \& {Wilms}, J.
  2014, \mnras, 444, L100, \dodoi{10.1093/mnrasl/slu125}

\bibitem[{{Dauser} {et~al.}(2016){Dauser}, {Garc{\'\i}a}, {Walton}, {Eikmann},
  {Kallman}, {McClintock}, \& {Wilms}}]{dauser2016}
{Dauser}, T., {Garc{\'\i}a}, J., {Walton}, D.~J., {et~al.} 2016, \aap, 590,
  A76, \dodoi{10.1051/0004-6361/201628135}

\bibitem[{{Dauser} {et~al.}(2010){Dauser}, {Wilms}, {Reynolds}, \&
  {Brenneman}}]{dauser2010}
{Dauser}, T., {Wilms}, J., {Reynolds}, C.~S., \& {Brenneman}, L.~W. 2010,
  \mnras, 409, 1534, \dodoi{10.1111/j.1365-2966.2010.17393.x}

\bibitem[{{De Marco} {et~al.}(2021){De Marco}, {Zdziarski}, {Ponti},
  {Migliori}, {Belloni}, {Segovia Otero}, {Dzie{\l}ak}, \& {Lai}}]{demarco2021}
{De Marco}, B., {Zdziarski}, A.~A., {Ponti}, G., {et~al.} 2021, \aap, 654, A14,
  \dodoi{10.1051/0004-6361/202140567}

\bibitem[{{Denisenko}(2018)}]{denisenko2018}
{Denisenko}, D. 2018, The Astronomer's Telegram, 11400, 1

\bibitem[{{Dewangan}(2021)}]{dewangan2021}
{Dewangan}, G.~C. 2021, Journal of Astrophysics and Astronomy, 42, 49,
  \dodoi{10.1007/s12036-021-09691-w}

\bibitem[{{Done} {et~al.}(2007){Done}, {Gierli{\'n}ski}, \&
  {Kubota}}]{done2007}
{Done}, C., {Gierli{\'n}ski}, M., \& {Kubota}, A. 2007, \aapr, 15, 1,
  \dodoi{10.1007/s00159-007-0006-1}

\bibitem[{{Dubus} {et~al.}(2019){Dubus}, {Done}, {Tetarenko}, \&
  {Hameury}}]{dubus2019}
{Dubus}, G., {Done}, C., {Tetarenko}, B.~E., \& {Hameury}, J.-M. 2019, \aap,
  632, A40, \dodoi{10.1051/0004-6361/201936333}

\bibitem[{{Dubus} {et~al.}(1999){Dubus}, {Lasota}, {Hameury}, \&
  {Charles}}]{dubus1999}
{Dubus}, G., {Lasota}, J.-P., {Hameury}, J.-M., \& {Charles}, P. 1999, \mnras,
  303, 139, \dodoi{10.1046/j.1365-8711.1999.02212.x}

\bibitem[{Dziełak {et~al.}(2021)Dziełak, De Marco, \&
  Zdziarski}]{dzielak2021}
Dziełak, M.~A., De Marco, B., \& Zdziarski, A.~A. 2021, Monthly Notices of
  the Royal Astronomical Society, 506, 2020, \dodoi{10.1093/mnras/stab1700}

\bibitem[{{Echibur{\'u}-Trujillo} {et~al.}(2023){Echibur{\'u}-Trujillo},
  {Tetarenko}, {Haggard}, {Russell}, {Koljonen}, {Bahramian}, {Wang}, {Bremer},
  {Bright}, {Casella}, {Russell}, {Altamirano}, {Baglio}, {Belloni},
  {Ceccobello}, {Corbel}, {Diaz Trigo}, {Maitra}, {Gabuya}, {Gallo}, {Heinz},
  {Homan}, {Kara}, {K{\"o}rding}, {Lewis}, {Lucchini}, {Markoff}, {Migliari},
  {Miller-Jones}, {Rodriguez}, {Saikia}, {Sarazin}, {Shahbaz}, {Sivakoff},
  {Soria}, {Testa}, {Tetarenko}, \& {Tudose}}]{trujillo2023}
{Echibur{\'u}-Trujillo}, C., {Tetarenko}, A.~J., {Haggard}, D., {et~al.} 2023,
  arXiv e-prints, arXiv:2311.11523, \dodoi{10.48550/arXiv.2311.11523}

\bibitem[{Fabian(2005)}]{fabian2005}
Fabian, A.~C. 2005, Broad iron lines in AGN and X-ray binaries, ed. T.~J.
  Maccarone, R.~P. Fender, \& L.~C. Ho (Dordrecht: Springer Netherlands),
  97--105, \dodoi{10.1007/1-4020-4085-7_12}

\bibitem[{{Fabian} {et~al.}(1989){Fabian}, {Rees}, {Stella}, \&
  {White}}]{fabian1989}
{Fabian}, A.~C., {Rees}, M.~J., {Stella}, L., \& {White}, N.~E. 1989, \mnras,
  238, 729, \dodoi{10.1093/mnras/238.3.729}

\bibitem[{{Fabian} {et~al.}(2020){Fabian}, {Buisson}, {Kosec}, {Reynolds},
  {Wilkins}, {Tomsick}, {Walton}, {Gandhi}, {Altamirano}, {Arzoumanian},
  {Cackett}, {Dyda}, {Garcia}, {Gendreau}, {Grefenstette}, {Homan}, {Kara},
  {Ludlam}, {Miller}, \& {Steiner}}]{fabian2020}
{Fabian}, A.~C., {Buisson}, D.~J., {Kosec}, P., {et~al.} 2020, \mnras, 493,
  5389, \dodoi{10.1093/mnras/staa564}

\bibitem[{{Frank} {et~al.}(2002){Frank}, {King}, \& {Raine}}]{fkr2002}
{Frank}, J., {King}, A., \& {Raine}, D.~J. 2002, {Accretion Power in
  Astrophysics: Third Edition}

\bibitem[{{F{\"u}rst} {et~al.}(2015){F{\"u}rst}, {Nowak}, {Tomsick}, {Miller},
  {Corbel}, {Bachetti}, {Boggs}, {Christensen}, {Craig}, {Fabian}, {Gandhi},
  {Grinberg}, {Hailey}, {Harrison}, {Kara}, {Kennea}, {Madsen}, {Pottschmidt},
  {Stern}, {Walton}, {Wilms}, \& {Zhang}}]{fuerst2015}
{F{\"u}rst}, F., {Nowak}, M.~A., {Tomsick}, J.~A., {et~al.} 2015, \apj, 808,
  122, \dodoi{10.1088/0004-637X/808/2/122}

\bibitem[{{Gandhi} {et~al.}(2018){Gandhi}, {Paice}, {Littlefair}, {Dhillon},
  {Chote}, \& {Marsh}}]{gandhi2018}
{Gandhi}, P., {Paice}, J.~A., {Littlefair}, S.~P., {et~al.} 2018, The
  Astronomer's Telegram, 11437, 1

\bibitem[{{Garc{\'\i}a} {et~al.}(2013){Garc{\'\i}a}, {Dauser}, {Reynolds},
  {Kallman}, {McClintock}, {Wilms}, \& {Eikmann}}]{garcia2013}
{Garc{\'\i}a}, J., {Dauser}, T., {Reynolds}, C.~S., {et~al.} 2013, \apj, 768,
  146, \dodoi{10.1088/0004-637X/768/2/146}

\bibitem[{{Garc{\'\i}a} {et~al.}(2014){Garc{\'\i}a}, {Dauser}, {Lohfink},
  {Kallman}, {Steiner}, {McClintock}, {Brenneman}, {Wilms}, {Eikmann},
  {Reynolds}, \& {Tombesi}}]{garcia2014}
{Garc{\'\i}a}, J., {Dauser}, T., {Lohfink}, A., {et~al.} 2014, \apj, 782, 76,
  \dodoi{10.1088/0004-637X/782/2/76}

\bibitem[{{Garc{\'\i}a} {et~al.}(2015){Garc{\'\i}a}, {Dauser}, {Steiner},
  {McClintock}, {Keck}, \& {Wilms}}]{garcia2015}
{Garc{\'\i}a}, J.~A., {Dauser}, T., {Steiner}, J.~F., {et~al.} 2015, \apjl,
  808, L37, \dodoi{10.1088/2041-8205/808/2/L37}

\bibitem[{{Garc{\'\i}a} {et~al.}(2016){Garc{\'\i}a}, {Fabian}, {Kallman},
  {Dauser}, {Parker}, {McClintock}, {Steiner}, \& {Wilms}}]{garcia2016}
{Garc{\'\i}a}, J.~A., {Fabian}, A.~C., {Kallman}, T.~R., {et~al.} 2016, \mnras,
  462, 751, \dodoi{10.1093/mnras/stw1696}

\bibitem[{Gendreau {et~al.}(2016)Gendreau, Arzoumanian, Adkins, Albert, Anders,
  Aylward, Baker, Balsamo, Bamford, Benegalrao, Berry, Bhalwani, Black,
  Blaurock, Bronke, Brown, Budinoff, Cantwell, Cazeau, Chen, Clement,
  Colangelo, Coleman, Coopersmith, Dehaven, Doty, Egan, Enoto, Fan, Ferro,
  Foster, Galassi, Gallo, Green, Grosh, Ha, Hasouneh, Heefner, Hestnes, Hoge,
  Jacobs, Jørgensen, Kaiser, Kellogg, Kenyon, Koenecke, Kozon, LaMarr,
  Lambertson, Larson, Lentine, Lewis, Lilly, Liu, Malonis, Manthripragada,
  Markwardt, Matonak, Mcginnis, Miller, Mitchell, Mitchell, Mohammed, Monroe,
  de~Garcia, Mulé, Nagao, Ngo, Norris, Norwood, Novotka, Okajima, Olsen,
  Onyeachu, Orosco, Peterson, Pevear, Pham, Pollard, Pope, Powers, Powers,
  Price, Prigozhin, Ramirez, Reid, Remillard, Rogstad, Rosecrans, Rowe, Sager,
  Sanders, Savadkin, Saylor, Schaeffer, Schweiss, Semper, Serlemitsos,
  Shackelford, Soong, Struebel, Vezie, Villasenor, Winternitz, Wofford, Wright,
  Yang, \& Yu}]{gendreau2016}
Gendreau, K.~C., Arzoumanian, Z., Adkins, P.~W., {et~al.} 2016, in Space
  Telescopes and Instrumentation 2016: Ultraviolet to Gamma Ray, ed. J.-W.~A.
  den Herder, T.~Takahashi, \& M.~Bautz, Vol. 9905, International Society for
  Optics and Photonics (SPIE), 420 -- 435, \dodoi{10.1117/12.2231304}

\bibitem[{{Gierli{\'n}ski} {et~al.}(2009){Gierli{\'n}ski}, {Done}, \&
  {Page}}]{done2009}
{Gierli{\'n}ski}, M., {Done}, C., \& {Page}, K. 2009, \mnras, 392, 1106,
  \dodoi{10.1111/j.1365-2966.2008.14166.x}

\bibitem[{{Gilfanov}(2010)}]{gilfanov2010}
{Gilfanov}, M. 2010, in Lecture Notes in Physics, Berlin Springer Verlag, ed.
  T.~{Belloni}, Vol. 794, 17, \dodoi{10.1007/978-3-540-76937-8_2}

\bibitem[{{Gilfanov} \& {Arefiev}(2005)}]{gilfanov2005}
{Gilfanov}, M., \& {Arefiev}, V. 2005, arXiv e-prints, astro,
  \dodoi{10.48550/arXiv.astro-ph/0501215}

\bibitem[{{Haardt} \& {Maraschi}(1991)}]{haardt1991}
{Haardt}, F., \& {Maraschi}, L. 1991, \apjl, 380, L51, \dodoi{10.1086/186171}

\bibitem[{{Harris} {et~al.}(2016){Harris}, {Jensen}, {Suzuki}, {Bautista},
  {Dawson}, {Vivek}, {Brownstein}, {Ge}, {Hamann}, {Herbst}, {Jiang}, {Moran},
  {Myers}, {Olmstead}, \& {Schneider}}]{harris2016}
{Harris}, D.~W., {Jensen}, T.~W., {Suzuki}, N., {et~al.} 2016, \aj, 151, 155,
  \dodoi{10.3847/0004-6256/151/6/155}

\bibitem[{{HI4PI Collaboration} {et~al.}(2016){HI4PI Collaboration}, {Ben
  Bekhti}, {Fl{\"o}er}, {Keller}, {Kerp}, {Lenz}, {Winkel}, {Bailin},
  {Calabretta}, {Dedes}, {Ford}, {Gibson}, {Haud}, {Janowiecki}, {Kalberla},
  {Lockman}, {McClure-Griffiths}, {Murphy}, {Nakanishi}, {Pisano}, \&
  {Staveley-Smith}}]{hi4pi2016}
{HI4PI Collaboration}, {Ben Bekhti}, N., {Fl{\"o}er}, L., {et~al.} 2016, \aap,
  594, A116, \dodoi{10.1051/0004-6361/201629178}

\bibitem[{{Homan} \& {Belloni}(2005)}]{homan2005}
{Homan}, J., \& {Belloni}, T. 2005, \apss, 300, 107,
  \dodoi{10.1007/s10509-005-1197-4}

\bibitem[{{Homan} {et~al.}(2001){Homan}, {Wijnands}, {van der Klis}, {Belloni},
  {van Paradijs}, {Klein-Wolt}, {Fender}, \& {M{\'e}ndez}}]{homan2001}
{Homan}, J., {Wijnands}, R., {van der Klis}, M., {et~al.} 2001, \apjs, 132,
  377, \dodoi{10.1086/318954}

\bibitem[{{Homan} {et~al.}(2018){Homan}, {Stevens}, {Altamirano}, {Gendreau},
  {Arzoumanian}, {Strohmayer}, {Uttley}, {Cackett}, {Kara}, {Pasham}, \& {Nicer
  Team}}]{homan2018}
{Homan}, J., {Stevens}, A.~L., {Altamirano}, D., {et~al.} 2018, The
  Astronomer's Telegram, 12068, 1

\bibitem[{{Homan} {et~al.}(2020){Homan}, {Bright}, {Motta}, {Altamirano},
  {Arzoumanian}, {Basak}, {Belloni}, {Cackett}, {Fender}, {Gendreau}, {Kara},
  {Pasham}, {Remillard}, {Steiner}, {Stevens}, \& {Uttley}}]{homan2020}
{Homan}, J., {Bright}, J., {Motta}, S.~E., {et~al.} 2020, \apjl, 891, L29,
  \dodoi{10.3847/2041-8213/ab7932}

\bibitem[{{Jiang} {et~al.}(2020){Jiang}, {Gallo}, {Fabian}, {Parker}, \&
  {Reynolds}}]{jiang2020}
{Jiang}, J., {Gallo}, L.~C., {Fabian}, A.~C., {Parker}, M.~L., \& {Reynolds},
  C.~S. 2020, \mnras, 498, 3888, \dodoi{10.1093/mnras/staa2625}

\bibitem[{{Kaastra} \& {Bleeker}(2016)}]{kaastra2016}
{Kaastra}, J.~S., \& {Bleeker}, J.~A.~M. 2016, \aap, 587, A151,
  \dodoi{10.1051/0004-6361/201527395}

\bibitem[{{Kajava} {et~al.}(2019){Kajava}, {Motta}, {Sanna}, {Veledina}, {Del
  Santo}, \& {Segreto}}]{kajava2019}
{Kajava}, J.~J.~E., {Motta}, S.~E., {Sanna}, A., {et~al.} 2019, \mnras, 488,
  L18, \dodoi{10.1093/mnrasl/slz089}

\bibitem[{{Kara} {et~al.}(2019){Kara}, {Steiner}, {Fabian}, {Cackett},
  {Uttley}, {Remillard}, {Gendreau}, {Arzoumanian}, {Altamirano}, {Eikenberry},
  {Enoto}, {Homan}, {Neilsen}, \& {Stevens}}]{kara2019}
{Kara}, E., {Steiner}, J.~F., {Fabian}, A.~C., {et~al.} 2019, \nat, 565, 198,
  \dodoi{10.1038/s41586-018-0803-x}

\bibitem[{{Kawamuro} {et~al.}(2018){Kawamuro}, {Negoro}, {Yoneyama}, {Ueno},
  {Tomida}, {Ishikawa}, {Sugawara}, {Isobe}, {Shimomukai}, {Mihara},
  {Sugizaki}, {Nakahira}, {Iwakiri}, {Yatabe}, {Takao}, {Matsuoka}, {Kawai},
  {Sugita}, {Yoshii}, {Tachibana}, {Harita}, {Morita}, {Yoshida}, {Sakamoto},
  {Serino}, {Kawakubo}, {Kitaoka}, {Hashimoto}, {Tsunemi}, {Nakajima},
  {Kawase}, {Sakamaki}, {Maruyama}, {Ueda}, {Hori}, {Tanimoto}, {Oda},
  {Morita}, {Yamada}, {Tsuboi}, {Nakamura}, {Sasaki}, {Kawai}, {Sato},
  {Yamauchi}, {Hanyu}, {Hidaka}, {Yamaoka}, \& {Shidatsu}}]{kawamuro2018}
{Kawamuro}, T., {Negoro}, H., {Yoneyama}, T., {et~al.} 2018, The Astronomer's
  Telegram, 11399, 1

\bibitem[{{Kimura} \& {Done}(2019)}]{kimura2019}
{Kimura}, M., \& {Done}, C. 2019, \mnras, 482, 626,
  \dodoi{10.1093/mnras/sty2736}

\bibitem[{{Kubota} {et~al.}(1998){Kubota}, {Tanaka}, {Makishima}, {Ueda},
  {Dotani}, {Inoue}, \& {Yamaoka}}]{kubota1998}
{Kubota}, A., {Tanaka}, Y., {Makishima}, K., {et~al.} 1998, \pasj, 50, 667,
  \dodoi{10.1093/pasj/50.6.667}

\bibitem[{{Kumar} {et~al.}(2023){Kumar}, {Dewangan}, {Singh}, {Gandhi},
  {Papadakis}, {Tripathi}, \& {Mallick}}]{kumar2023}
{Kumar}, S., {Dewangan}, G.~C., {Singh}, K.~P., {et~al.} 2023, \apj, 950, 90,
  \dodoi{10.3847/1538-4357/acc941}

\bibitem[{{Lewis}(2018)}]{lewis2018}
{Lewis}, F. 2018, Robotic Telescope, Student Research and Education
  Proceedings, 1, 237, \dodoi{10.48550/arXiv.1807.00762}

\bibitem[{{Li} {et~al.}(2005){Li}, {Zimmerman}, {Narayan}, \&
  {McClintock}}]{li2005}
{Li}, L.-X., {Zimmerman}, E.~R., {Narayan}, R., \& {McClintock}, J.~E. 2005,
  \apjs, 157, 335, \dodoi{10.1086/428089}

\bibitem[{{Liska} {et~al.}(2018){Liska}, {Hesp}, {Tchekhovskoy}, {Ingram}, {van
  der Klis}, \& {Markoff}}]{liska2018}
{Liska}, M., {Hesp}, C., {Tchekhovskoy}, A., {et~al.} 2018, \mnras, 474, L81,
  \dodoi{10.1093/mnrasl/slx174}

\bibitem[{{Littlefield}(2018)}]{littlefield2018}
{Littlefield}, C. 2018, The Astronomer's Telegram, 11421, 1

\bibitem[{{Makishima} {et~al.}(1986){Makishima}, {Maejima}, {Mitsuda}, {Bradt},
  {Remillard}, {Tuohy}, {Hoshi}, \& {Nakagawa}}]{makishima1986}
{Makishima}, K., {Maejima}, Y., {Mitsuda}, K., {et~al.} 1986, \apj, 308, 635,
  \dodoi{10.1086/164534}

\bibitem[{{Mandal} {et~al.}(2018){Mandal}, {Singh}, {Stalin}, {Chandra}, \&
  {Gandhi}}]{mandal2018}
{Mandal}, A.~K., {Singh}, A., {Stalin}, C.~S., {Chandra}, S., \& {Gandhi}, P.
  2018, The Astronomer's Telegram, 11458, 1

\bibitem[{{Marino} {et~al.}(2021){Marino}, {Barnier}, {Petrucci}, {Del Santo},
  {Malzac}, {Ferreira}, {Marcel}, {Segreto}, {Motta}, {D'A{\`\i}}, {Di Salvo},
  {Guillot}, \& {Russell}}]{marino2021}
{Marino}, A., {Barnier}, S., {Petrucci}, P.~O., {et~al.} 2021, \aap, 656, A63,
  \dodoi{10.1051/0004-6361/202141146}

\bibitem[{{Markoff} {et~al.}(2020){Markoff}, {Russell}, {Dexter}, {Pfuhl},
  {Eisenhauer}, {Abuter}, {Miller-Jones}, \& {Russell}}]{markoff2020}
{Markoff}, S., {Russell}, D.~M., {Dexter}, J., {et~al.} 2020, \mnras, 495, 525,
  \dodoi{10.1093/mnras/staa1193}

\bibitem[{{McClintock} {et~al.}(2014){McClintock}, {Narayan}, \&
  {Steiner}}]{mcclintock2014}
{McClintock}, J.~E., {Narayan}, R., \& {Steiner}, J.~F. 2014, \ssr, 183, 295,
  \dodoi{10.1007/s11214-013-0003-9}

\bibitem[{{Meshcheryakov} {et~al.}(2018){Meshcheryakov}, {Tsygankov},
  {Khamitov}, {Shakura}, {Bikmaev}, {Eselevich}, {Vlasyuk}, \&
  {Pavlinsky}}]{alex2018}
{Meshcheryakov}, A.~V., {Tsygankov}, S.~S., {Khamitov}, I.~M., {et~al.} 2018,
  \mnras, 473, 3987, \dodoi{10.1093/mnras/stx2565}

\bibitem[{{Miko{\l}ajewska} {et~al.}(2022){Miko{\l}ajewska}, {Zdziarski},
  {Zi{\'o}{\l}kowski}, {Torres}, \& {Casares}}]{joanna2022}
{Miko{\l}ajewska}, J., {Zdziarski}, A.~A., {Zi{\'o}{\l}kowski}, J., {Torres},
  M. A.~P., \& {Casares}, J. 2022, \apj, 930, 9,
  \dodoi{10.3847/1538-4357/ac6099}

\bibitem[{{Mitsuda} {et~al.}(1984){Mitsuda}, {Inoue}, {Koyama}, {Makishima},
  {Matsuoka}, {Ogawara}, {Shibazaki}, {Suzuki}, {Tanaka}, \&
  {Hirano}}]{mitsuda1984}
{Mitsuda}, K., {Inoue}, H., {Koyama}, K., {et~al.} 1984, \pasj, 36, 741

\bibitem[{{Morton}(2003)}]{morton2003}
{Morton}, D.~C. 2003, \apjs, 149, 205, \dodoi{10.1086/377639}

\bibitem[{{Mu{\~n}oz-Darias} {et~al.}(2019){Mu{\~n}oz-Darias},
  {Jim{\'e}nez-Ibarra}, {Panizo-Espinar}, {Casares}, {Mata S{\'a}nchez},
  {Ponti}, {Fender}, {Buckley}, {Garnavich}, {Torres}, {Armas Padilla},
  {Charles}, {Corral-Santana}, {Kajava}, {Kotze}, {Littlefield},
  {S{\'a}nchez-Sierras}, {Steeghs}, \& {Thomas}}]{munoz2019}
{Mu{\~n}oz-Darias}, T., {Jim{\'e}nez-Ibarra}, F., {Panizo-Espinar}, G.,
  {et~al.} 2019, \apjl, 879, L4, \dodoi{10.3847/2041-8213/ab2768}

\bibitem[{{{\"O}zbey Arabac{\i}} {et~al.}(2022){{\"O}zbey Arabac{\i}},
  {Kalemci}, {Din{\c{c}}er}, {Bailyn}, {Altamirano}, \& {Ak}}]{arabaci2022}
{{\"O}zbey Arabac{\i}}, M., {Kalemci}, E., {Din{\c{c}}er}, T., {et~al.} 2022,
  \mnras, 514, 3894, \dodoi{10.1093/mnras/stac1574}

\bibitem[{{Paice} {et~al.}(2019){Paice}, {Gandhi}, {Shahbaz}, {Uttley},
  {Arzoumanian}, {Charles}, {Dhillon}, {Gendreau}, {Littlefair}, {Malzac},
  {Markoff}, {Marsh}, {Misra}, {Russell}, \& {Veledina}}]{paice2019}
{Paice}, J.~A., {Gandhi}, P., {Shahbaz}, T., {et~al.} 2019, \mnras, 490, L62,
  \dodoi{10.1093/mnrasl/slz148}

\bibitem[{{Pirbhoy} {et~al.}(2020){Pirbhoy}, {Baglio}, {Russell}, {Bramich},
  {Saikia}, {Yazeedi}, \& {Lewis}}]{pirbhoy2020}
{Pirbhoy}, S.~F., {Baglio}, M.~C., {Russell}, D.~M., {et~al.} 2020, The
  Astronomer's Telegram, 13451, 1

\bibitem[{Postma \& Leahy(2017)}]{postma2017}
Postma, J.~E., \& Leahy, D. 2017, Publications of the Astronomical Society of
  the Pacific, 129, 115002, \dodoi{10.1088/1538-3873/aa8800}

\bibitem[{{Poutanen} {et~al.}(2018){Poutanen}, {Veledina}, \&
  {Zdziarski}}]{poutanen2018}
{Poutanen}, J., {Veledina}, A., \& {Zdziarski}, A.~A. 2018, \aap, 614, A79,
  \dodoi{10.1051/0004-6361/201732345}

\bibitem[{{Poutanen} {et~al.}(2022){Poutanen}, {Veledina}, {Berdyugin},
  {Berdyugina}, {Jermak}, {Jonker}, {Kajava}, {Kosenkov}, {Kravtsov},
  {Piirola}, {Shrestha}, {Perez Torres}, \& {Tsygankov}}]{poutanen2022}
{Poutanen}, J., {Veledina}, A., {Berdyugin}, A.~V., {et~al.} 2022, Science,
  375, 874, \dodoi{10.1126/science.abl4679}

\bibitem[{{Reis} {et~al.}(2010){Reis}, {Fabian}, \& {Miller}}]{reis2010}
{Reis}, R.~C., {Fabian}, A.~C., \& {Miller}, J.~M. 2010, \mnras, 402, 836,
  \dodoi{10.1111/j.1365-2966.2009.15976.x}

\bibitem[{{Reis} {et~al.}(2009){Reis}, {Miller}, \& {Fabian}}]{reis2009}
{Reis}, R.~C., {Miller}, J.~M., \& {Fabian}, A.~C. 2009, \mnras, 395, L52,
  \dodoi{10.1111/j.1745-3933.2009.00640.x}

\bibitem[{{Remillard} \& {McClintock}(2006)}]{remillard2006}
{Remillard}, R.~A., \& {McClintock}, J.~E. 2006, \araa, 44, 49,
  \dodoi{10.1146/annurev.astro.44.051905.092532}

\bibitem[{{Remillard} {et~al.}(2021){Remillard}, {Loewenstein}, {Steiner},
  {Prigozhin}, {LaMarr}, {Enoto}, {Gendreau}, {Arzoumanian}, {Markwardt},
  {Basak}, {Stevens}, {Ray}, {Altamirano}, \& {Buisson}}]{remillard2021}
{Remillard}, R.~A., {Loewenstein}, M., {Steiner}, J.~F., {et~al.} 2021, arXiv
  e-prints, arXiv:2105.09901.
\newblock \doarXiv{2105.09901}

\bibitem[{{Rodi} {et~al.}(2021){Rodi}, {Tramacere}, {Onori}, {Bruni},
  {S{\`a}nchez-Fern{\`a}ndez}, {Fiocchi}, {Natalucci}, \&
  {Ubertini}}]{rodi2021}
{Rodi}, J., {Tramacere}, A., {Onori}, F., {et~al.} 2021, \apj, 910, 21,
  \dodoi{10.3847/1538-4357/abdfd0}

\bibitem[{{Ross} \& {Fabian}(2005)}]{ross2005}
{Ross}, R.~R., \& {Fabian}, A.~C. 2005, \mnras, 358, 211,
  \dodoi{10.1111/j.1365-2966.2005.08797.x}

\bibitem[{{Ross} \& {Fabian}(2007)}]{ross2007}
---. 2007, \mnras, 381, 1697, \dodoi{10.1111/j.1365-2966.2007.12339.x}

\bibitem[{{Russell} {et~al.}(2019{\natexlab{a}}){Russell}, {Baglio}, \&
  {Lewis}}]{russell2019}
{Russell}, D.~M., {Baglio}, M.~C., \& {Lewis}, F. 2019{\natexlab{a}}, The
  Astronomer's Telegram, 12534, 1

\bibitem[{{Russell} {et~al.}(2006){Russell}, {Fender}, {Hynes}, {Brocksopp},
  {Homan}, {Jonker}, \& {Buxton}}]{russell2006}
{Russell}, D.~M., {Fender}, R.~P., {Hynes}, R.~I., {et~al.} 2006, \mnras, 371,
  1334, \dodoi{10.1111/j.1365-2966.2006.10756.x}

\bibitem[{{Russell} {et~al.}(2019{\natexlab{b}}){Russell}, {Bramich}, {Lewis},
  {AlMannaei}, {Al Qaissieh}, {Al Qasim}, {Al Yazeedi}, {Baglio}, {Bernardini},
  {Elgalad}, {Gabuya}, {Lasota}, {Palado}, {Roche}, {Shivkumar}, {Udrescu}, \&
  {Zhang}}]{russel2019}
{Russell}, D.~M., {Bramich}, D.~M., {Lewis}, F., {et~al.} 2019{\natexlab{b}},
  Astronomische Nachrichten, 340, 278, \dodoi{10.1002/asna.201913610}

\bibitem[{{Sako} {et~al.}(2018){Sako}, {Ohsawa}, {Ichiki}, {Maehara}, {Morii},
  \& {Tanaka}}]{sako2018}
{Sako}, S., {Ohsawa}, R., {Ichiki}, M., {et~al.} 2018, The Astronomer's
  Telegram, 11426, 1

\bibitem[{{Shakura} \& {Sunyaev}(1973)}]{shakura1973}
{Shakura}, N.~I., \& {Sunyaev}, R.~A. 1973, \aap, 24, 337

\bibitem[{{Shidatsu} {et~al.}(2019){Shidatsu}, {Nakahira}, {Murata}, {Adachi},
  {Kawai}, {Ueda}, \& {Negoro}}]{shidatsu2019}
{Shidatsu}, M., {Nakahira}, S., {Murata}, K.~L., {et~al.} 2019, \apj, 874, 183,
  \dodoi{10.3847/1538-4357/ab09ff}

\bibitem[{{Shimura} \& {Takahara}(1995)}]{shimura1995}
{Shimura}, T., \& {Takahara}, F. 1995, \apj, 445, 780, \dodoi{10.1086/175740}

\bibitem[{Singh {et~al.}(2017)Singh, Stewart, Westergaard, Bhattacharayya,
  Chandra, Chitnis, Dewangan, Kothare, Mirza, Mukerjee,
  {et~al.}}]{singh2017soft}
Singh, K., Stewart, G., Westergaard, N., {et~al.} 2017, Journal of Astrophysics
  and Astronomy, 38, 1

\bibitem[{Singh {et~al.}(2014)Singh, Tandon, Agrawal, Antia, Manchanda, Yadav,
  Seetha, Ramadevi, Rao, Bhattacharya, Paul, Sreekumar, Bhattacharyya, Stewart,
  Hutchings, Annapurni, Ghosh, Murthy, Pati, Rao, Stalin, Girish,
  Sankarasubramanian, Vadawale, Bhalerao, Dewangan, Dedhia, Hingar, Katoch,
  Kothare, Mirza, Mukerjee, Shah, Shah, Mohan, Sangal, Nagabhusana, Sriram,
  Malkar, Sreekumar, Abbey, Hansford, Beardmore, Sharma, Murthy, Kulkarni,
  Meena, Babu, \& Postma}]{singh2014}
Singh, K.~P., Tandon, S.~N., Agrawal, P.~C., {et~al.} 2014, in Space Telescopes
  and Instrumentation 2014: Ultraviolet to Gamma Ray, ed. T.~Takahashi,
  J.-W.~A. den Herder, \& M.~Bautz, Vol. 9144, International Society for Optics
  and Photonics (SPIE), 517 -- 531, \dodoi{10.1117/12.2062667}

\bibitem[{Singh {et~al.}(2016)Singh, Stewart, Chandra, Mukerjee, Kotak,
  Beardmore, Chitnis, Dewangan, Bhattacharyya, Mirza,
  {et~al.}}]{singh2016orbit}
Singh, K.~P., Stewart, G.~C., Chandra, S., {et~al.} 2016, in Space Telescopes
  and Instrumentation 2016: Ultraviolet to Gamma Ray, Vol. 9905, International
  Society for Optics and Photonics, 99051E

\bibitem[{{Sunyaev} \& {Titarchuk}(1980)}]{sunyaev1980}
{Sunyaev}, R.~A., \& {Titarchuk}, L.~G. 1980, \aap, 86, 121

\bibitem[{Tandon {et~al.}(2017)Tandon, Subramaniam, Girish, Postma,
  Sankarasubramanian, Sriram, Stalin, Mondal, Sahu, Joseph,
  {et~al.}}]{tandon2017orbit}
Tandon, S., Subramaniam, A., Girish, V., {et~al.} 2017, The Astronomical
  Journal, 154, 128

\bibitem[{Tandon {et~al.}(2020)Tandon, Postma, Joseph, Devaraj, Subramaniam,
  Barve, George, Ghosh, Girish, Hutchings, {et~al.}}]{tandon2020additional}
Tandon, S., Postma, J., Joseph, P., {et~al.} 2020, The Astronomical Journal,
  159, 158

\bibitem[{{Tetarenko} {et~al.}(2020){Tetarenko}, {Dubus}, {Marcel}, {Done}, \&
  {Clavel}}]{tetarenko2020}
{Tetarenko}, B.~E., {Dubus}, G., {Marcel}, G., {Done}, C., \& {Clavel}, M.
  2020, \mnras, 495, 3666, \dodoi{10.1093/mnras/staa1367}

\bibitem[{{Thomas} {et~al.}(2022){Thomas}, {Charles}, {Buckley}, {Kotze},
  {Lasota}, {Potter}, {Steiner}, \& {Paice}}]{thomas2022}
{Thomas}, J.~K., {Charles}, P.~A., {Buckley}, D. A.~H., {et~al.} 2022, \mnras,
  509, 1062, \dodoi{10.1093/mnras/stab3033}

\bibitem[{{Tomsick} {et~al.}(2018){Tomsick}, {Parker}, {Garc{\'\i}a},
  {Yamaoka}, {Barret}, {Chiu}, {Clavel}, {Fabian}, {F{\"u}rst}, {Gandhi},
  {Grinberg}, {Miller}, {Pottschmidt}, \& {Walton}}]{tomsick2018}
{Tomsick}, J.~A., {Parker}, M.~L., {Garc{\'\i}a}, J.~A., {et~al.} 2018, \apj,
  855, 3, \dodoi{10.3847/1538-4357/aaaab1}

\bibitem[{{Tonry} {et~al.}(2018){Tonry}, {Denneau}, {Flewelling}, {Heinze},
  {Onken}, {Smartt}, {Stalder}, {Weiland}, \& {Wolf}}]{tonry2018}
{Tonry}, J.~L., {Denneau}, L., {Flewelling}, H., {et~al.} 2018, \apj, 867, 105,
  \dodoi{10.3847/1538-4357/aae386}

\bibitem[{{Torres} {et~al.}(2020){Torres}, {Casares}, {Jim{\'e}nez-Ibarra},
  {{\'A}lvarez-Hern{\'a}ndez}, {Mu{\~n}oz-Darias}, {Armas Padilla}, {Jonker},
  \& {Heida}}]{torres2020}
{Torres}, M.~A.~P., {Casares}, J., {Jim{\'e}nez-Ibarra}, F., {et~al.} 2020,
  \apjl, 893, L37, \dodoi{10.3847/2041-8213/ab863a}

\bibitem[{{Torres} {et~al.}(2019){Torres}, {Casares}, {Jim{\'e}nez-Ibarra},
  {Mu{\~n}oz-Darias}, {Armas Padilla}, {Jonker}, \& {Heida}}]{torres2019}
---. 2019, \apjl, 882, L21, \dodoi{10.3847/2041-8213/ab39df}

\bibitem[{{Trushkin} {et~al.}(2018){Trushkin}, {Nizhelskij}, {Tsybulev}, \&
  {Erkenov}}]{trushkin2018}
{Trushkin}, S.~A., {Nizhelskij}, N.~A., {Tsybulev}, P.~G., \& {Erkenov}, A.
  2018, The Astronomer's Telegram, 11439, 1

\bibitem[{{Uttley} {et~al.}(2018){Uttley}, {Gendreau}, {Markwardt},
  {Strohmayer}, {Bult}, {Arzoumanian}, {Pottschmidt}, {Ray}, {Remillard},
  {Pasham}, {Steiner}, {Neilsen}, {Homan}, {Miller}, {Iwakiri}, \&
  {Fabian}}]{uttley2018}
{Uttley}, P., {Gendreau}, K., {Markwardt}, C., {et~al.} 2018, The Astronomer's
  Telegram, 11423, 1

\bibitem[{Vadawale {et~al.}(2016)Vadawale, Rao, Bhattacharya, Bhalerao,
  Dewangan, Vibute, S., Chattopadhyay, \& Sreekumar}]{czti}
Vadawale, S.~V., Rao, A.~R., Bhattacharya, D., {et~al.} 2016, SPIE Proceedings,
  9905, 409, \dodoi{10.1117/12.2235373}

\bibitem[{{van Paradijs} \& {McClintock}(1994)}]{vanparadijs1994}
{van Paradijs}, J., \& {McClintock}, J.~E. 1994, \aap, 290, 133

\bibitem[{{Vanden Berk} {et~al.}(2001){Vanden Berk}, {Richards}, {Bauer},
  {Strauss}, {Schneider}, {Heckman}, {York}, {Hall}, {Fan}, {Knapp},
  {Anderson}, {Annis}, {Bahcall}, {Bernardi}, {Briggs}, {Brinkmann}, {Brunner},
  {Burles}, {Carey}, {Castander}, {Connolly}, {Crocker}, {Csabai}, {Doi},
  {Finkbeiner}, {Friedman}, {Frieman}, {Fukugita}, {Gunn}, {Hennessy},
  {Ivezi{\'c}}, {Kent}, {Kunszt}, {Lamb}, {Leger}, {Long}, {Loveday}, {Lupton},
  {Meiksin}, {Merelli}, {Munn}, {Newberg}, {Newcomb}, {Nichol}, {Owen}, {Pier},
  {Pope}, {Rockosi}, {Schlegel}, {Siegmund}, {Smee}, {Snir}, {Stoughton},
  {Stubbs}, {SubbaRao}, {Szalay}, {Szokoly}, {Tremonti}, {Uomoto}, {Waddell},
  {Yanny}, \& {Zheng}}]{berk2001}
{Vanden Berk}, D.~E., {Richards}, G.~T., {Bauer}, A., {et~al.} 2001, \aj, 122,
  549, \dodoi{10.1086/321167}

\bibitem[{{Verner} {et~al.}(1996){Verner}, {Ferland}, {Korista}, \&
  {Yakovlev}}]{vern1996}
{Verner}, D.~A., {Ferland}, G.~J., {Korista}, K.~T., \& {Yakovlev}, D.~G. 1996,
  \apj, 465, 487, \dodoi{10.1086/177435}

\bibitem[{Wang {et~al.}(2020)Wang, Ji, Zhang, Méndez, Qu, Maggi, Ge, Qiao,
  Tao, Zhang, Altamirano, Zhang, Ma, Lu, Li, Huang, Zheng, Chen, Chang, Tuo,
  Güngör, Song, Xu, Cao, Chen, Liu, Bu, Cai, Chen, Chen, Chen, Chen, Cui,
  Cui, Deng, Dong, Du, Fu, Gao, Gao, Gao, Gu, Guan, Guo, Han, Huo, Jia, Jiang,
  Jiang, Jin, Jin, Kong, Li, Li, Li, Li, Li, Li, Li, Li, Li, Li, Liang, Liao,
  Liu, Liu, Liu, Liu, Lu, Lu, Luo, Luo, Meng, Nang, Nie, Ou, Sai, Shang, Song,
  Sun, Tan, Wang, Wang, Wang, Wang, Wang, Wen, Wu, Wu, Wu, Xiao, Xiao, Xiong,
  Yang, Yang, Yang, Yang, Yi, Yin, You, Zhang, Zhang, Zhang, Zhang, Zhang,
  Zhang, Zhang, Zhang, Zhang, Zhang, Zhang, Zhang, Zhang, Zhang, Zhang, Zhang,
  Zhao, Zhao, Zhou, Zhou, Zhuang, Zhu, Zhu, \& Wang}]{wang2020}
Wang, Y., Ji, L., Zhang, S.~N., {et~al.} 2020, The Astrophysical Journal, 896,
  33, \dodoi{10.3847/1538-4357/ab8db4}

\bibitem[{{Weisskopf} {et~al.}(2010){Weisskopf}, {Guainazzi}, {Jahoda},
  {Shaposhnikov}, {O'Dell}, {Zavlin}, {Wilson-Hodge}, \&
  {Elsner}}]{2010ApJ...713..912W}
{Weisskopf}, M.~C., {Guainazzi}, M., {Jahoda}, K., {et~al.} 2010, \apj, 713,
  912, \dodoi{10.1088/0004-637X/713/2/912}

\bibitem[{{Wilms} {et~al.}(2000){Wilms}, {Allen}, \& {McCray}}]{wilms2000}
{Wilms}, J., {Allen}, A., \& {McCray}, R. 2000, \apj, 542, 914,
  \dodoi{10.1086/317016}

\bibitem[{{Wood} {et~al.}(2021){Wood}, {Miller-Jones}, {Homan}, {Bright},
  {Motta}, {Fender}, {Markoff}, {Belloni}, {K{\"o}rding}, {Maitra}, {Migliari},
  {Russell}, {Russell}, {Sarazin}, {Soria}, {Tetarenko}, \&
  {Tudose}}]{woods2021}
{Wood}, C.~M., {Miller-Jones}, J.~C.~A., {Homan}, J., {et~al.} 2021, \mnras,
  505, 3393, \dodoi{10.1093/mnras/stab1479}

\bibitem[{Yadav {et~al.}(2016)Yadav, Agrawal, Antia, Chauhan, Dedhia, Katoch,
  Madhwani, Manchanda, Misra, Pahari, {et~al.}}]{yadav2016large}
Yadav, J., Agrawal, P., Antia, H., {et~al.} 2016, in Space Telescopes and
  Instrumentation 2016: Ultraviolet to Gamma Ray, Vol. 9905, SPIE, 374--388

\bibitem[{{Zdziarski} \& {De Marco}(2020)}]{zdziarskidisk}
{Zdziarski}, A.~A., \& {De Marco}, B. 2020, \apjl, 896, L36,
  \dodoi{10.3847/2041-8213/ab9899}

\bibitem[{{Zdziarski} {et~al.}(2021{\natexlab{a}}){Zdziarski}, {Dzie{\l}ak},
  {De Marco}, {Szanecki}, \& {Nied{\'z}wiecki}}]{zdziarski2021}
{Zdziarski}, A.~A., {Dzie{\l}ak}, M.~A., {De Marco}, B., {Szanecki}, M., \&
  {Nied{\'z}wiecki}, A. 2021{\natexlab{a}}, \apjl, 909, L9,
  \dodoi{10.3847/2041-8213/abe7ef}

\bibitem[{{Zdziarski} {et~al.}(1996){Zdziarski}, {Johnson}, \&
  {Magdziarz}}]{zdziarski1996}
{Zdziarski}, A.~A., {Johnson}, W.~N., \& {Magdziarz}, P. 1996, \mnras, 283,
  193, \dodoi{10.1093/mnras/283.1.193}

\bibitem[{{Zdziarski} {et~al.}(2020){Zdziarski}, {Szanecki}, {Poutanen},
  {Gierli{\'n}ski}, \& {Biernacki}}]{zdziarski2020}
{Zdziarski}, A.~A., {Szanecki}, M., {Poutanen}, J., {Gierli{\'n}ski}, M., \&
  {Biernacki}, P. 2020, \mnras, 492, 5234, \dodoi{10.1093/mnras/staa159}

\bibitem[{{Zdziarski} {et~al.}(2022{\natexlab{a}}){Zdziarski}, {Tetarenko}, \&
  {Sikora}}]{tetarenko2022}
{Zdziarski}, A.~A., {Tetarenko}, A.~J., \& {Sikora}, M. 2022{\natexlab{a}},
  \apj, 925, 189, \dodoi{10.3847/1538-4357/ac38a9}

\bibitem[{{Zdziarski} {et~al.}(2022{\natexlab{b}}){Zdziarski}, {You},
  {Szanecki}, {Li}, \& {Ge}}]{zdziarski2022}
{Zdziarski}, A.~A., {You}, B., {Szanecki}, M., {Li}, X.-B., \& {Ge}, M.
  2022{\natexlab{b}}, \apj, 928, 11, \dodoi{10.3847/1538-4357/ac54a7}

\bibitem[{{Zdziarski} {et~al.}(2021{\natexlab{b}}){Zdziarski}, {Jourdain},
  {Lubi{\'n}ski}, {Szanecki}, {Nied{\'z}wiecki}, {Veledina}, {Poutanen},
  {Dzie{\l}ak}, \& {Roques}}]{zdziarski2021hybrid}
{Zdziarski}, A.~A., {Jourdain}, E., {Lubi{\'n}ski}, P., {et~al.}
  2021{\natexlab{b}}, \apjl, 914, L5, \dodoi{10.3847/2041-8213/ac0147}

\bibitem[{{Zhang} {et~al.}(2020){Zhang}, {Altamirano}, {C{\'u}neo}, {Alabarta},
  {Enoto}, {Homan}, {Remillard}, {Uttley}, {Vincentelli}, {Arzoumanian},
  {Bult}, {Gendreau}, {Markwardt}, {Sanna}, {Strohmayer}, {Steiner}, {Basak},
  {Neilsen}, \& {Tombesi}}]{zhang2020}
{Zhang}, L., {Altamirano}, D., {C{\'u}neo}, V.~A., {et~al.} 2020, \mnras, 499,
  851, \dodoi{10.1093/mnras/staa2842}

\bibitem[{{Zhao} {et~al.}(2021){Zhao}, {Gou}, {Dong}, {Tuo}, {Liao}, {Li},
  {Jia}, {Feng}, \& {Steiner}}]{zhao2021}
{Zhao}, X., {Gou}, L., {Dong}, Y., {et~al.} 2021, \apj, 916, 108,
  \dodoi{10.3847/1538-4357/ac07a9}

\bibitem[{{Zhu} {et~al.}(2017){Zhu}, {Tian}, {Li}, \& {Zhang}}]{zhu2017}
{Zhu}, H., {Tian}, W., {Li}, A., \& {Zhang}, M. 2017, \mnras, 471, 3494,
  \dodoi{10.1093/mnras/stx1580}

\bibitem[{{{\.Z}ycki} {et~al.}(1999){{\.Z}ycki}, {Done}, \&
  {Smith}}]{zycki1999}
{{\.Z}ycki}, P.~T., {Done}, C., \& {Smith}, D.~A. 1999, \mnras, 309, 561,
  \dodoi{10.1046/j.1365-8711.1999.02885.x}

\end{thebibliography}
\bibliographystyle{aasjournal}


\begin{table*}
\caption{Best-fit parameter values and the corresponding errors at 90$\%$ confidence level for the Model 1A (hard state). In \texttt{XSPEC} notation, this model reads as \texttt{tbabs*constant*(diskbb+nthcomp(1)+nthcomp(2)+reflionxhd(2)+relconv*reflionxhd(1)}). 
\label{tab:hardx}}
\medskip
\renewcommand{\arraystretch}{0.98}
\footnotesize
\centering
\begin{tabular}{clcccc}
\hline \hline
Spectral Components & Parameters & Model ~1A \\
\hline
{\sc Constant}& C$_{\rm SXT}$&$1.0^f$\\

&C$_{\rm LAXPC}$&$0.736\pm0.004$\\

&C$_{\rm CZTI-0}$&$0.788\pm0.007$\\

&C$_{\rm CZTI-1}$&$0.796\pm0.007$\\

&C$_{\rm CZTI-2}$&$0.715\pm0.007$\\

&C$_{\rm CZTI-3}$&$0.771\pm0.007$\\

&C$_{\rm NICER-1}$&$0.899\pm0.002$\\

&C$_{\rm NICER-2}$&$0.909\pm0.002$\\

{\sc tbabs} &$N\rm_{H}~(10^{22}~cm^{-2})$ &$0.13^{f}$\\

{\sc diskbb} &$kT\rm_{in}$~(keV)&$0.19\pm0.01$\\

&$N\rm_{\rm disk}$~($10^{4}$)&$32.51_{-1.56}^{+1.73}$\\


{\sc nthcomp(1)} &$\Gamma$&$1.59\pm0.01$\\

&$kT\rm_{e}~(keV)$&$15.39\pm0.35$\\

&$\rm Norm$&$4.08\pm0.03$\\

{\sc nthcomp(2)} &$\Gamma$ &$1.17\pm0.01$\\

&$kT\rm_{e}~(keV)$&$30.55\pm0.97$\\

&$\rm Norm$&$0.15\pm0.01$\\




{\sc relconv*reflionxhd(1)} &$\mathcal{R}_{\rm in}$ ($R_{\rm ISCO}$)&$50.43^{+12.30}_{-9.18}$\\

&$q$ & $3^f$ \\

&$i$ (degree) & $64^f$ \\

& $a$ &$0.998^f$ \\

&$\rm log (\rm n_e)$&$20.31\pm0.03$\\

&$\xi$&$2365.49_{-46.93}^{+64.48}$\\

&$A_{\rm Fe}$ ($A_{\rm Fe, solar}$)&$1.54\pm0.04$\\

&$\rm Norm$&$9.14\pm0.36$\\

{\sc reflionxhd(2)} &$\xi$&$488.14_{-12.04}^{+13.84}$\\

&$\rm Norm$ &$5.03\pm0.16$\\
\hline
$\chi^2/\text{d.o.f}$ &&1904.2/2388\\
\hline
\end{tabular}
\begin{flushleft}
{\bf Note:} In this table, $f$ means that the parameter is fixed during the fit and Norm refers to normalization. See Section \ref{sec:hardx} for more details.
\end{flushleft}
\end{table*}
\begin{table*}
\caption{Best-fit parameter values and the corresponding errors at 90$\%$ confidence level for the Model 1B (hard state). 
In \texttt{XSPEC}, this model reads as: \texttt{redden*(gauss(\ion{He}{2})+gauss(\ion{C}{4})+gauss(\ion{Si}{4})+bbodyrad(UV))}.
\label{tab:harduv}}
\medskip
\renewcommand{\arraystretch}{0.92}
\footnotesize
\centering
\begin{tabular}{clcc}
\hline \hline
Spectral Components & Parameters & Values\\
\hline

{\sc redden} &$E(B-V)$ &$0.17^f$\\

{\sc bbodyrad(UV)} &$kT\rm_{\rm uv}$~(eV)&$3.27\pm0.08$ &\\

&$\rm Norm$ ($10^{12}$)& $2.42_{-0.02}^{+0.09}$\\

{\sc gauss} (Si IV) &$E$~(eV)& $8.90\pm0.01$\\

&$\sigma$~($10^{-2}$ eV)&$7.63\pm1.26$\\

&$\rm Norm$&$0.45\pm0.05$\\

{\sc gauss} (C IV) &$E$~(eV)& $7.99\pm0.01$\\

&$\sigma$~($10^{-2}$ eV)&$1.68\pm0.73$\\

&$\rm Norm$&$0.34\pm0.04$\\

{\sc gauss} (He II) &$E$~(eV)& $7.55\pm0.01$\\

&$\sigma$~($10^{-2}$ eV)&$1.68\pm0.73$\\

&$\rm Norm$&$0.29\pm0.04$\\

\hline
$\chi^2/\text{d.o.f}$ &&  222.1/165\\
\hline
\end{tabular}
\begin{flushleft}
{\bf Note:} In this table, $f$ means that the parameter is fixed during the fit and Norm refers to normalization. The Gaussian width ($\sigma$) of the emission lines \ion{C}{4} and \ion{He}{2} are tied in this model. The component \texttt{bbodyrad} is normalized in the unit of $R^2_{\rm km}/D^2_{10}$, where $R_{\rm km}$ is the source radius in km. See Section \ref{sec:harduv} for more details.
\end{flushleft}
\end{table*}
\begin{table*}
\caption{Best-fit parameter values and the corresponding errors at 90$\%$ confidence level for the Model 1C (hard state). In \texttt{XSPEC}, this model reads as: \texttt{tbabs*redden*constant*(diskbb+nthcomp(1)+nthcomp(2)+reflionxhd(2)+relconv*reflionxhd(1)\\+gauss(\ion{He}{2})+gauss(\ion{C}{4})+gauss(\ion{Si}{4})+bbodyrad(UV)+bbodyrad(optical)}).
\label{tab:hardtotal}}
\medskip
\footnotesize
\centering
\begin{tabular}{clcc}
\hline \hline
Spectral Components & Parameters & Values\\
\hline

{\sc Constant}& C$_{\rm SXT}$&$1.0^f$\\

&C$_{\rm LAXPC}$&$0.737\pm0.004$\\

&C$_{\rm CZTI-0}$&$0.789\pm0.007$\\

&C$_{\rm CZTI-1}$&$0.798\pm0.007$\\

&C$_{\rm CZTI-2}$&$0.717\pm0.007$\\

&C$_{\rm CZTI-3}$&$0.772\pm0.007$\\

&C$_{\rm NICER-1}$&$0.899\pm0.002$\\

&C$_{\rm NICER-2}$&$0.909\pm0.002$\\

{\sc redden} &$E(B-V)$ &$0.17^f$\\

{\sc tbabs} &$N\rm_{H}~(10^{22}~cm^{-2})$ &$0.13^{f}$\\

{\sc diskbb} &$kT\rm_{in}$~(keV)& $0.19\pm0.01$\\

&$N\rm_{\rm disk}$~($10^{5}$)& $3.26\pm0.18$\\

{\sc nthcomp(1)} &$\Gamma$&$1.60\pm0.01$\\

&$kT\rm_{e}~(keV)$&$15.37\pm0.34$\\

&$\rm Norm$&$4.09\pm0.08$\\

{\sc nthcomp(2)} &$\Gamma$&$1.16\pm0.01$\\

&$kT\rm_{e}~(keV)$&$30.41^{+0.84}_{+1.18}$\\

&$\rm Norm$&$0.14\pm0.02$\\

{\sc relconv*reflionxhd(1)} &$\mathcal{R}_{\rm in}$ ($R_{\rm ISCO}$)&$49.39^{+12.48}_{+8.70}$\\

&$i$ (degree) & $64^f$\\

& $a$ &  $0.998^f$\\

&$\rm log (\rm n_e)$&$20.31\pm0.02$\\

&$\xi$&$2358.39_{-45.02}^{+391.54}$\\

&$A_{\rm Fe}$ ($A_{\rm Fe, solar}$)&$1.56\pm0.04$\\

&$\rm Norm$&$9.21\pm0.44$\\

{\sc reflionxhd(2)} &$\xi$&$486.02_{-09.54}^{+82.31}$\\

&$\rm Norm$ &$4.95\pm0.15$\\

{\sc bbodyrad(UV)}  &$kT\rm_{\rm uv}$~(eV)&$3.27^f$ &\\

&$\rm Norm$ ($10^{11}$)& $22.30_{-0.17}^{+0.15}$\\

{\sc bbodyrad(optical)} &$kT\rm_{\rm optical}$~(eV)&$0.80\pm0.03$\\

&$\rm Norm$ ($10^{14}$)& $1.31\pm0.12$\\

{\sc gauss} (\ion{Si}{4}) &$\rm Norm$&$0.49\pm0.05$\\

{\sc gauss} (\ion{C}{4}) &$\rm Norm$&$0.33\pm0.04$\\

{\sc gauss} (\ion{He}{2}) &$\rm Norm$&$0.26\pm0.04$\\

\hline
Flux (0.1-200.0 keV)  &&$18.87$\\
Flux (0.5-10.0 eV)  &&$0.18$\\
$\chi^2/\text{d.o.f}$ && 2157.7/2562\\
\hline
\end{tabular}
\begin{flushleft}
{\bf Note:} In this table, $f$ means that the parameter is fixed during the fit and Norm refers to normalization.  All the unabsorbed fluxes are in units of $10^{-8}\ \rm erg~cm^{-2}s^{-1}$. In this model, we fix $kT_{\rm uv}$, the energy and width of emission lines (described by Gaussian line profiles) at their best-fit values as found in Model 1B. See Section \ref{sec:hardtotal} for more details.
\end{flushleft}
\end{table*}
\begin{table*}
\caption{Best-fit parameter values and the corresponding errors at 90$\%$ confidence level for the Models 2A, 2B, and 2C (soft state). 
In \texttt{XSPEC} notation, these models read as follows, Model 2A: \texttt{tbabs*constant*thcomp*}(\texttt{diskbb+bbodyrad}), Model 2B: \texttt{tbabs*constant*thcomp*}(\texttt{kerrbb+bbodyrad}), and Model 2C: \texttt{tbabs*constant}*(\texttt{diskir+bbodyrad}).
\label{tab:softx}}
\medskip
\renewcommand{\arraystretch}{0.95}
\footnotesize
\centering
\begin{tabular}{clccccc}
\hline \hline
Spectral Components & Parameters & Model ~2A &\multicolumn{2}{c}{Model ~2B}& Model ~2C\\
 & & & $i$ (Free) & $i$ (Fixed) & \\ 
\hline
{\sc Constant}& C$_{\rm SXT}$&$1.0^f$&$1.0^f$ &$1.0^f$ &$1.0^f$\\

&C$_{\rm LAXPC}$&$1.21\pm0.05$&$1.26\pm0.05$&$1.25\pm0.06$&$1.21\pm0.06$\\

{\sc tbabs} &$N\rm_{H}~(10^{20}~cm^{-2})$ &$8.03_{-0.67}^{+0.98}$&$11.00_{-0.87}^{+0.50}$&$10.34\pm0.86$&$8.07_{-0.71}^{+0.92}$\\

{\sc diskbb} &$kT\rm_{in}$~(keV)&$0.58\pm0.02$ &&&\\

&$N\rm_{\rm disk}$ ($10^{4}$)& $3.10_{-0.27}^{+0.50}$ &&&\\


{\sc bbodyrad} &$kT\rm_{\rm BB}$&$0.79\pm0.02$&$0.87\pm0.04$&$0.84\pm0.04$&$0.79\pm0.02$\\

&$\rm Norm$ ($10^{3}$) &$2.89_{-0.49}^{+0.85}$&$0.90_{-0.22}^{+0.64}$&$1.31_{-0.53}^{+1.16}$&$2.91_{-0.53}^{+0.78}$\\

{\sc thcomp} &$\Gamma$&$2.19\pm0.05$ &$2.15\pm0.04$&$2.16_{-0.11}^{+0.05}$&\\

&$kT\rm_{\rm e}~(keV)$&$>36.5$&$>56.2$&$>17.04$&\\

&$cov\_frac~(10^{-3})$&$4.83\pm0.99$&$3.90\pm0.84$&$4.11\pm0.98$&\\

{\sc kerrbb} &$M\ (\rm M_{\rm \sun})$&&$9.73_{-2.52}^{+2.25}$ &$8.72_{-2.82}^{+3.28p}$&\\

&$i\ (\rm degree)$&&$46.83_{-10.14}^{+4.68}$&$64^f$&\\

&$a$&&$0.998_{-0.157}^{+0.000p}$&$0.85_{-0.25}^{+0.10}$&\\

&$\kappa$& & $1.7^f$&$1.7^f$& \\

&$\dot{M}$~$(10^{17}\rm g/s)$&&$0.45_{-0.28}^{+0.36}$&$0.80_{-0.68}^{+11.30}$&\\

&$\rm Norm$&&$7.11_{-3.03}^{+58.31}$ &$9.79_{-2.94}^{+11.07}$\\

{\sc diskir} &$kT_{\rm disk}$~(keV)&& & &$0.58\pm0.02$\\

&$\Gamma$&&&&$2.19\pm0.05$\\

&$kT\rm_{e}~(keV)$&&&&$>41.6$\\

&$L_c/L_d$~($10^{-3}$)&&&&$8.76\pm0.76$\\

&$\rm Norm$~($10^{4}$)&&&&$3.08_{-0.29}^{+0.46}$\\

\hline
$\chi^2/\text{d.o.f}$ && 490.1/413&464.1/410&467.2/411&490.1/413\\
\hline
\end{tabular}
\begin{flushleft}
{\bf Note:} In this table, $f$ means that the parameter is fixed during the fit, $p$ denotes that the parameter is pegged at its limit, and Norm refers to normalization. See Section \ref{sec:softx} for more details.
\end{flushleft}
\end{table*}
\begin{table*}
\caption{Best-fit parameter values and the corresponding errors at 90$\%$ confidence level for the Model 2D (soft state). 
In \texttt{XSPEC}, this model reads as: \texttt{redden*(bbodyrad(UV)+gauss(\ion{N}{4})+gauss(\ion{He}{2})+gauss(\ion{C}{4})
+gauss(\ion{N}{4]})+gauss(\ion{Si}{4}))}. \label{tab:softuv}}
\medskip
\renewcommand{\arraystretch}{0.9}
\footnotesize
\centering
\begin{tabular}{clcc}
\hline \hline
Spectral Components & Parameters & Values\\
\hline

{\sc redden} &$E(B-V)$ &$0.12^f$\\

{\sc bbodyrad(UV)} &$kT\rm_{\rm uv}$~(eV)&$3.87\pm0.24$ &\\

&$\rm Norm$ ($10^{11}$)& $3.69^{+0.61}_{-0.44}$\\

{\sc gauss} (\ion{Si}{4}) &$E$~(eV)& $8.88\pm0.02$\\

&$\sigma$~($10^{-2}$ eV)&$8.52\pm1.73$\\

&$\rm Norm$&$0.22\pm0.03$\\

{\sc gauss} (\ion{N}{4]}) &$E$~(eV)& $8.34\pm0.02$\\

&$\sigma$~($10^{-2}$ eV)&$3.19\pm0.56$\\

&$\rm Norm$&$0.07\pm0.02$\\

{\sc gauss} (\ion{C}{4}) &$E$~(eV)& $7.99\pm0.01$\\

&$\sigma$~($10^{-2}$ eV)&$3.19\pm0.56$\\

&$\rm Norm$&$0.16\pm0.02$\\

{\sc gauss} (\ion{He}{2}) &$E$~(eV)& $7.56\pm0.01$\\

&$\sigma$~($10^{-2}$ eV)&$3.19\pm0.56$\\

&$\rm Norm$&$0.13\pm0.02$\\

{\sc gauss} (\ion{N}{4}) &$E$~(eV)& $7.22\pm0.02$\\

&$\sigma$~($10^{-2}$ eV)&$3.19\pm0.56$\\

&$\rm Norm$&$0.08\pm0.02$\\

\hline
$\chi^2/\text{d.o.f}$ && 408.6/333\\
\hline
\end{tabular}
\begin{flushleft}
{\bf Note:} In this table, $f$ means that the parameter is fixed during the fit and Norm refers to normalization. The Gaussian width ($\sigma$) of the emission lines N IV], C IV, He II, and N IV are tied in this model. See Section \ref{sec:softuv} for more details.
\end{flushleft}
\end{table*}
\begin{table*}
\caption{Best-fit parameter values and the corresponding errors at 90$\%$ confidence level for the Models 2E, 2F, and 2G (soft state). 
In \texttt{XSPEC} notation, these models read as follows, Model 2E: \texttt{tbabs*redden*con*(bbodyrad(UV)+bbodyrad(optical)+gauss(\ion{N}{4})+gauss(\ion{He}{2})+gauss(\ion{C}{4})+gauss(\ion{N}{4]})+gauss(Si IV)+thcomp*(bbodyrad+diskbb))}, Model 2F: \texttt{tbabs*redden*con*(bbodyrad(UV)+bbodyrad(optical)+gauss(N IV)+gauss(\ion{He}{2})+gauss(\ion{C}{4})+gauss(\ion{N}{4]})+gauss(Si IV)+thcomp*(bbodyrad+kerrbb))}, and Model 2G: \texttt{tbabs*redden*con*(gauss(\ion{He}{2})+gauss(\ion{C}{4})+gauss(\ion{N}{4]})+gauss(\ion{Si}{4})+bbodyrad+diskir)}.\label{tab:softtotal}}
\medskip
\footnotesize
\centering
\begin{tabular}{clcccc}
\hline 
\hline
Spectral Components & Parameters & Model ~2E  & Model ~2F & Model ~2G \\
\hline
{\sc Constant}& C$_{\rm SXT}$&$1.0^f$&$1.0^f$&$1.0^f$\\

&C$_{\rm LAXPC}$&$1.18\pm0.05$&$1.28\pm0.05$&$1.18\pm0.03$\\

{\sc tbabs} &$N\rm_{H}~(10^{20}~cm^{-2})$&$9.0^f$&$9.0^f$&$9.0^f$\\

{\sc redden} &$E(B-V)$ &$0.12^f$&$0.12^f$&$0.12^f$\\

{\sc bbodyrad} &$kT\rm_{\rm BB}$&$0.79\pm0.01$&$0.89\pm0.03$&$0.79\pm0.01$\\

&$\rm Norm~(10^3)$&$3.41_{-0.39}^{+0.57}$&$0.68_{-0.28}^{+0.35}$&$3.35_{-0.20}^{+0.40}$\\

{\sc diskbb} &$kT\rm_{\rm in}$&$0.56\pm0.01$&&\\

&$N\rm_{\rm disk}$ ($10^{4}$)&$3.49_{-0.18}^{+0.26}$&&\\

{\sc thcomp} &$\Gamma$&$2.20\pm0.05$&$2.14_{-0.15}^{+0.05}$&&\\

&$kT\rm_{\rm e}~(keV)$&$>40.96$&$>17.42$&&\\

&$\rm cov\_frac$~($10^{-3}$)&$5.02\pm0.92$&$3.86\pm1.01$&&\\

{\sc kerrbb} &$M\ (\rm M_{\rm \sun})$&&$6.75^f$ &&\\

&$i\ (\rm degree)$&&$64^f$&&\\

&$a$&&$0.75^f$ &&\\

&$\kappa$& & $1.7^f$& \\

&$\dot{M}$~$(10^{17}\rm g/s)$&&$1.48_{-0.09}^{+0.13}$ &&\\

&$\rm Norm$&&$8.68\pm0.57$ &&\\

{\sc diskir} &$kT_{\rm disk}$~(keV)&&&$0.57\pm0.01$\\

&$\Gamma$&&&$2.20\pm0.04$\\

&$kT\rm_{e}~(keV)$&&&$>45.3$\\

&$L_c/L_d$~($10^{-3}$)&&&$9.11\pm0.56$\\

&$f_{\rm out}$~($10^{-3}$)&&&$1.98\pm0.03$\\

&$\rm log(r_{\rm out})$&&&$4.38\pm0.02$\\

&$\rm Norm$~($10^{4}$)&&&$3.42_{-0.10}^{+0.19}$\\

{\sc bbodyrad(UV)} &$kT\rm_{\rm uv}$~(eV)&$3.87^f$ & $3.87^f$ & \\

&$\rm Norm$ ($10^{11}$)&$2.55_{-0.05}^{+0.04}$ &$2.32\pm0.05$&\\

{\sc bbodyrad(optical)} &$kT\rm_{\rm optical}$~(eV)&$0.75\pm0.04$&$0.75\pm0.04$ &\\

&$\rm Norm$ ($10^{13}$)&$3.51_{-0.50}^{+0.55}$&$3.35_{-0.50}^{+0.56}$&\\

{\sc gauss} (\ion{Si}{4}) &$\rm Norm$&$0.25\pm0.03$&$0.26\pm0.03$&$0.38\pm0.03$\\

{\sc gauss} (\ion{N}{4]}) &$\rm Norm$ &$0.09\pm0.02$&$0.09\pm0.02$&$0.11\pm0.02$&\\

{\sc gauss} (\ion{C}{4}) &$\rm Norm$&$0.18\pm0.02$&$0.18\pm0.02$&$0.16\pm0.02$\\

{\sc gauss} (\ion{He}{2}) &$\rm Norm$&$0.14\pm0.02$&$0.14\pm0.02$&$0.09\pm0.02$\\

{\sc gauss} (\ion{N}{4}) &$\rm Norm$&$0.07\pm0.03$&$0.07\pm0.03$&&\\

\hline
Flux (0.1-200.0 keV)& &10.50&11.53&10.56\\
Flux (0.5-10.0 eV)& &0.049&0.053&0.056\\
$\chi^2/\text{d.o.f}$& &900.5/756&881.9/756&1230.6/758\\
\hline
\end{tabular}
\begin{flushleft}
{\bf Note:} In this table, Norm refers to the normalization of the associated spectral component, $f$ means that the parameter is fixed during the fit, and $p$ denotes that the parameter is pegged at its limit. All the unabsorbed fluxes are in units of $10^{-8}\ \rm erg~cm^{-2}s^{-1}$. In this model, we fix $kT_{\rm uv}$, the energy and width of emission lines (described by Gaussian line profiles) at their best-fit values as found in Model 2D.  See Section \ref{sec:softtotal} for more details.
\end{flushleft}
\end{table*}


\end{document}